\documentclass[3p,times,twocolumn]{elsarticle}

\usepackage{ecrc}

\volume{00}

\firstpage{1}

\journalname{Nuclear Physics B Proceedings Supplement}

\runauth{R.J.\ Furnstahl}

\jid{nuphbp}

\jnltitlelogo{Nuclear Physics B Proceedings Supplement}

\usepackage{amssymb}
\usepackage{amsmath}
\usepackage{bm}
\usepackage[figuresright]{rotating}

\newcommand{\Lra}[0]{$\Longrightarrow$\ }

\newcommand{\beq}[0]{\begin{equation}}
\newcommand{\eeq}[0]{\end{equation}}
\newcommand{\beqfn}[0]{\begin{footnotesize}\[}
\newcommand{\eeqfn}[0]{\]\end{footnotesize}}
\newcommand{\beqa}[0]{\begin{small}\begin{eqnarray*}}
\newcommand{\eeqa}[0]{\end{eqnarray*}\end{small}}
\newcommand{\bsm}[0]{\begin{small}}
\newcommand{\esm}[0]{\end{small}}
\newcommand{\bfn}[0]{\begin{footnotesize}}
\newcommand{\efn}[0]{\end{footnotesize}}
\newcommand{\bst}[0]{\begin{small}}
\newcommand{\est}[0]{\end{small}}

\newcommand{\be}{\begin{enumerate}}
\newcommand{\ee}{\end{enumerate}}
\newcommand{\bi}{\begin{itemize}}
\newcommand{\ei}{\end{itemize}}
\newcommand{\I}{\item}

\newcommand{\vlowk}{V_{{\rm low\,}k}}

\newcommand{\ad}{a^{\dagger}}

\newcommand{\flow}{s}
\newcommand{\Hzero}{\Trel}

\newcommand{\bea}{\begin{eqnarray}}
\newcommand{\eea}{\end{eqnarray}}
\newcommand{\fmi}{\mbox{\,fm}^{-1}}

\newcommand{\la}{\langle}
\newcommand{\ra}{\rangle}

\newcommand{\amps}[1]{\!\!\!&#1&\!\!\!}
\newcommand{\ampseq}{\amps{=}}
\newcommand{\Tr}{\rm Tr\,}
\newcommand{\Nmax}{N_{\rm max}}
\newcommand{\Trel}{T_{\rm rel}}

\newcommand{\vnn}{V_{\rm NN}}

\newcommand{\lm}{\Lambda}
\newcommand{\Vtwo}[2]{V_{#1#2}}
\newcommand{\Vthree}{V_{123}}
\newcommand{\Ttwo}[2]{T_{#1#2}}

\newcommand{\nthreelo}{N$^3$LO}
\newcommand{\wt}{\widetilde}

\newcommand{\xvec}{{\bf x}}
\newcommand{\yvec}{{\bf y}}
\newcommand{\kf}[0]{k_{\rm F}}

\begin{document}

\begin{frontmatter}


\ead{furnstahl.1@osu.edu}

\dochead{}

\title{The Renormalization Group in Nuclear Physics}


\author{R. J. Furnstahl}

\address{Department of Physics, Ohio State University, Columbus, OH 43210, USA}

\begin{abstract}
Modern techniques of the renormalization group (RG) combined with effective field theory (EFT) methods are revolutionizing nuclear many-body physics. In these lectures we will explore the motivation for RG in low-energy nuclear systems and its implementation in systems ranging from the deuteron to neutron stars, both formally and in practice.  Flow equation approaches applied to Hamiltonians both in free space and in the medium will be emphasized.
This is a conceptually simple technique to transform interactions
to more perturbative and universal forms.
An unavoidable complication for nuclear systems 
from both the EFT and flow equation perspective
is the need to treat many-body forces and operators, so we will
consider these aspects in some detail.  We'll finish with 
a survey of current developments and open problems in nuclear RG.  
\end{abstract}


\end{frontmatter}

% \linenumbers


\section{Overview}
\label{sec:overview}

The topic of these lectures is the use of renormalization group (RG)
methods in low-energy nuclear systems, which include the full range
of atomic nuclei as well as astrophysical systems such as neutron
stars.
We will examine why the RG has become an increasingly useful tool
for nuclear physics theory over the last ten years and consider
how to apply  RG technology both formally and in practice.  Of particular emphasis will be flow equation approaches applied to Hamiltonians both in free space and in the medium, which are an accessible but powerful method to make nuclear physics computationally more
like quantum chemistry.  We will see how interactions are evolved to increasingly universal form and become more amenable to perturbative methods. A key element in nuclear systems is the role of many-body forces and operators; dealing with their evolution is an important on-going challenge. 

The expected background for these lectures is a thorough knowledge of nonrelativistic quantum mechanics, including scattering, the basics of quantum field theory,
and linear algebra (it's all matrices!).
We will not assume a knowledge of nuclear structure or reactions,
or even many-body physics beyond Hartree-Fock.
No advanced computing experience is assumed (although Mathematica or MATLAB
knowledge will be very helpful in exploring simple RG examples). 

By necessity, we will only scratch the surface in these lectures.
For a thorough treatment of flow equations for many-body systems
not including nuclei,
see the book by Kehrein~\cite{Kehrein:2006}.  For more details on 
applications
of flow-equation and similar renormalization group methods to
low-energy nuclear physics, the review article~\cite{Bogner:2009bt} 
and references therein are recommended.

%%%%%%%%%%%%%%%%%%%%%%%%%%%%%%%%%%%%%%%%%%%%%%%%%%%%%%%%%%%%%%%%%%%%%%%%
%%%%%%%%%%%%%%%%%%%%%%%%%%%%%%%%%%%%%%%%%%%%%%%%%%%%%%%%%%%%%%%%%%%%%%%%

\section{Atomic nuclei at low resolution via RG}
\label{sec:resolution}

\subsection{Goals and scope of low-energy nuclear physics}

\begin{figure}[!tbp]
   \includegraphics[width=3in]{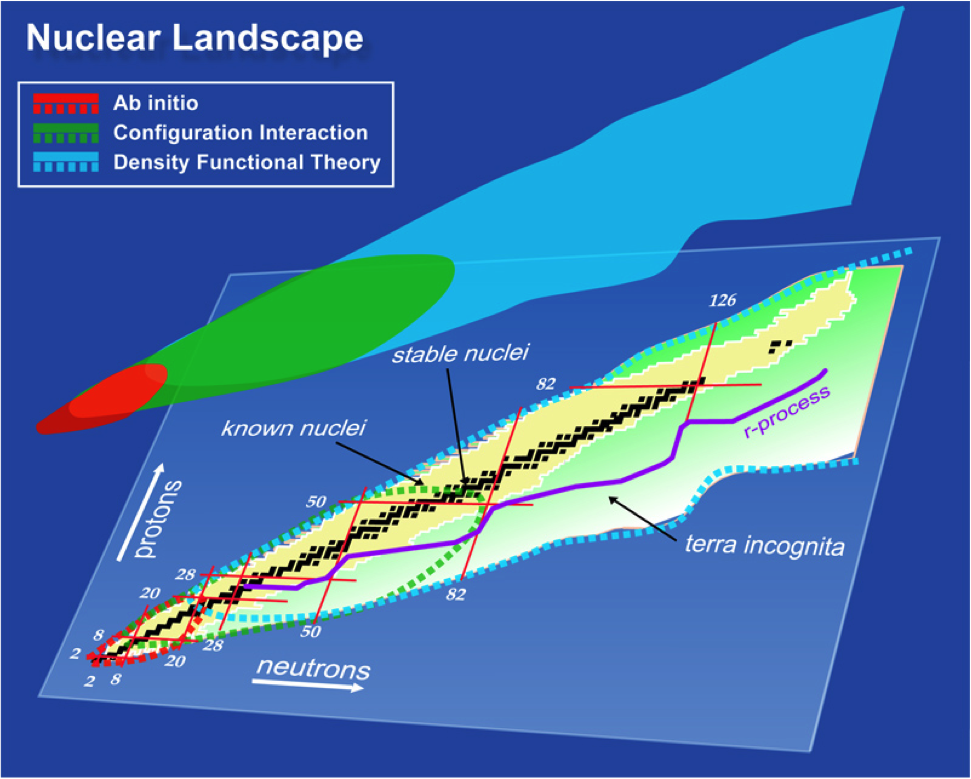}
   \caption{The nuclear landscape.  A nuclide is specified by the number
   of protons and neutrons~\cite{LRP:2007}.}
   \label{fig:landscape}
\end{figure}

The playing field for low-energy nuclear physics is the table of the nuclides,
shown in Fig.~\ref{fig:landscape}.
There are several hundred stable nuclei (black squares) 
but also several thousand \emph{unstable}
nuclei are known through experimental measurements.  The total 
number of nuclides is still unknown (see the region marked ``terra incognita''), with theoretical
estimates suggesting it could be as high as ten thousand!
These unstable nuclei are the object of scrutiny for new and planned
experimental facilities around the world. 
The challenge for low-energy nuclear theory is to describe their structure and
reactions.
We'll return in the final lecture to discuss the overlapping regions where
the theoretical many-body methods listed in the figure can be applied.

Let's start with some of the questions that drive low-energy nuclear physics
research.  These include general questions about the physics of
nuclei~\cite{LRP:2007}:
  \bi
  \I
  How do protons and neutrons make stable nuclei and rare isotopes?
  Where are the limits?
  %\I
  %What are the heaviest nuclei that can exist?
  \I
  What is the equation of state of nucleonic matter?
  \I
  What is the origin of simple patterns observed in complex nuclei? 
  \I How do we describe fission, fusion, and other nuclear reactions?
  \ei
These topics inform and are in turn illuminated by applications
to other fields, such as astrophysics, where one can ask:
  \bi
  \I
  { How did the elements from iron to uranium originate? }
  \I
  { How do stars explode? }
  \I
  { What is the nature of neutron star matter?}
  \ei
or of fundamental symmetries:    
  \bi
  \I Why is there now more matter than antimatter in the
 universe?
  \I What is the nature of the neutrinos,
  what are their masses, 
  and how have they shaped the evolution of the universe?
  \ei
Finally, there are applications, for which
we are led to ask:  How can our knowledge of nuclei and our ability to produce them  benefit  humankind?  The impact is very broad, encompassing the
Life Sciences, Material Sciences, Nuclear Energy, and National Security.

\begin{figure}[!tbp]
   \includegraphics[width=3in]{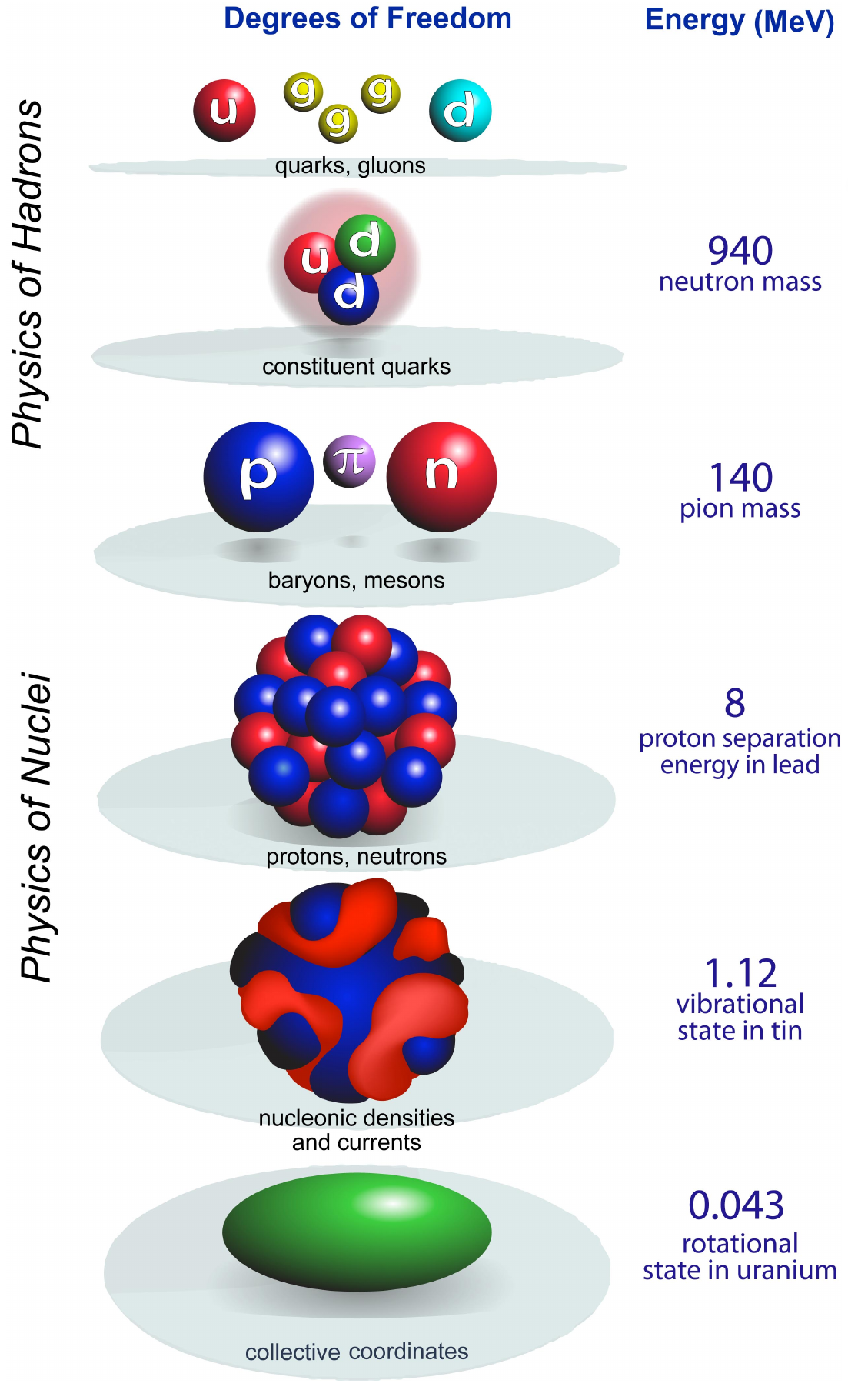}
   \caption{Hierarchy of nuclear degrees of freedom and associated
   energy scales~\cite{LRP:2007}.}
   \label{fig:scales}
\end{figure}

In Figure~\ref{fig:scales}, the energy scales of nuclear physics are illustrated.  There is an extended hierarchy, which is a challenge, but also an opportunity to make use of effective field theory (EFT) and renormalization group (RG) techniques.  The ratio of scales can become 
an expansion parameter, leading to a systematic treatment at lower
energies.  
The progression from top to bottom can be viewed as a reduction in resolution.
In these lectures, our focus is on the intermediate region only, where protons and neutrons are the relevant degrees of freedom.  But even within this 
limited scope, the concept of reducing resolution by RG methods is extremely powerful.

\subsection{Lowering the resolution with RG}

\begin{figure}[t!]
   \includegraphics[width=3in]{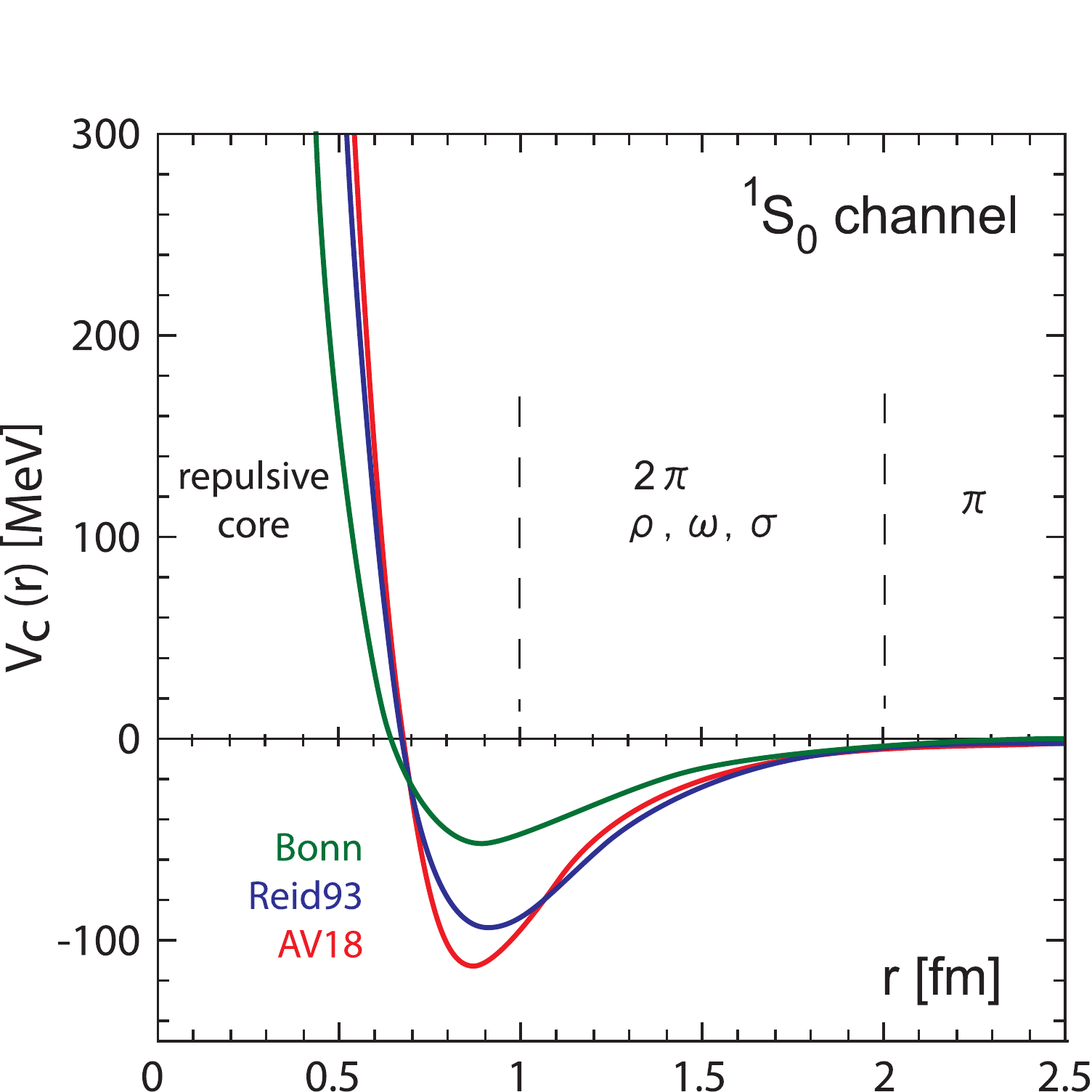}
   \caption{Some phenomenological potentials that accurately describe
   proton-neutron scattering up to laboratory energies
   of 300\,MeV~\cite{Aoki:2008hh}.}
   \label{fig:phenpots}
\end{figure}

What do we mean by resolution?  Even the general public these days
is familiar with the concept of digital resolution for computer screens,
cellphones, televisions.  High resolution is associated with more
pixels, which follows because as pixel size becomes small compared to
characteristic scales in an image, greater detail is seen.  In our
discussion, we associate resolution with the Fourier transform space and 
the phenomenon of diffraction.

Recall the basic physics:  if the wavelength of light is comparable
to or larger than an aperture, then diffraction is significant.  If there are
two sources, we say we can resolve them if the diffracted images
don't overlap too much.  For a fixed angle between sources or details
in the object being observed, we find that the wavelength determines
whether or not we resolve the details.
Being unable to resolve details at long wavelength is generally
considered to be a disadvantage (e.g., for astronomical observations), 
but we turn it to an advantage.

A fundamental principle of \emph{any} effective low-energy description
(not restricted to nuclear physics!) is that if a system is probed
at low energies, fine details are not resolved, and one can instead use
low-energy variables for low-energy processes.  Renormalization theory
tells us that the short-distance structure can be replaced by something
simpler without distorting low-energy \emph{observables}.  The familiar
analog from classical electrodynamics is the replacement of a complicated
charge or current distribution with a truncated multipole expansion.
In the quantum case, the replacement can
be done by constructing a model, or in a systematic way using
effective field theory.  We emphasize that while observable quantities
(such as cross sections) do not change, the physics interpretation
can (and generally does) change with resolution.
What if there is no external probe?  Then the particles still probe each other
with resolution set by their de Broglie wavelengths.  Low-density
nuclear systems would seem to imply low resolution.
But the picture is complicated by the nature of traditional
internucleon potentials.

\begin{figure}[t]
   \includegraphics[width=3in]{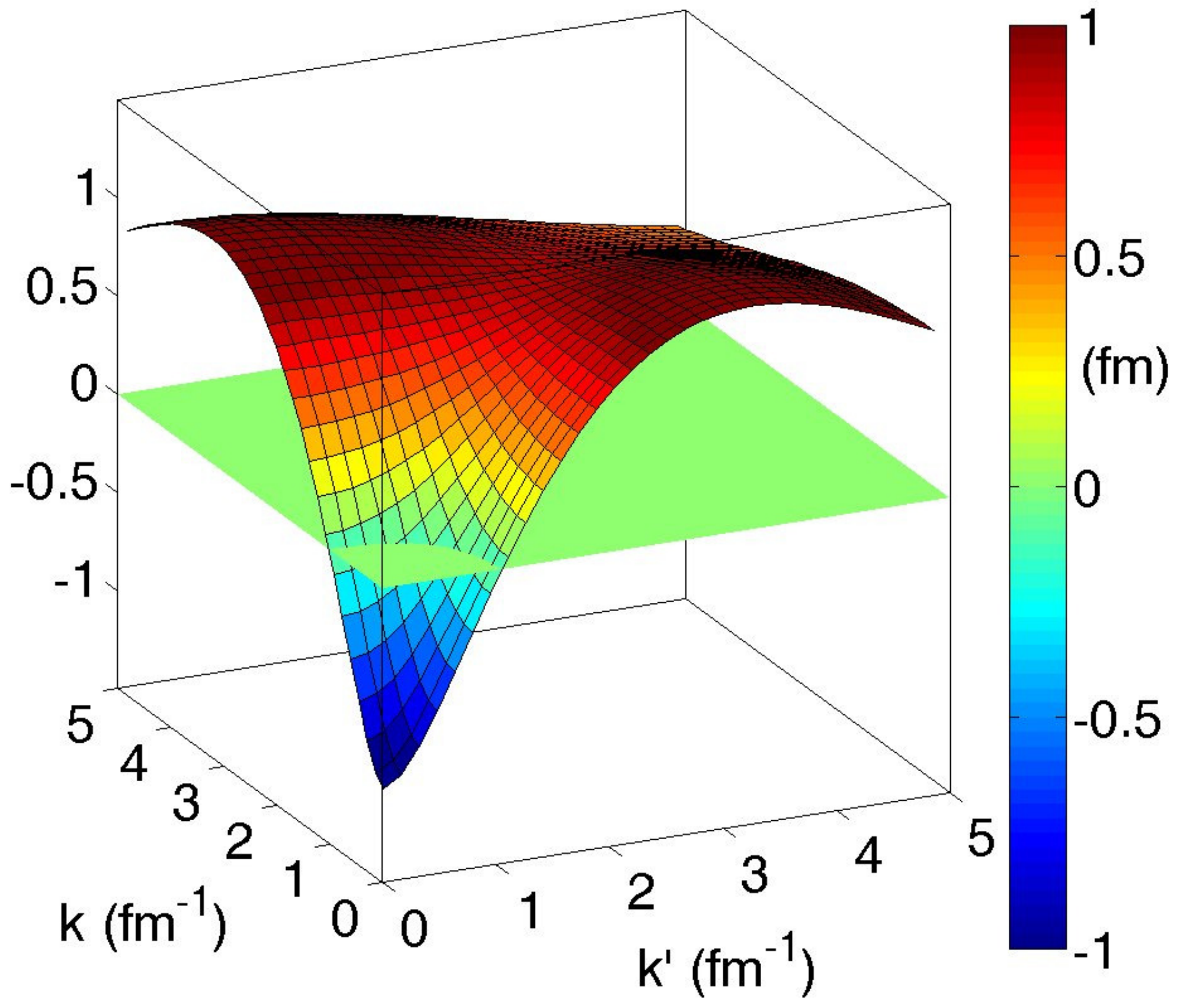}
   \caption{Momentum space representation of the Argonne $v_{18}$ (AV18)
   NN potential in the $^1$S$_0$ channel~\cite{Bogner:2009bt}.}
   \label{fig:vnnsurf}
\end{figure}

\begin{figure}[t]
   \includegraphics[width=3in]{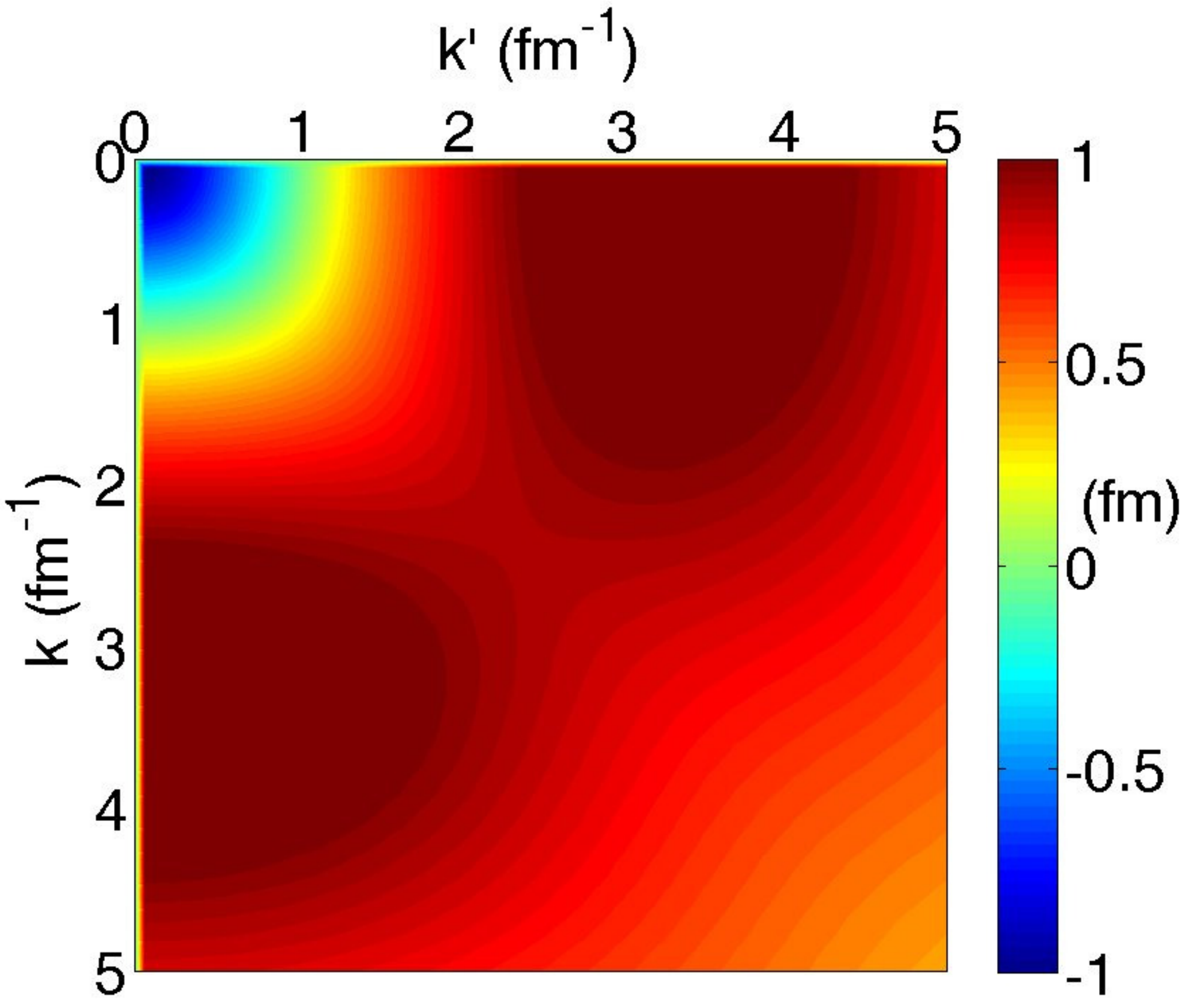}
   \caption{Alternative momentum space representation of the AV18 potential
   in the $^1$S$_0$ channel~\cite{Bogner:2009bt}.}
   \label{fig:vnncntr}
\end{figure}

Figure~\ref{fig:phenpots} shows several phenomenological potentials
that reproduce nucleon-nucleon scattering phase shifts up to about
300\,MeV in lab energy.  They are characterized by a long-range
attractive tail from one-pion exchange, intermediate attraction,
and a strongly repulsive short-range ``core''.
For our purposes, the partial-wave momentum-space representation, such as
\beq
        \langle k | V_{L=0} | k' \rangle 
  \propto \int\! r^2\,dr  \, j_0(kr)\, V(r)\, j_0(k'r)
\eeq
%V_{L=0}(k,k')
for S-waves is more useful.
Here $k$ and $k'$ are the relative momenta of the two nucleons.
This is shown for the AV18 potential in Fig.~\ref{fig:vnnsurf}
in the $^1$S$_0$ channel.
(The spin and isospin dependence of the nuclear interactions is very
important, unlike the situation in quantum chemistry.)  
In the momentum basis, the potentials in Fig.~\ref{fig:phenpots}
are no longer diagonal, so we
need three-dimensional information, but we generally use the flat
contour representation of the same information, as in
Fig.~\ref{fig:vnncntr}.  The RG evolution of potentials will be
visualized as changes in such pictures.

We work in units for which $\hbar = c = 1$.  Then the typical
relative momentum in the Fermi sea of any large nucleus
is of order $1\fmi$ or 200\,MeV.  However, it is evident from
Figs.~\ref{fig:vnnsurf} and \ref{fig:vnncntr} that there are large matrix elements connecting
such momenta to much larger momenta.  This is directly associated with
the repulsive core of the potential.  For our discussion,
we will adopt $2\fmi$ as the (arbitrary but reasonable)
dividing line between low and high momentum for nuclei.

\begin{figure}
   \includegraphics[width=3in]{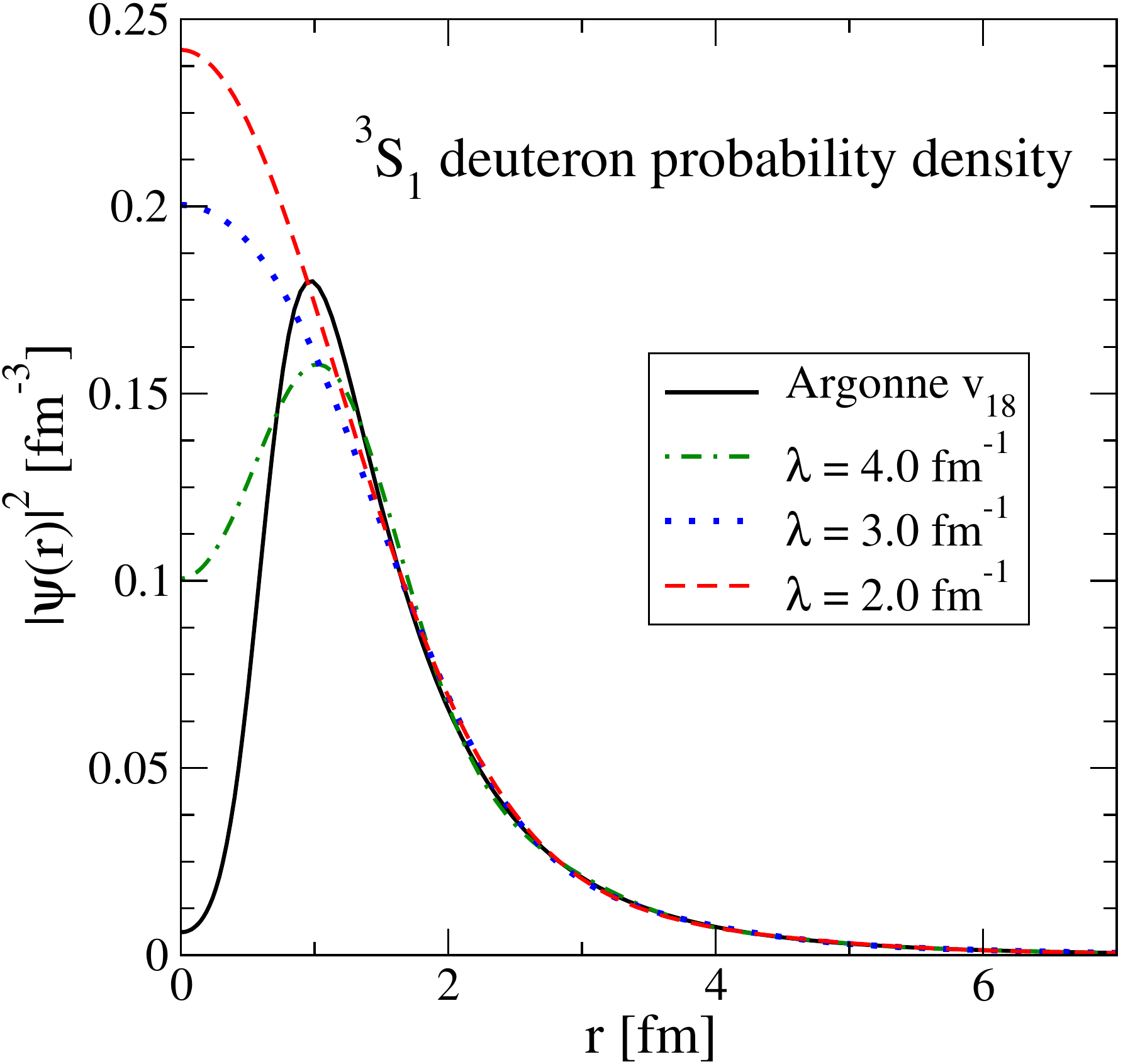}
   \caption{Short-range correlations in the deuteron $^3\mbox{S}_1$
   probability density from the AV18 potential (solid line).  They
   are essentially eliminated by the RG evolution to lower flow
   parameter $\lambda$ (see Section~\ref{subsec:floweq}).}
   \label{fig:deutprob}
\end{figure}

The consequences of the coupling to high momentum are readily seen
in the probability density of the only two-body bound state,
the deuteron.  
Consider the Argonne $v_{18}$~\cite{Wiringa:1994wb} curve in Fig.~\ref{fig:deutprob}.
The probability at small separations is significantly suppressed
as a result of high-momentum components in the
wave function.  
This suppression, called ``short-range correlations'' in this context, 
carries over to many-body wave functions and greatly complicates
basis expansions.
For example, in a harmonic oscillator basis, which is frequently
the choice for self-bound nuclei because it readily allows removal
of center-of-mass contamination, convergence is greatly slowed
by the need to accommodate high-momentum components.
The factorial growth of the basis size with the number of
nucleons then greatly limits the reach of calculations.

The underlying problem is that the resolution scale
induced by the potential is mismatched with the basic scale of the
low-energy nuclear states (given, for example, by the Fermi momentum).
A solution is to eliminate the coupling to high momentum.
This is readily accomplished for spatial images (i.e., photographs)
by Fourier transforming
and then applying a low-pass filter---simply set the short wavelength
parts to zero---and then transforming back.  Let's try that
for our Hamiltonian by setting to zero all of the matrix elements
in Fig.~\ref{fig:vnncntr} for $k > 2\fmi$.
We test the implications by seeing the effect on scattering phase shifts
in the region of laboratory energies corresponding to $k \leq 2\fmi$.

\begin{figure}[tb!]
   \includegraphics[width=2.8in]{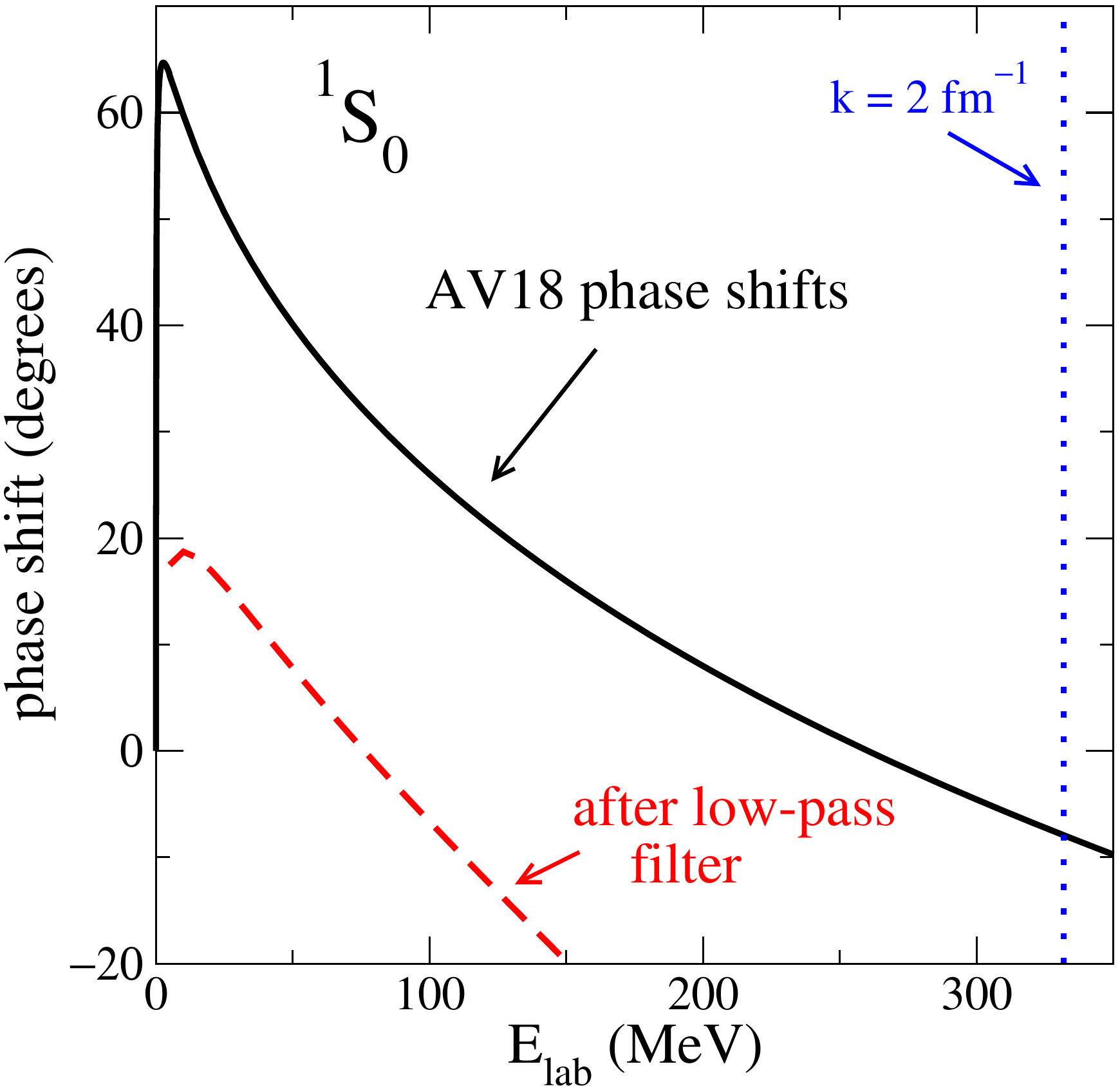}
   \caption{Effect of a low-pass filter on observables: the $^1$S$_0$
   phase shifts.  Note that the AV18 phase shifts reproduce experimentally extracted phase shifts
   in this energy range.}
   \label{fig:lowpass}
\end{figure}

The result is shown in Fig.~\ref{fig:lowpass}.
It would be unsurprising that our filtered Hamiltonian fails close
to the cut-off, but it is evident that there is a failure at all
energies.  What happened?  The basic problem is that low momentum
and high momentum are coupled when solving quantum mechanically
for observables.  For example, consider perturbation theory for the (tangent)
of the phase shift (represented schematically here):
\beq 
   \langle k | V | k \rangle
      + \sum_{k'}\frac{\langle k | V | k' \rangle \langle k' | V | k \rangle}
              {(k^2 - {k'}^2)/m} + \cdots 
	      \label{eq:ptphase}
\eeq
where a low momentum $k$ is mixed with all other momenta at second order
to a degree based on the size of the off-diagonal matrix elements.
(As a computational aside, although momentum is continuous
in principle, in practice we work on a discrete grid.  This means
that Eq.~\eqref{eq:ptphase} becomes a matrix equation.  For two-body
potentials, roughly $100\times100$ matrices are sufficient.)  
The phase shift even at low energy or $k$ will get large contributions from
high $k'$ if the coupling matrix elements $\langle k | V | k' \rangle$
are large.

How can we fix this?  Our solution is to use a (short-distance)
unitary transformation of the Hamiltonian
matrix to \emph{decouple} low and high energies.
That is, insert the operator $U^\dagger U = 1$ repeatedly:
\bea
      E_n \ampseq \langle \Psi_n | H | \Psi_n \rangle 
        = \langle \Psi_n | U^\dagger ) U H U^\dagger
	   U | \Psi_n \rangle ) \nonumber\\
	 \amps{\equiv} \langle \widetilde\Psi_n | \widetilde H | \widetilde\Psi_n \rangle \;.
\eea
In doing so we have modified operators \emph{and} wavefunctions but 
observables (measurable quantities) are unchanged.
An appropriate choice of the unitary transformation can, in principle, 
achieve the
desired decoupling.  This general approach has long been used in
nuclear structure physics and for other many-body applications.
The new feature here is the use of renormalization group flow
equations to create the net unitary transformation via a series
of infinitesimal transformations.

The renormalization group is well suited to this purpose.
The common features of RG for critical phenomena and high-energy
scattering are discussed by Steven Weinberg in an essay in
Ref.~\cite{Guth:1984rq}.  He summarizes:
\begin{quote}
``The method in its most general form can I think be understood
as a way to arrange in various theories that the degrees of freedom
that you're talking about are the relevant degrees of freedom for the
problem at hand.''
\end{quote}
This is the essence of what is done with the low-momentum interaction
approaches considered here: arrange for the degrees of freedom for nuclear structure
to be the relevant ones.  This does not mean that other degrees of
freedom cannot be used, but to again quote
Weinberg~\cite{Guth:1984rq}: ``You can use any degrees of freedom you
want, but if you use the wrong ones, you'll be sorry.''

The consequences of using RG for high-energy (particle) physics include improving perturbation
theory, e.g., in QCD.  
A mismatch of energy scales can generate large logarithms
that ruins perturbative convergence even when couplings by
themselves are small.
The RG shifts strength between loop integrals and coupling constants
to reduce these logs.  For critical phenomena in condensed matter
systems, the RG reveals the
nature of observed
universal behavior by filtering out short-distance degrees
of freedom.  We will see both these aspects in our calculations
of nuclear structure and reactions.  The end result can be said
to make nuclear physics look more like quantum chemistry, opening
the door to a wider variety of techniques (such as many-body perturbation
theory) and simplifying calculations (e.g., by improving convergence
of basis expansions).

\subsection{Summary points}

Low-energy nuclear physics
is made difficult by a mismatch
of energy scales inherent in conventional phenomenological potentials
and those of nuclear structure and reactions.
The renormalization group offers a way out by lowering the resolution,
which means decoupling low- from high-momentum degrees of freedom. 

%%%%%%%%%%%%%%%%%%%%%%%%%%%%%%%%%%%%%%%%%%%%%%%%%%%%%%%%%%%%%%%%%%%%%%%%
%%%%%%%%%%%%%%%%%%%%%%%%%%%%%%%%%%%%%%%%%%%%%%%%%%%%%%%%%%%%%%%%%%%%%%%%

\section{Overview of flow equations}
\label{sec:flow}

\begin{figure*}[t!]
   \begin{center}
   \includegraphics[width=2.2in]{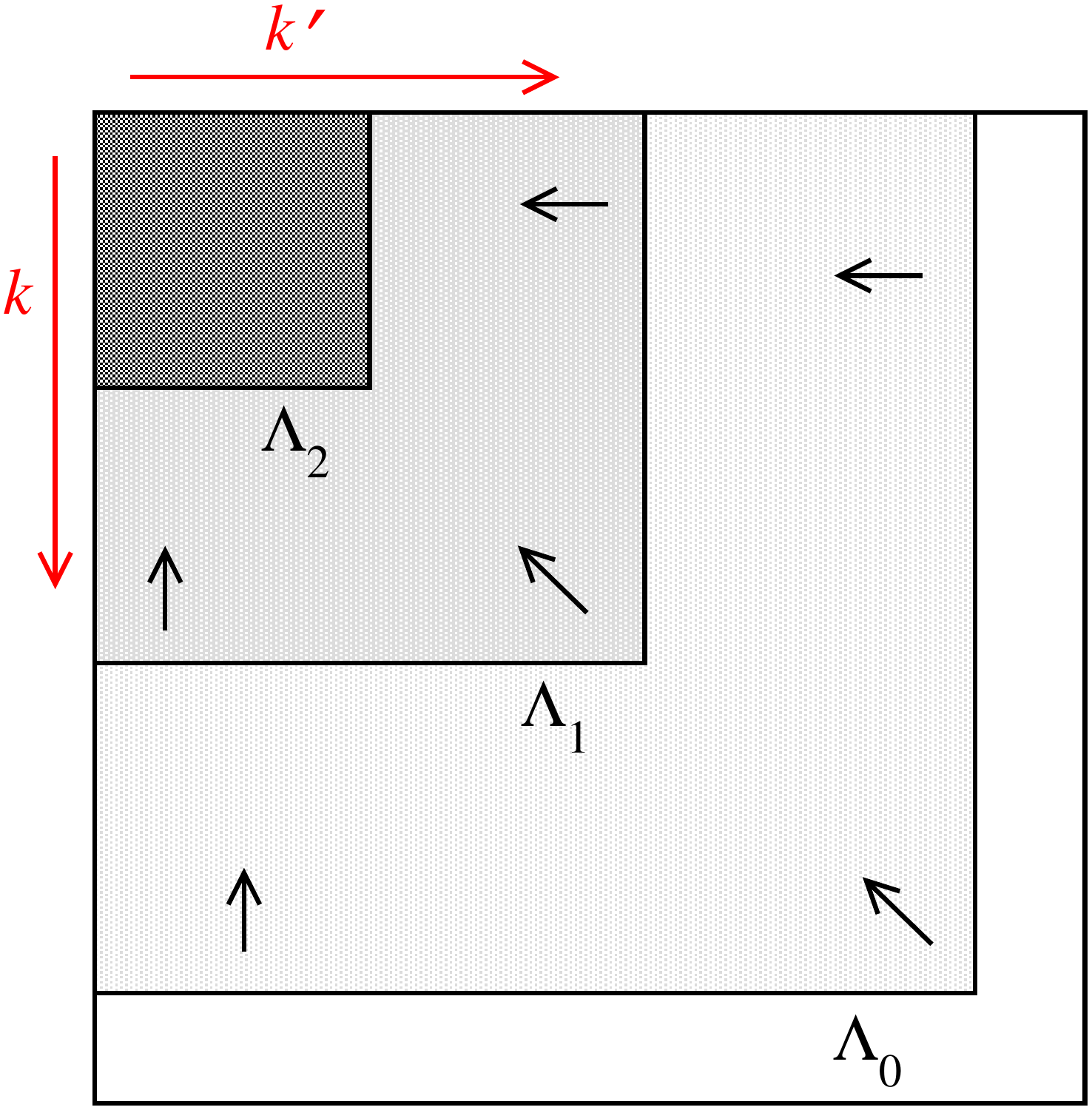}
   \hspace*{.5in}
   \includegraphics[width=2.2in]{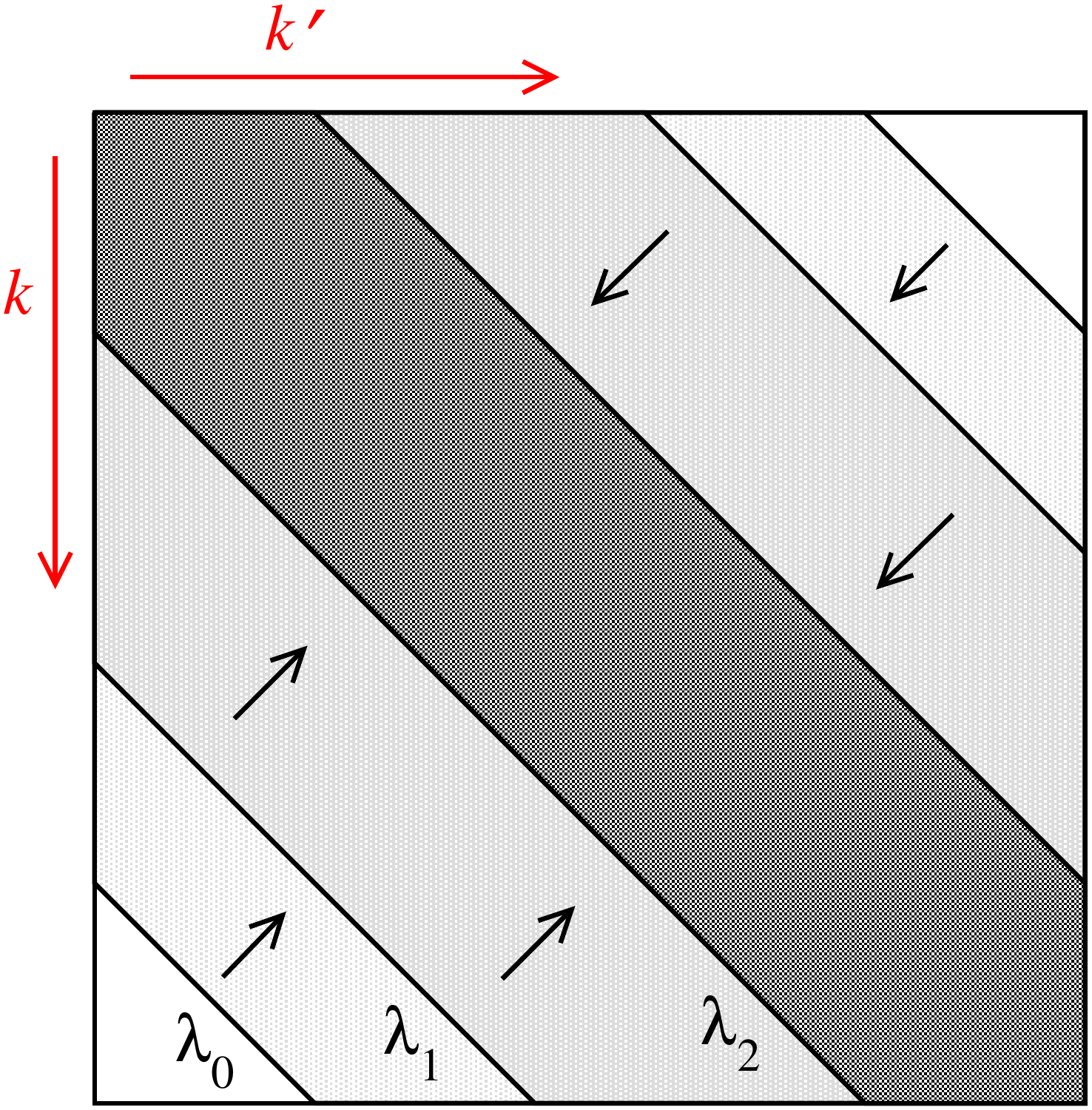}
   \caption{Schematic $\vlowk$ RG evolution (left) and
   flow equation RG evolution (right).}
   \label{fig:vlowkschematic}
  \end{center}
\end{figure*}

In Fig.~\ref{fig:vlowkschematic}, we show
schematically two options for how the RG can be used to decouple
a Hamiltonian matrix.  The more conventional approach on the left in
Fig.~\ref{fig:vlowkschematic} lowers a cutoff $\Lambda$ in momentum
in small steps, with the matrix elements adjusted by requiring some
quantity such as the on-shell T-matrix to be invariant.
(In practice this may be carried out by enforcing that the half-on-shell
T-matrix is independent of $\Lambda$.)
Matrix elements above $\Lambda$ are zero and are therefore trivially
decoupled.
When adapted to low-energy nuclear physics, this approach is typically
referred to as ``$\vlowk$'' \cite{Bogner:2003wn,Bogner:2009bt}.

\begin{figure*}[t!]
\begin{center}
   \includegraphics[width=5.8in]{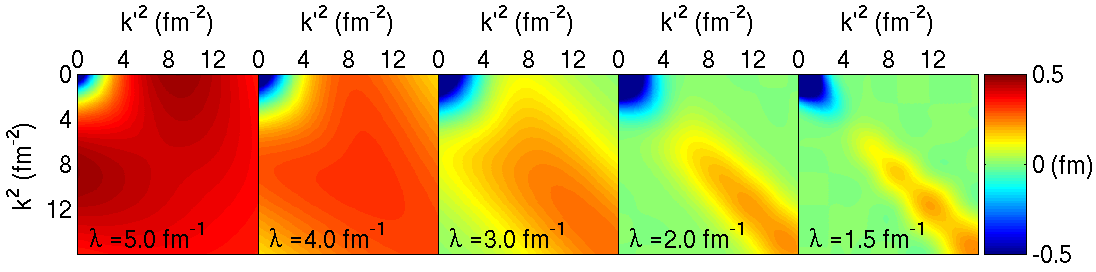}
   \caption{SRG flow of the AV18 NN potential in the $^3$S$_1$ channel
   at selected values of the flow parameter $\lambda$~\cite{Bogner:2009bt}.}
   \label{fig:srgAV183S1}
\end{center}
\end{figure*}

The approach we will focus on is illustrated in Fig.~\ref{fig:vlowkschematic}
on the right,
in which the matrix is driven toward band-diagonal form, achieving
decoupling again but without truncating the matrix.
The corresponding RG was developed in the early 1990's
by Wegner~\cite{Wegner:1994,Wegner:2000gi,Wegner:2006zz} for condensed matter applications
under the name ``flow equations'' and independently by
Glazek and Wilson~\cite{Glazek:1993rc} for solving quantum chromodynamics
in light-front formalism under the name ``similarity renormalization group''
(SRG).  
Only in the last five years was it realized that the band-diagonal
approach is
particular well suited for low-energy nuclear physics, where
it is technically simpler and more versatile than other methods~\cite{Bogner:2006pc,Bogner:2009bt}.
We will apply formalism closer to the flow equation formulation,
but generally use the simple abbreviation SRG.
We have introduced a cutoff-like parameter $\lambda$ that serves
as a momentum decoupling scale.  Elsewhere we will also use the natural
flow parameter $s = 1/\lambda^4$.

\subsection{Flow equation basics}
\label{subsec:floweq}

The basic flow equation is a set of coupled differential equations
for the (discrete) matrix elements
of the Hamiltonian matrix (or the potential in practice, because
we fix the kinetic energy matrix by construction).  
Let's see an example in action before stepping back and considering details.
For the two-body
potential in a partial-wave momentum basis, the flow equation takes the form
  \bea
  && \frac{dV_\lambda}{d\lambda}(k,k') \propto
      - (\epsilon_k - \epsilon_{k'})^2 V_\lambda(k,k') \nonumber \\
    && \quad \null +  \sum_q (\epsilon_k + \epsilon_{k'} - 2\epsilon_q)
        V_\lambda(k,q) V_\lambda(q,k') 
	\;,
  \eea
where $k$ and $k'$ are the relative momenta and 
$\epsilon_k \equiv \hbar^2 k^2/M$ and we have omitted an inessential constant.
The evolution is continuous, but snapshots at selected $\lambda$ values
are shown in Fig.~\ref{fig:srgAV183S1}.  
(Note that the axes are the kinetic energy $k^2$.)
The evolution toward diagonalization is evident, with the width
of the band in $k^2$ roughly given by $\lambda^2$.

\begin{figure*}[t]
\begin{center}
   \includegraphics[width=4.8in]{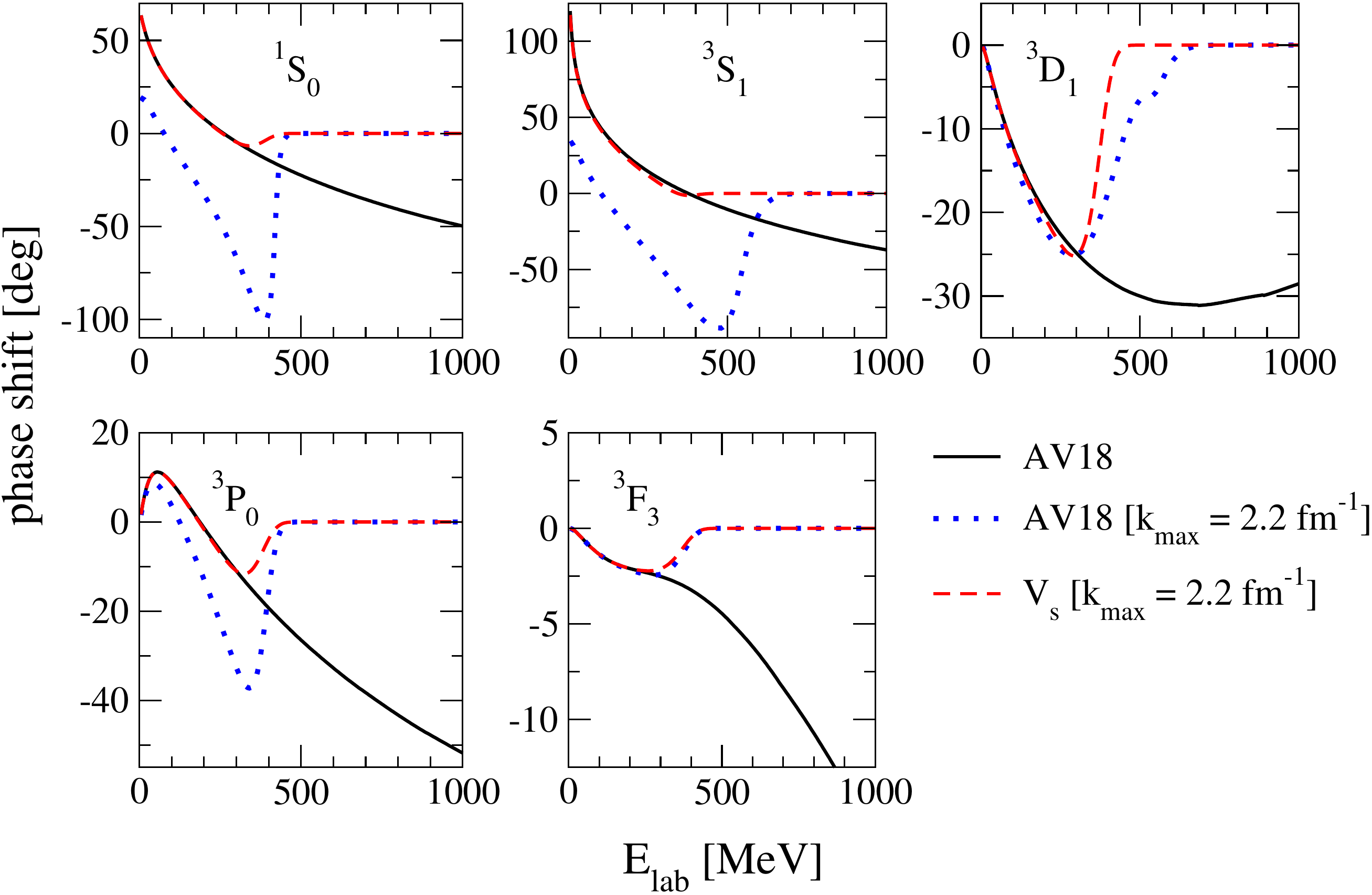}
   \caption{Nucleon-nucleon (NN) 
   phase shifts showing the effect of a low-pass filter
   at $k_{\rm max} = 2.2\fmi$ in various partial-wave 
   channels~\cite{Bogner:2007jb,Jurgenson:2007td}.  In each
   panel, the solid line is for the original AV18 potential while
   the dotted line is the result when the potential is set to zero
   for $\{k,k'\}>k_{\rm max}$.  The dashed line is the results
   when the potential is evolved first and then cut at $k_{\rm max}$.}
   \label{fig:lowpassworks}
\end{center}
\end{figure*}

It is evident that the off-diagonal matrix elements are driven toward
zero, so we expect that a low pass filter will now be effective.
Indeed it is, as shown in Fig.~\ref{fig:lowpassworks}, where NN phase
shifts for the AV18 potential in a variety of channels are compared
with the results from applying a low-pass filter to the original 
(dotted) and evolved (dashed) potentials.  The evolved result agrees
up to the low-pass cutoff.  Note that if this cutoff is not applied, the phase
shifts for the evolved Hamiltonian agree precisely \emph{at all energies},
because the transformation is unitary in the two-body system
(and this is preserved to high numerical precision).

If we now revisit the consequence of a repulsive core for the deuteron
probability density, we see in Fig.~\ref{fig:deutprob} that the
``wound'' in the wave function at small $r$ is filled in
as the core is transformed away.
(Note: it may look like the normalization is not conserved, but
if we multiplied by $r^2$, we would see that the area under the curves
is the same, as are the large-$r$ tails.) 
Thus the short-range correlations in the wave function are
drastically reduced.  This means that the physics interpretation of
various phenomena is altered, even though the observable quantities
such as energies and cross sections are unchanged.
(The long-range part of the wave function \emph{is} preserved;
this is related to the asymptotic normalization constants, which can
be extracted from experiment.)
We cannot immediately visualize the changes in the potential in coordinate
space in a conventional plot like Fig.~\ref{fig:phenpots}, however,
because it is now \emph{non-local}, which is to say it is not diagonal
in coordinate representation:
\beq
V({\bf r})\psi({\bf r}) \longrightarrow
         \int\! d^3{\bf r'}\, V({\bf r},{\bf r'})\psi({\bf r'})
	 \;.
\eeq
This is a technical problem for certain quantum many-body methods (such
as Green's function Monte Carlo) but not for methods using harmonic
oscillator matrix elements.

\begin{figure*}[t!]
\begin{center}
   \includegraphics[width=5.8in]{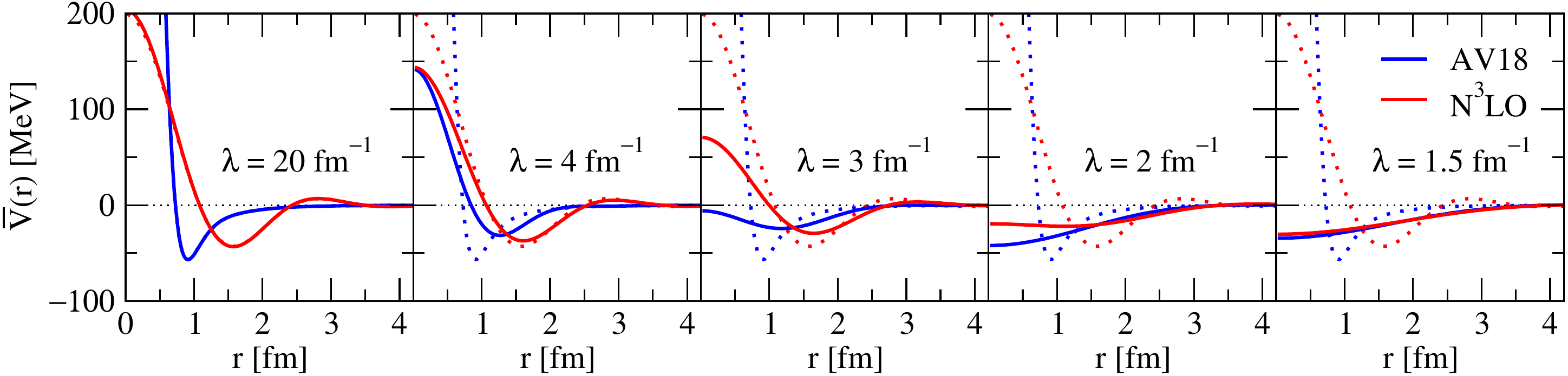}
   \caption{Local projection of AV18 and N3LO(500\,MeV) potentials 
   in $^3$S$_1$ channel
   at different resolutions~\cite{Wendt:2011private}.}
   \label{fig:LPsideSwave}
\end{center}
\end{figure*}

\begin{figure*}[t!]
\begin{center}
   \includegraphics[width=5.8in]{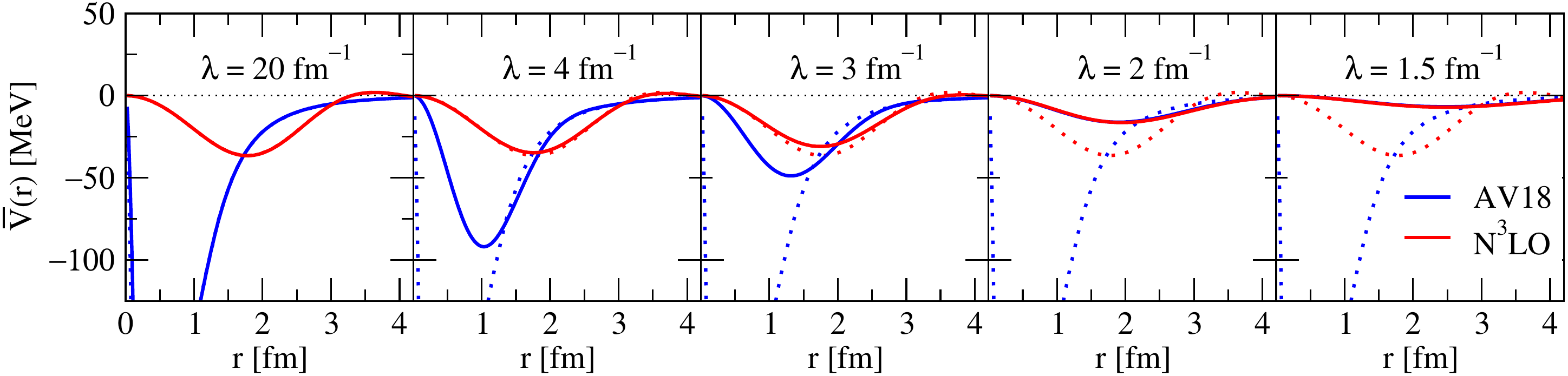}
   \caption{Local projection of AV18 and N3LO(500\,MeV) potentials  in mixed
   $^3$S$_1$--$^3$D$_1$ channel
   at different resolutions~\cite{Wendt:2011private}.}
   \label{fig:LPsideDwave}
\end{center}
\end{figure*}

We can visualize the evolution \emph{approximately} by considering a local
projection of the potential:
\beq
\overline V_\lambda(r) = \int\! d^3r'\, V_\lambda(r,r')
\eeq
which leaves a local $V(r)$ unchanged.
For a non-local potential, this roughly gives the action of
the potential on long-wavelength nucleons.
This is shown in Figs.~\ref{fig:LPsideSwave} and \ref{fig:LPsideDwave},
where in the former we see the central potential dissolving and
in the latter similar effects on the tensor part of the potential~\cite{Wendt:2011private}.
Also evident is the flow of the two potentials, initially quite
different (potentials are not observables!), toward a universal
flow at the lower values of $\lambda$.
Very recent work suggests that such a local projection may
capture most of the physics of the full evolved potential, with
the effects of the residual potential calculable in perturbation
theory.  This may open the door to using RG-evolved potential
with quantum Monte Carlo methods.

\begin{figure}[t!]
   \begin{center}
   \includegraphics[width=2.5in]{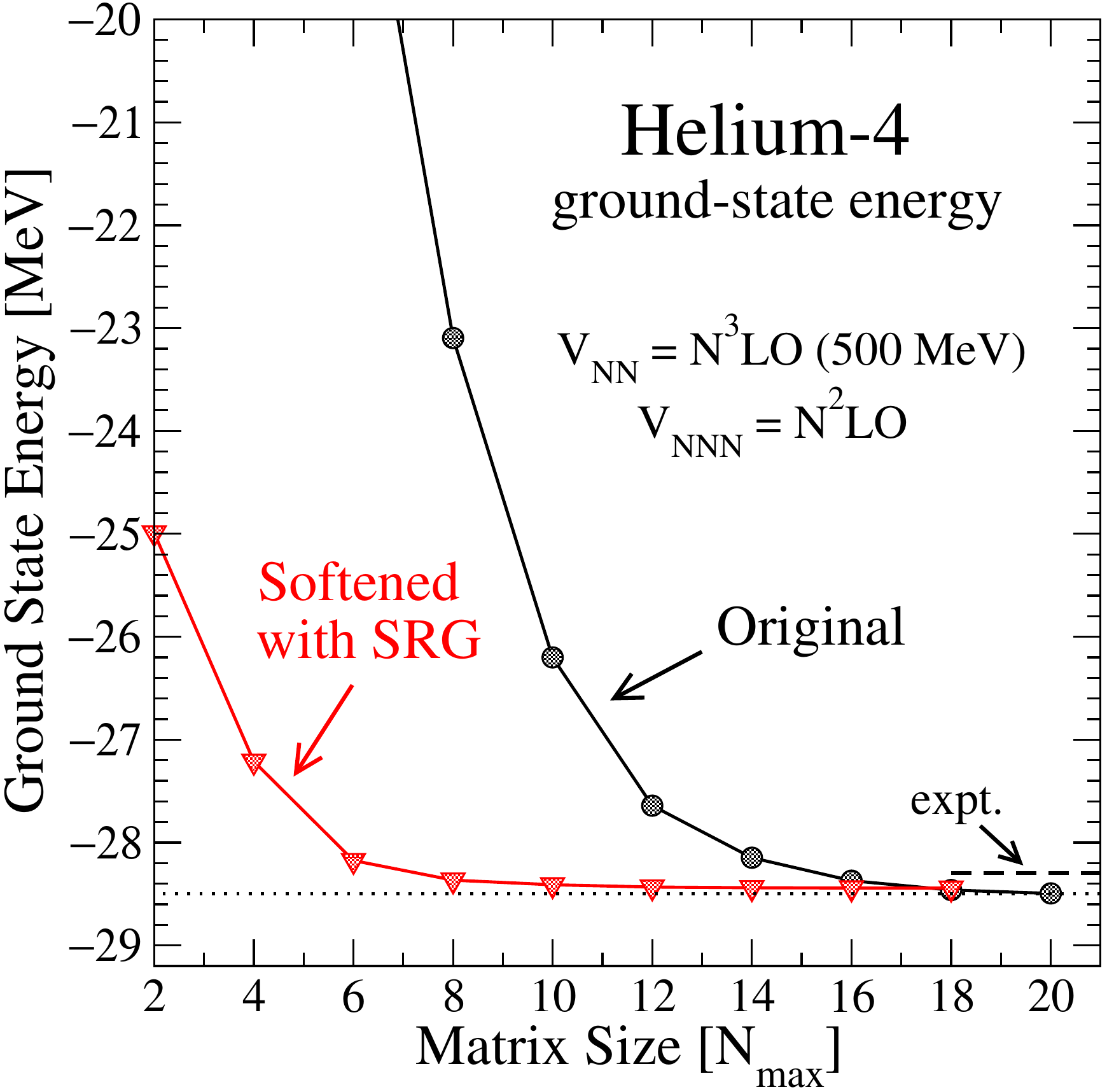}
   \caption{Convergence of NCFC calculations of the ${^4}$He ground state
   energy with basis size at SRG different resolutions.  The initial
   potential includes both two- and three-body components~\cite{Jurgenson:2009qs,Jurgenson:2010wy}.}
   \label{fig:He4vsNmax}
   \end{center}
\end{figure}

\begin{figure}[t!]
   \begin{center}
   \includegraphics[width=2.55in]{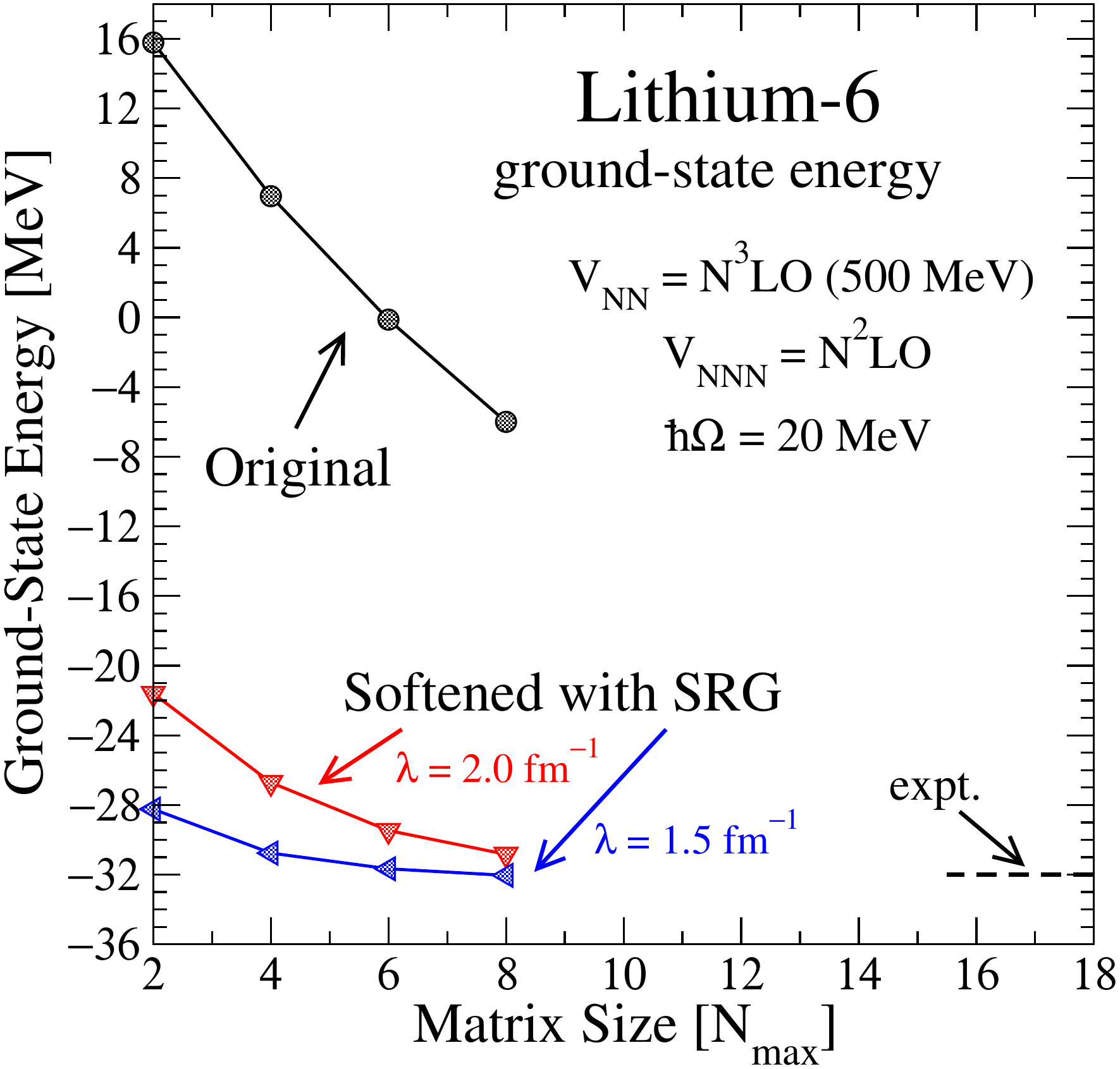}
   \caption{Convergence of NCFC calculations of the ${^6}$Li ground state
   energy with basis size at SRG different resolutions. The initial
   potential includes both two- and three-body components~\cite{Jurgenson:2009qs,Jurgenson:2010wy}.}
   \label{fig:Li6vsNmax}
   \end{center}
\end{figure}

\begin{figure}[t!]
   \begin{center}
   \includegraphics[width=2.3in]{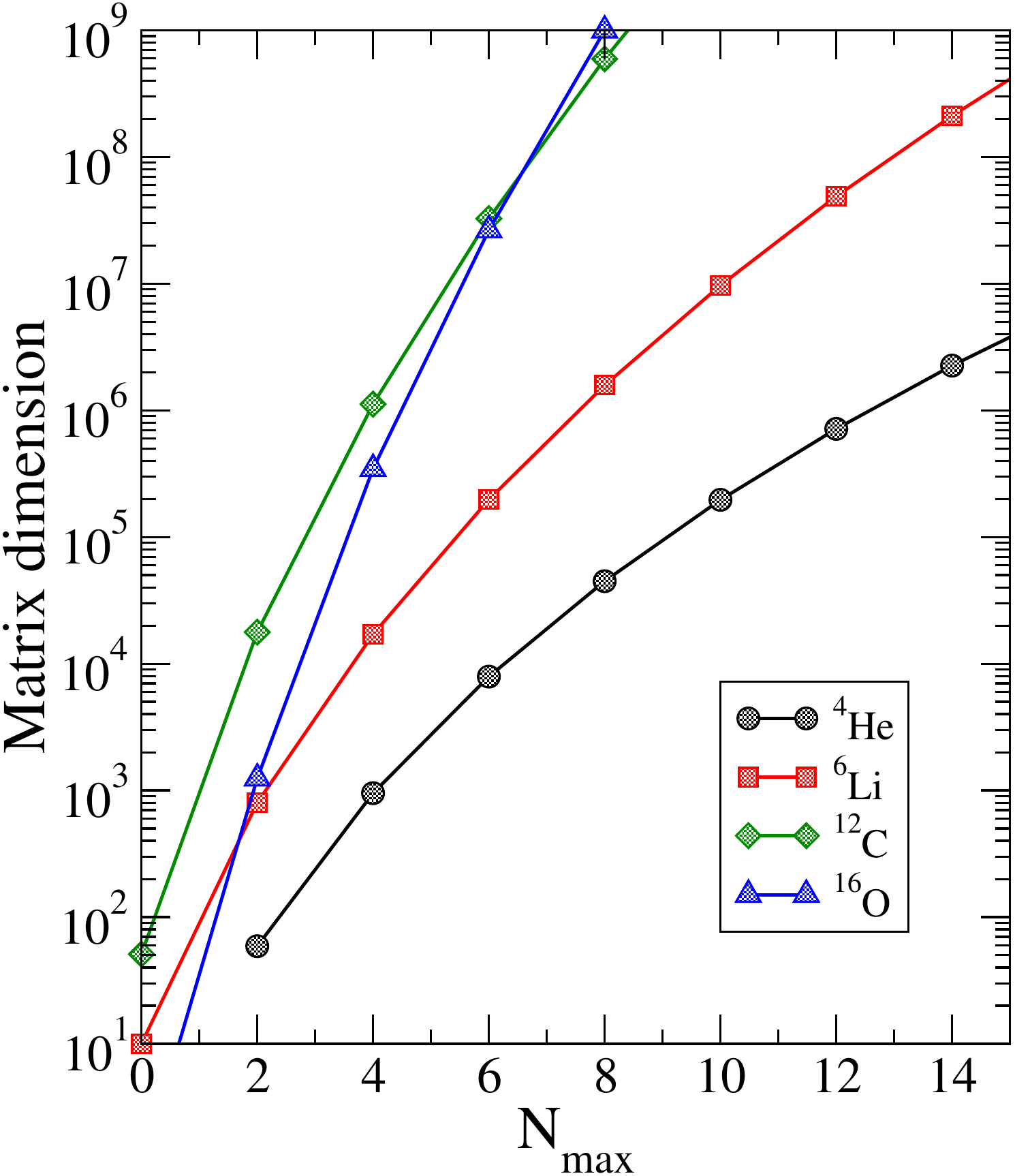}
   \caption{Dimension of the Hamiltonian matrix for NCFC/NCSM calculations
   as a function of the harmonic oscillator basis size ($\Nmax$ shells)
   for selected nuclei.}
   \label{fig:ncsmsize}
   \end{center}
\end{figure}

The consequences of low-momentum potentials
for harmonic-oscillator-basis calculations are
illustrated in Figs.~\ref{fig:He4vsNmax} and \ref{fig:Li6vsNmax}.
(We'll refer to the method used, which is a direct diagonalization of
the Hamiltonian matrix and therefore is variational, as no-core full configuration, or NCFC.)
The original potential in this case is already soft (that is, there
is much
less coupling to high momentum than in the AV18 potential), but
convergence in a harmonic oscillator basis with $\Nmax$ shells for
excitation is slow (these are the ``Original'' curves).  
Note that the matrix dimension grows rapidly with $\Nmax$ \emph{and} 
the number of nucleons $A$;
for example $\Nmax = 8$ has dimension about 50,000 for $^4$He
but over one million for $^6$Li (see Fig.~\ref{fig:ncsmsize} for
other examples).
But with SRG evolution, there is vastly improved convergence.
The convergence is also smooth, which makes it possible to 
reliably extrapolate
partly converged results to the $\Nmax \rightarrow \infty$ limit.
Nevertheless, the rapid growth of the basis with $A$ is still
a major hindrance to calculating larger nuclei.  One solution 
being explored by Roth and collaborators~\cite{Roth:2007sv,Roth:2011ar} is 
to use importance sampling of matrix elements, in which only
a fraction of the full matrix is used.  This technique
is enabled by the RG softening of the potential, which allows the
importance to be evaluated perturbatively. 

Let's now return to the basics of SRG flow equations.
We wish to transform an initial hamiltonian, $ H = T + V$
in a series of steps, labeled by the flow parameter $s$:
\beq
   H_\flow = U_\flow H U^\dagger_\flow \equiv T + V_\flow 
\eeq
with   
\beq
   \quad U^\dagger_\flow U_\flow = 
   U_\flow U^\dagger_\flow = 1 \;.
\eeq
%where $\flow$ is the \emph{flow parameter}.  
Note that the kinetic energy $T$ is taken to be independent of $s$.  Differentiating 
with respect to $s$:
\bea
  \frac{dH_\flow}{d\flow}
    \amps{=} \frac{dU_\flow}{d\flow} U^\dagger_\flow U_\flow H
       U^\dagger_\flow + U_\flow H U^\dagger_\flow U_\flow
    \frac{dU^\dagger_\flow}{ds} \nonumber \\
    \amps{=} [\eta_\flow,H_\flow]
\eea
with     
\beq
   \eta_\flow \equiv \frac{dU_\flow}{d\flow} U^\dagger_\flow 
          = -\eta^\dagger_\flow
   \ .
\eeq
The anti-Hermitian generator $\eta_\flow$ can be specified by a commutator 
of $H_\flow$ with
a Hermitian operator   $G_s$:
\beq
     \eta_\flow =  [G_s, H_\flow] \ ,
\eeq
   which yields the flow equation (with $T$ held fixed!),
  \beq
    \frac{dH_\flow}{d\flow} 
    = \frac{dV_s}{d\flow} 
    =  [ [G_s, H_\flow], H_\flow] \ .
  \eeq
The operator $G_s$ determines the flow and there are many choices
one can consider.

\begin{figure}[b!]
   \begin{center}
  \includegraphics[width=2.55in]{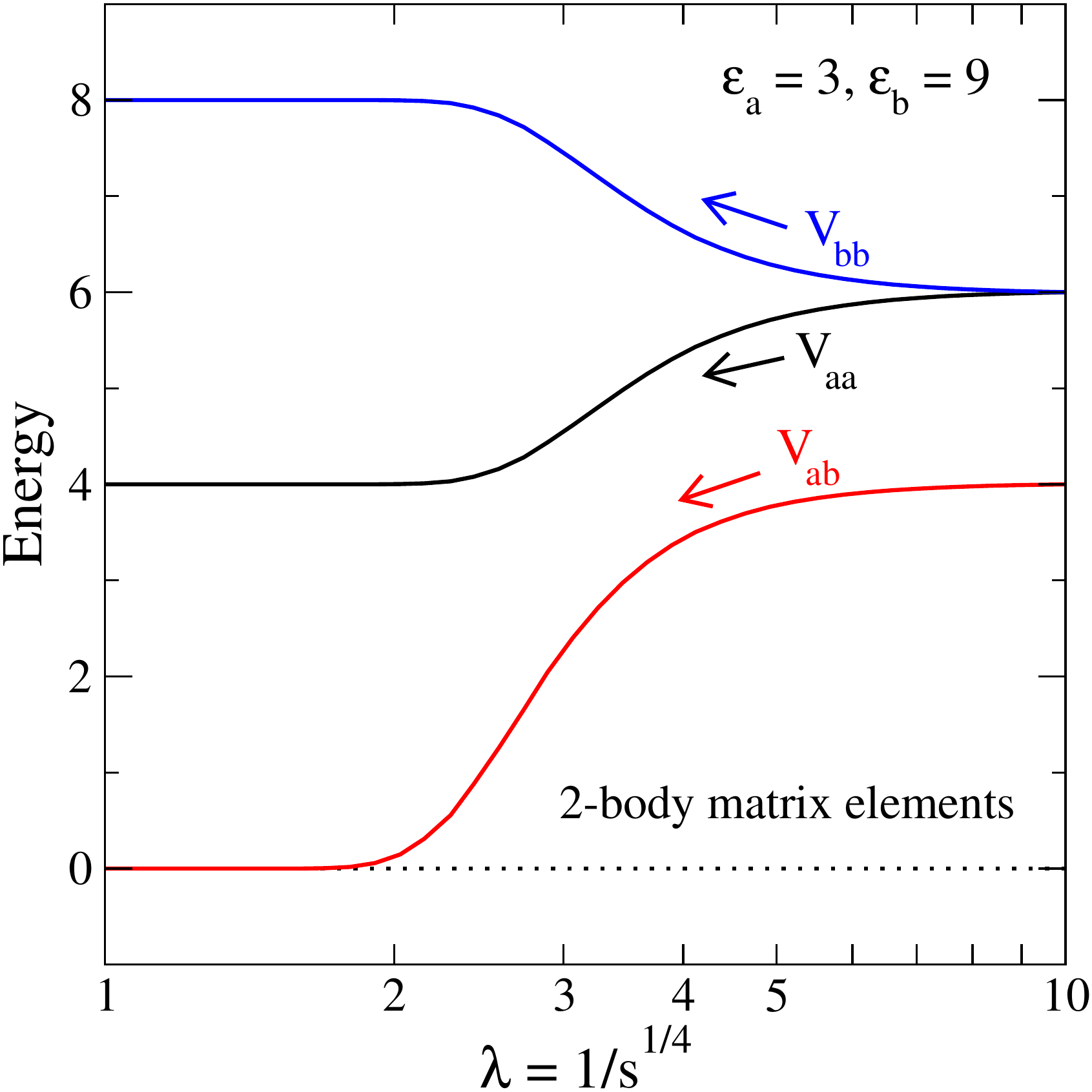}
   \caption{SRG flow for a simple two-state system (see text).}
   \label{fig:two-state}
   \end{center}
\end{figure}

\begin{figure*}[thb!]
\begin{center}  
\includegraphics[width=1.5in]{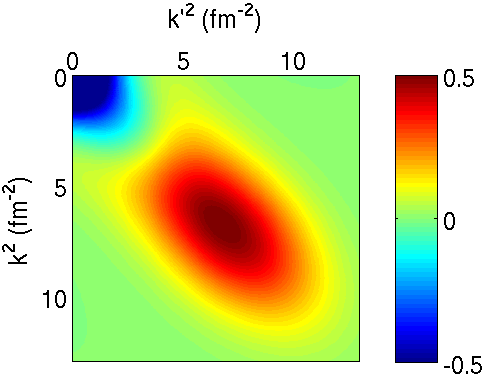}
\includegraphics[width=1.13in]{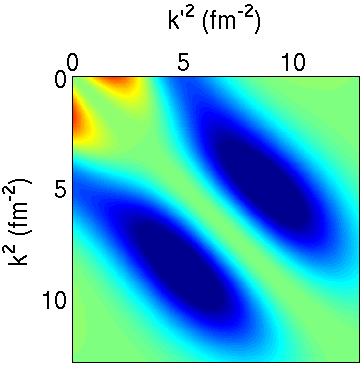}
\includegraphics[width=1.5in]{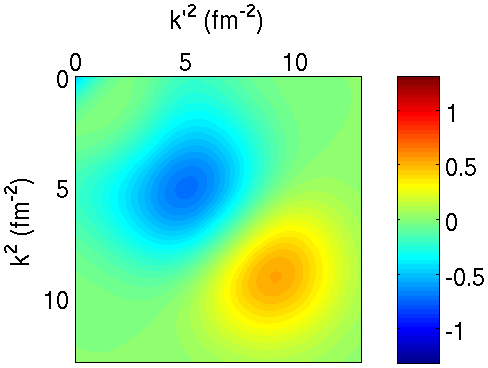}
\includegraphics[width=1.5in]{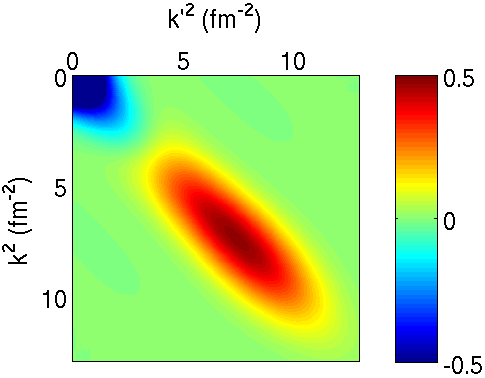}
  \caption{SRG flow in stages.  On the far left and far right are potentials
  in the $^1$S$_0$ channel evolved to $\lambda = 2.0\fmi$ and $\lambda=1.5\fmi$, respectively.  The middle panels show the first (left) and second (right)
  terms on the right side of Eq.~\eqref{eq:srgeq2}.}
  \label{fig:srgstages}
\end{center}
\end{figure*}

Probably the simplest example we can consider is just a two-state
system~\cite{Bogner:2007qb}.  Let $H = T + V$, where 
\beq
 T|i\rangle = \epsilon_i | i\rangle 
    \quad\mbox{and}\quad
      V_{ij} \equiv \langle i|V|j\rangle
      \;.
\eeq
Then we can choose $G_s = T$ and 
\beq
   \frac{dH_s}{ds} = [[T,H_s],H_s]
\eeq
becomes (with the $s$ dependence implicit)
\bea
       \frac{d}{ds} V_{ij} \ampseq 
       -(\epsilon_i-\epsilon_j)^2 V_{ij} 
       \nonumber \\ \amps {} \null
       + \sum_k (\epsilon_i+\epsilon_j-2 \epsilon_k) V_{ik} V_{kj}
       \;.       
\eea
For a two-level system with  $i = \{a,b\}$, we can express the 
flowing Hamiltonian in terms of Pauli matrices:
\beq
  T = \frac12(\epsilon_a + \epsilon_b) I + \frac12(\epsilon_a - \epsilon_b)
   \sigma_z
\eeq   
and       
\bea 
  V_s \ampseq \frac12 (V_{aa}+V_{bb})I 
  \nonumber \\ \amps{} \null + \frac12 (V_{aa}-V_{bb})\sigma_z
  + V_{ab}\sigma_x \;.
\eea
The solution to the SRG flow equation is easily found.  
It is convenient to parametrize
the result in terms of $\theta(s)$ with constant $\omega$: 
\beq
  \frac{d\theta}{ds} = -2(\epsilon_a-\epsilon_b) \omega \sin\theta(s)
\eeq
where
\beq 
   \theta(s) = 2 \tan^{-1}[\tan(\theta(0)/2) e^{-2(\epsilon_a-\epsilon_b)
        \omega s}] 
	\;,
\eeq
\beq \omega \cos\theta = (\epsilon_a-\epsilon_b+ V_{aa}-V_{bb})/2
\;,
\eeq
and
\beq
   \omega \sin\theta =  V_{ab} \;.
\eeq
The resulting flow is plotted for sample energies $\epsilon_a$ and $\epsilon_b$ in Fig.~\ref{fig:two-state}.  We clearly see the off-diagonal matrix
element $V_{ab}$ driven to zero. (Try reproducing
this in Mathematica!)

For a nuclear two-body (NN) potential
in a partial-wave momentum basis with  $\eta_s = [T,H_\flow]$,
we project on relative momentum states  $|k\rangle$
using $1 = \frac{2}{\pi}\int_0^\infty|q\rangle q^2\,dq \langle q |$
with $\hbar^2/M = 1$. The flow equation reduces to:
  \beq
     \frac{dV_s}{d\flow} 
    =  [ [\Hzero, V_\flow], H_\flow] 
    \quad \mbox{with} \quad
    \Hzero | k \rangle =  \epsilon_k | k \rangle     
  \eeq
and $\lambda^2 = 1/\sqrt{s}$.
$\Trel$ is the relative kinetic energy of the nucleons.
Then
  \bea
   {\frac{dV_\lambda}{d\lambda}(k,k')} \amps{\propto}
     { - (\epsilon_k - \epsilon_{k'})^2 V_\lambda(k,k') }
  \nonumber \\   \amps{\null} \hspace*{-.4in}
   \null  +  { \sum_q (\epsilon_k + \epsilon_{k'} - 2\epsilon_q)
        V_\lambda(k,q) V_\lambda(q,k')} \;.
	\label{eq:srgeq2}
  \eea	
This particular equation is for $A=2$, but the results
are generic if one lets $k$ represent a set of Jacobi momenta.
The first term in Eq.~\eqref{eq:srgeq2} drives $V_\lambda$ toward the diagonal:
    \beq 
      V_\lambda(k,k') = V_{\lambda=\infty}(k,k') 
          \,  e^{-[(\epsilon_k-\epsilon_{k'})/\lambda^2]^2} + \cdots
	 \;, \label{eq:diagonalize}
    \eeq
which can be visualized in Fig.~\ref{fig:srgstages}.
These panels represent a sequence from $\lambda = 2.0$ to $\lambda = 1.5$.
The potentials at the beginning and the end are on the outside, while the
two middle panels are the first and second term of Eq.~\eqref{eq:srgeq2}.
For off-diagonal matrix elements, the first term is numerically dominant
and each element is driven to zero 
according to Eq.~\eqref{eq:diagonalize}, with further off-diagonal
elements changing more rapidly.
Note that the width of the diagonal is given roughly by $\lambda^2$,
in accord with Eq.~\eqref{eq:diagonalize}.

A more general proof follows if we use the generator advocated by
Wegner, which includes the diagonal part of the Hamiltonian, $H_d$.
Call the diagonal elements $H_{ii} = e_i$, then (with $\eta_s = [H_d,H_s]$),
   \bea
   \frac{dH_{ij}}{ds} \ampseq \langle i | [[H_d,H_s],H_s] | j \rangle
   \nonumber \\ \ampseq \sum_k (e_i + e_j - 2 e_k) H_{ik}H_{kj}
   \nonumber \\ \ampseq 2\sum_k (e_i - e_k) |H_{ik}|^2
   \;.
   \eea
But   
   \bea
      \frac{d}{ds} \sum_i |H_{ii}|^2
      \ampseq 2\sum_i H_{ii}\frac{dH_{ii}}{ds} \nonumber\\
      \ampseq 4 \sum_{i\neq k} e_i(e_i-e_k)|H_{ik}|^2 \nonumber\\
      \ampseq 2\sum_{i\neq k}(e_i-e_k)^2 |H_{ik}|^2 \geq 0
      \;.
   \eea
Now use: 
\beq
  \Tr H_s^2  = \mbox{const.} = \sum_{ij} |H_{ij}|^2  =
   \sum_{i} |H_{ii}|^2 + \sum_{i\neq j} |H_{ij}|^2 
   \;,
\eeq
to obtain
   \bea
    \frac{d}{ds}\sum_{i\neq j}|H_{ij}|^2
    \ampseq -\frac{d}{ds}\sum_{i}|H_{ii}|^2  \nonumber \\
      \ampseq -2\sum_{i\neq k}(e_i-e_k)^2 |H_{ik}|^2 \leq 0 \;.
   \eea
Thus, in the absence of degeneracies, the off-diagonal matrix 
elements will decrease (or at least remain unchanged)~\cite{Kehrein:2006}. 

This feature of the diagonal generator is desirable, but we also
note that for nuclear Hamiltonians in a momentum basis, the
diagonal is completely dominated by the kinetic energy, so
$H_d \approx \Trel$ is a very good approximation.  
Can this break down?  Glazek and Perry~\cite{Glazek:2008pg}
showed that it can (see also Wendt et al.~\cite{Wendt:2011qj}). 
 Reconsider the proof of diagonalization, but now with $G_s = \Hzero$.
Now we have $H_{ii} = e_i$ and 
   $\Hzero | i\rangle = \epsilon_i | i \rangle$, so that
   \beq
   \frac{d H_{ii}}{ds} = 2\sum_k (\epsilon_i - \epsilon_k) |H_{ik}|^2
   \;.
   \eeq
So consider   
   \beq
      \frac{d}{ds} \sum_i |H_{ii}|^2
      = 2\sum_i H_{ii}\frac{dH_{ii}}{ds}
      = 4 \sum_{i\neq k} e_i(\epsilon_i-\epsilon_k)|H_{ik}|^2
      \;,
    %  = 2\sum_{i\neq k}(e_i-e_k)^2 |H_{ik}|^2 
   \eeq
from which we conclude   
   \beq
    \frac{d}{ds}\sum_{i\neq j}|H_{ij}|^2
      = -2\sum_{i\neq k}(e_i-e_k)(\epsilon_i-\epsilon_k) |H_{ik}|^2 
      \;.
   \eeq
Thus the off-diagonal decrease depends on 
  $ e_i - e_k \approx \epsilon_i - \epsilon_k $.
But there is the possibility of this not being true,
e.g., if there are spurious bound states as in large-cutoff EFT~\cite{Wendt:2011qj}.

\subsection{Alternative generators}

Other choices of generator are also possible.  Recent work by
Shirley Li while an undergraduate physics major at Ohio State
explored choices designed to accelerate evolution.
In particular, one can choose $G_s$ as
\beq
  G_s = -\frac{\Lambda^2}{1+\Trel/\Lambda^2} \approx c + \Trel
  + \cdots
\eeq
or
\beq
  G_s = -{\Lambda^2}e^{-\Trel/\Lambda^2} \approx c + \Trel
  + \cdots
\eeq
The expansions show that these reduce to the conventional $\Trel$
for momenta that are small compared to the cutoff parameter $\Lambda$
(not to be confused with $\lambda$).  For $\Lambda = 2\fmi$, the low 
energy part of the potential is still decoupled but there is much
less evolution at high energy, which makes it computationally much faster
and allows evolution to low $\lambda$.  See Ref.~\cite{Li:2011sr} for details.

\begin{figure}[t]
\begin{center}
   \includegraphics[width=2.5in]{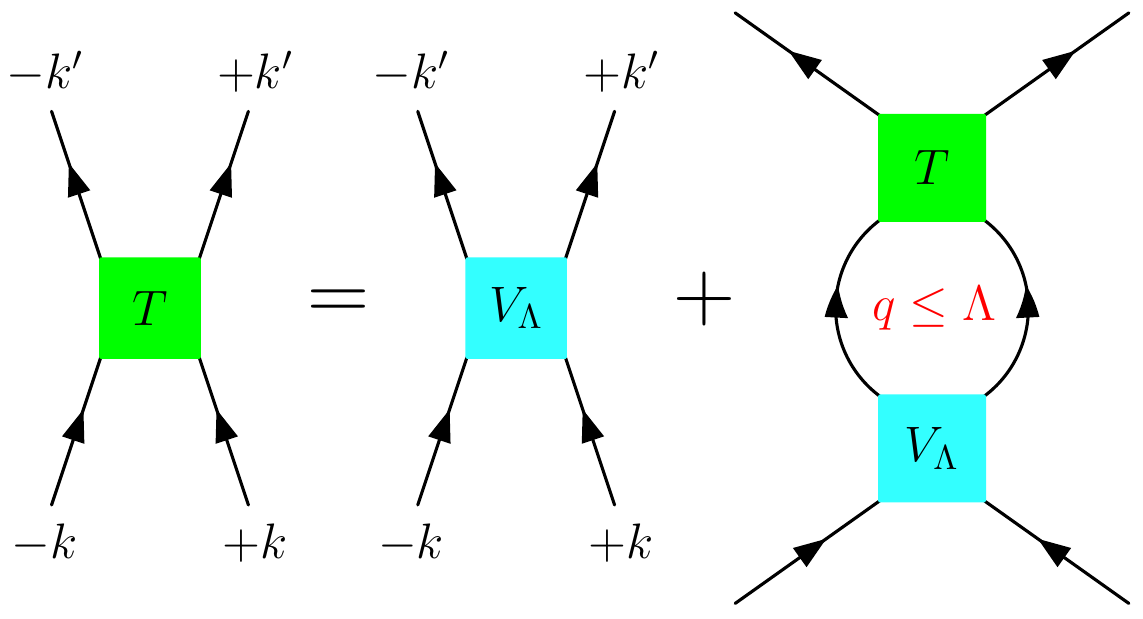}
   \caption{Schematic version of the Lippmann-Schwinger equation
   for the T-matrix, with cutoff $\Lambda$ on the intermediate
   states.  The $\vlowk$ potential $V_{\Lambda}$ is determined
   by requiring half-on-shell matrix elements of this equation
   to be invariant under changes in 
   $\Lambda$\cite{Bogner:2003wn,Bogner:2009bt}.}
   \label{fig:LSvlowk}
\end{center}
\end{figure}

The usual approach to $\vlowk$ RG evolution 
(left diagram in Fig.~\ref{fig:vlowkschematic})  is based on the Lippmann-Schwinger
equation for the half-on-shell T-matrix illustrated in Fig.~\ref{fig:LSvlowk}.
A cutoff $\Lambda$ is imposed on the integral in the second term
and we demand that matrix elements of T are invariant with an
infinitesimal reduction of $\Lambda$.  That is, we require
$dT(k,k';E_k)/d\Lambda = 0$, which establishes an RG equation
for $V_\Lambda$.  In contrast to the SRG equation, which is second
order in the running potential, the $\vlowk$ RG equation has the
T-matrix on the right side. Thus
\bea
T(k',k;k^{2}) \ampseq \vnn(k',k) \nonumber \\
  \amps{} \hspace*{-.3in} \null +\frac{2}{\pi} \, \mathcal{P} 
\int_{0}^{\lm_\infty} 
\frac{\vnn(k',p) \, T(p,k;k^{2})}{k^{2}-p^{2}} \, p^{2} dp  \nonumber \\
\ampseq 
  \vlowk^\lm(k',k) \nonumber \\
 \amps{} \hspace*{-.3in} \null +  \frac{2}{\pi} \, \mathcal{P} 
\int_{0}^{\lm} \frac{\vlowk^\lm(k',p) \, T(p,k;k^{2})}{k^{2}-p^{2}} \, 
p^{2} dp 
 \nonumber \\
\eea
for all $k,k' < \Lambda$.
(Note: we are using standing-wave boundary conditions for numerical
reasons; this is often called the K-matrix.) 
From $dT/d\Lambda = 0$, we get the $\vlowk$ RG equation:
\beq
\frac{d}{d \lm} \vlowk^\lm(k',k) = \frac{2}{\pi} \frac{\vlowk^\lm(k',\lm) \,
T^\lm(\lm,k;\lm^{2})}{1-(k / \lm)^{2}} \;.
\eeq
Note that the full T~matrix appears on the right side,
in contrast to the partial-wave SRG flow equation (with $G_s = \Hzero$),
%  \beq
%    \re
%    \frac{dH_\flow}{d\flow} 
%    =  [ [G_s, H_\flow], H_\flow] \ .
%  \eeq
  \bea
   {\frac{d}{d\lambda}V_\lambda(k,k')} \amps{\propto}
     { - (\epsilon_k - \epsilon_{k'})^2 V_\lambda(k,k') } \nonumber \\
    && 
    \hspace*{-.6in}
    \null + { \sum_q (\epsilon_k + \epsilon_{k'} - 2\epsilon_q)
        V_\lambda(k,q) V_\lambda(q,k')}
	\;,
  \eea	
which is only second-order in the potential.  
We can also define smooth regulators for $\vlowk$ as in 
Ref.~\cite{Bogner:2006vp}.

\begin{figure*}[t]
\begin{center}
   \includegraphics[width=5.6in]{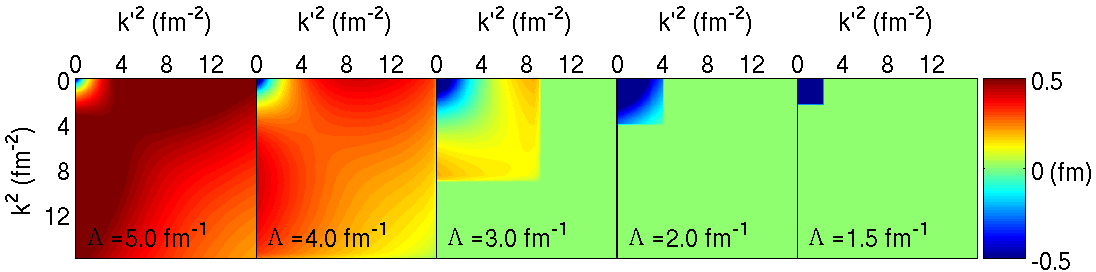}
   \includegraphics[width=5.6in]{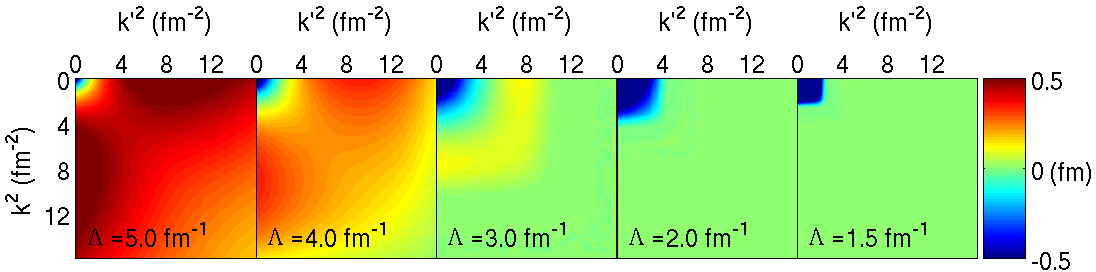}
   \vspace*{-.1in}
   \caption{$\vlowk$ flow of AV18 in the $^3$S$_1$ channel for a sharp
   (top) and smooth (bottom) regulator~\cite{Bogner:2006vp}.}
   \label{fig:vlowkAV183S1}
\end{center}
\end{figure*}

The evolution of NN potentials using the $\vlowk$ method is illustrated
for a sharp and smooth regulator in Fig.~\ref{fig:vlowkAV183S1}.
Comparing the $\vlowk$ flow in these figures to the
SRG flow in Fig.~\ref{fig:srgAV183S1}, we see the same decoupling
of low and high momentum. 
Other non-RG unitary transformations (which perform the transformation
all at once, rather than in steps) also decouple;
an example is the UCOM method~\cite{Feldmeier:1997zh},
which has close
connections to the SRG (see Refs.~\cite{Hergert:2007wp,Roth:2008km}).

\begin{figure*}[t!]
\begin{center}
 \includegraphics[width=5.8in]{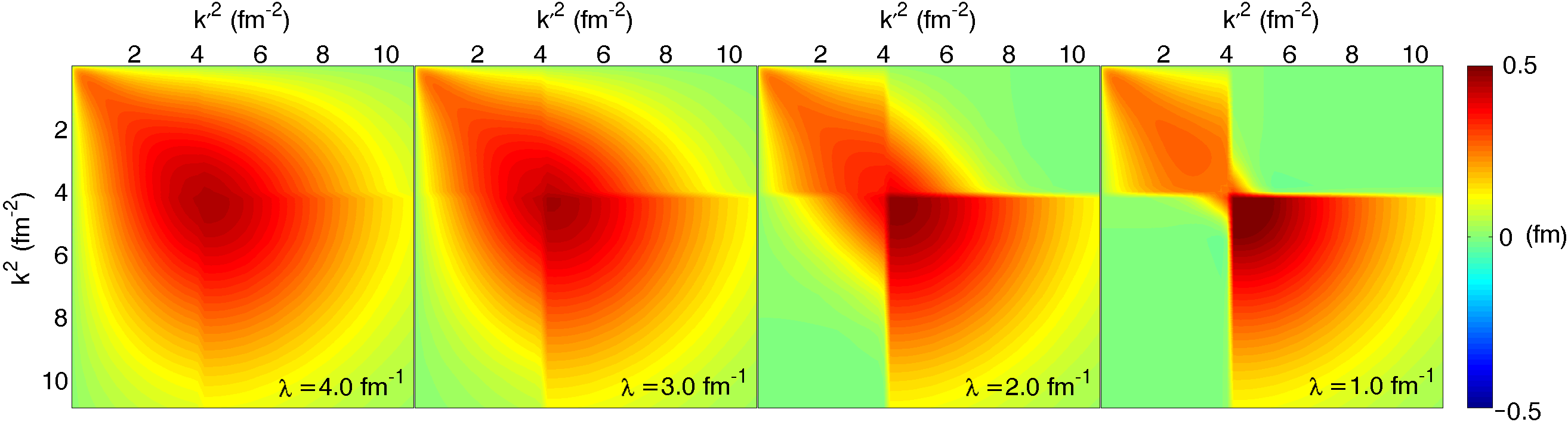}
   \vspace*{-.1in}
   \caption{SRG sharp block-diagonal flow of AV18 in 
   the $^1$P$_1$ channel at $\lambda = 4$, $3$, $2$, and $1\fmi$~\cite{Anderson:2008mu}.
   The initial N$^3$LO potential is from Ref.~\cite{Entem:2003ft}
   and $\Lambda = 2\fmi$.  The axes are in units of $k^2$ from 0 to 11\,fm$^{-2}$
   and the color scale is from $-0.5$ to $+0.5\,$fm.}
   \label{fig:srgbd1P1}
\end{center}
\end{figure*}

It is also possible to choose a generator that reproduces the
block diagonal (as opposed to band diagonal) form of the $\vlowk$ RG
shown schematically in Fig.~\ref{fig:vlowkschematic}, except that
the transformation will be unitary.
In particular, we can use
\beq
  \frac{dH_\flow}{d\flow} = [ [G_s, H_\flow], H_\flow] 
  \label{eq:srgbd3}
\eeq
with
\beq
 G_s =  \left( 
        \begin{array}{cc}
          PH_{s}P & 0   \\
          0     & QH_{s}Q
        \end{array}
        \right) 
	\;,
\eeq
where projection operators $P$ and $Q = 1 - P$ are simply step functions
at a given $\Lambda$ in partial-wave momentum representation.
An example of the subsequent
flow is shown in Fig.~\ref{fig:srgbd1P1} for the
$^1$P$_1$ channel~\cite{Anderson:2008mu}.  To get the fully block-diagonal form, one would
have to evolve to $\lambda = \infty$.  But in practice, going to
$\lambda = 1\fmi$ is sufficient for essentially complete decoupling
at $\Lambda = 2\fmi$.

The proof of block diagonalization (see Gubankova et al.~\cite{Gubankova:1997ha,Gubankova:1997mq}) goes
as follow.
The generator $\eta_s = [G_s,H_s]$ is non-zero only where $G_s$ is zero,
which means in the off-diagonal blocks.
This will then evolve the potential in this same pattern (this is
generically true, if one desires a different pattern~\cite{Anderson:2008mu}).
A measure of off-diagonal coupling of $H_s$ is
\beq
{\rm Tr}[(Q H_s P)^{\dagger} (Q H_s P)]
= {\rm Tr}[P H_s Q H_s P] \geqslant 0
\;. 
\eeq
Now we can calculate the derivative of this expression 
by applying the SRG equation \eqref{eq:srgbd3}:
\bea
\frac{d}{ds} \, {\rm Tr}[P H_s Q H_s P] 
  \ampseq
  \nonumber \\
  \amps{} \hspace*{-.55in}
{\rm Tr}[P\eta_s  Q(Q H_s Q H_s P - Q H_s P H_s P)]
\nonumber \\  
 \amps{} \hspace*{-.55in}
 \null + {\rm Tr}[(P H_s P H_s Q - P H_s Q H_s Q) Q\eta_s  P]  
 \nonumber \\
 \amps{} \hspace*{-.55in} =
  -2 \, {\rm Tr} [(Q \eta_s P)^{\dagger} (Q \eta_s  P)] \leqslant 0
  \;.
\eea
Thus the off-diagonal $Q H_s P$ block will decrease as $s$
increases.

\begin{figure}[ht!]
   \begin{center}   \includegraphics[width=2.9in]{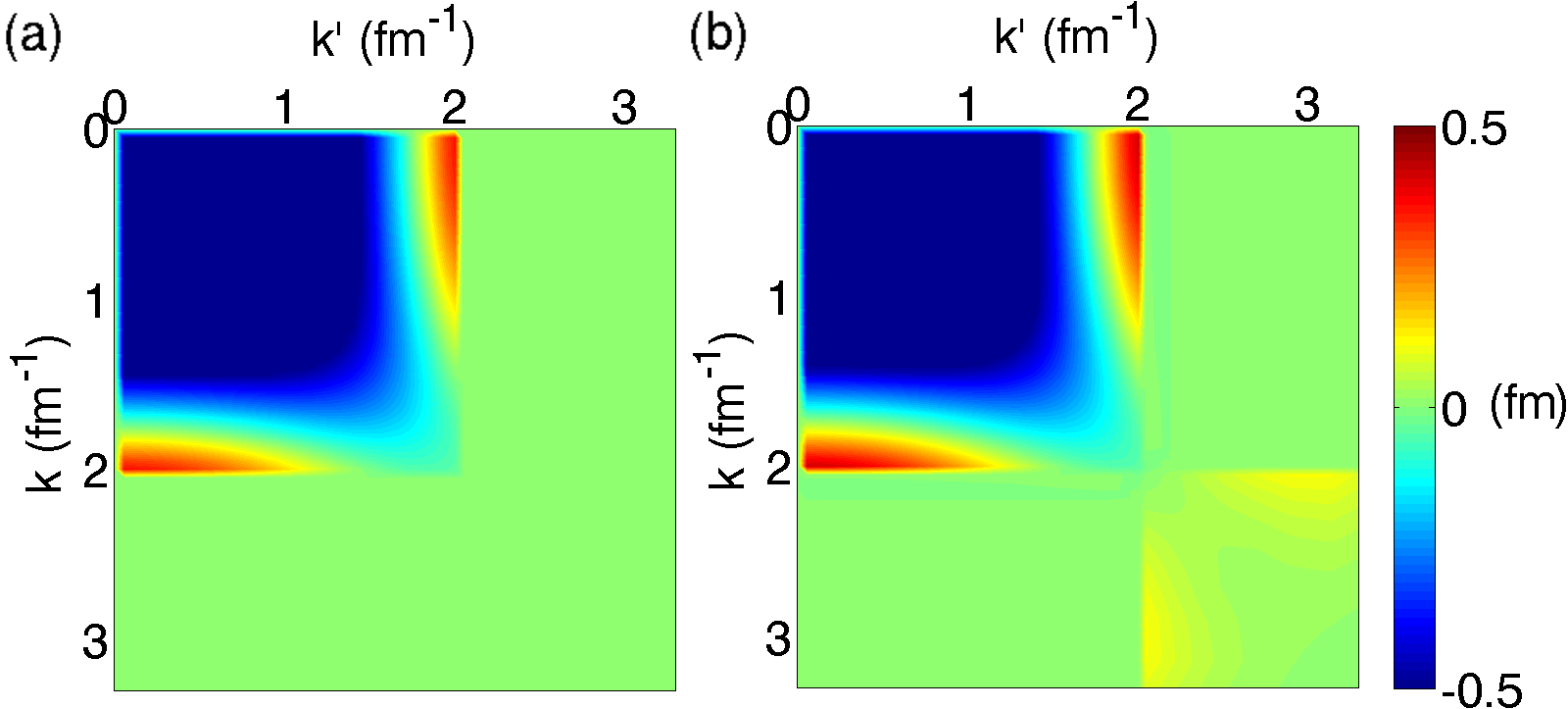}
    \vspace*{-.1in}
   \caption{Comparison of momentum-space (a) $\vlowk$ and (b) SRG 
   block-diagonal $^3$S$_1$ potentials with $\Lambda = 2\fmi$~\cite{Anderson:2008mu}.  }
   \label{fig:srgbd2}
%   \end{center}
%\end{figure}
%
%
%\begin{figure}[h!]
%   \begin{center} 
  \vspace*{.2in}
\includegraphics[width=2.9in]{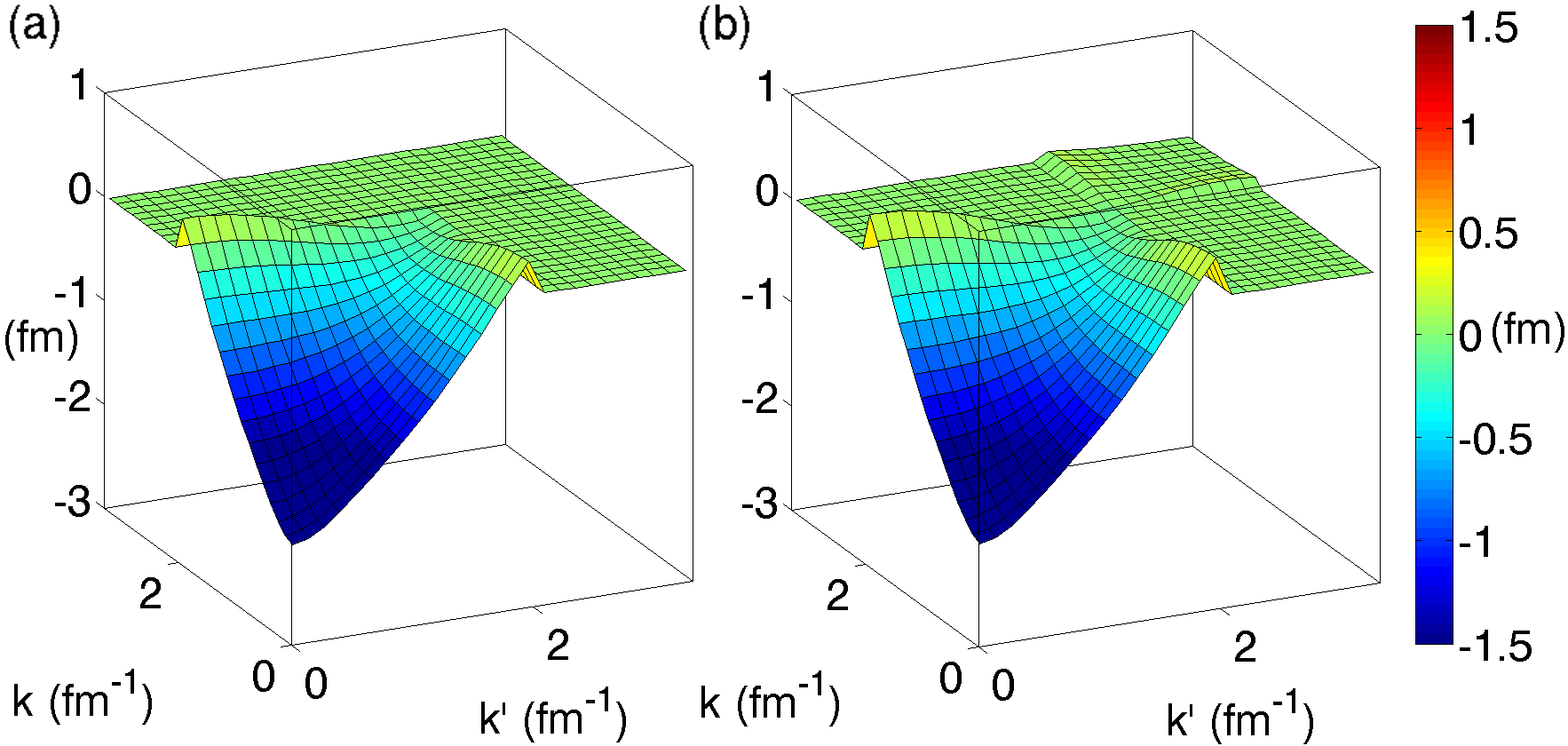}
    \vspace*{-.1in}
   \caption{Same as Fig.~\ref{fig:srgbd2}, only a surface plot.}.  
   \label{fig:srgbd3}
   \end{center}
\end{figure}

The low-momentum block of this SRG is found to be remarkably similar
to the corresponding $\vlowk$ RG potential.  
Examples are shown in Figs.~\ref{fig:srgbd2} and \ref{fig:srgbd3}.
However, \emph{proving} that these different RG approaches should
yield the same potential remains an open problem.

\subsection{Many-body forces}

\begin{figure}[t!]
\begin{center}
   \includegraphics[width=2.7in]{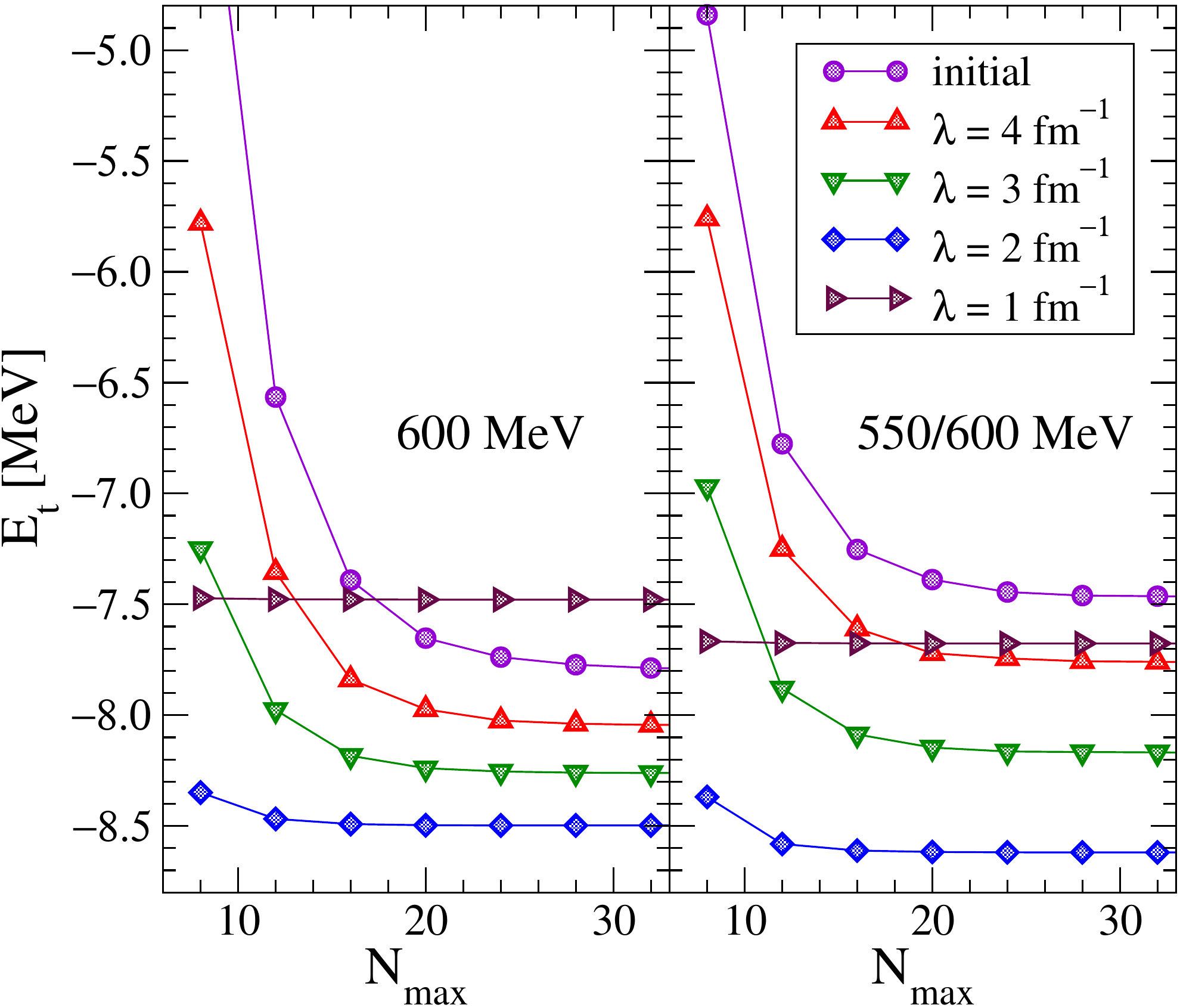}
   \caption{Convergence of three-body energies for two potentials
   and several sets of $\lambda$'s.}
   \label{fig:threebodyenergies}
\end{center}
\end{figure}

\begin{figure}[bh!]
\begin{center}
   \includegraphics[width=2.7in]{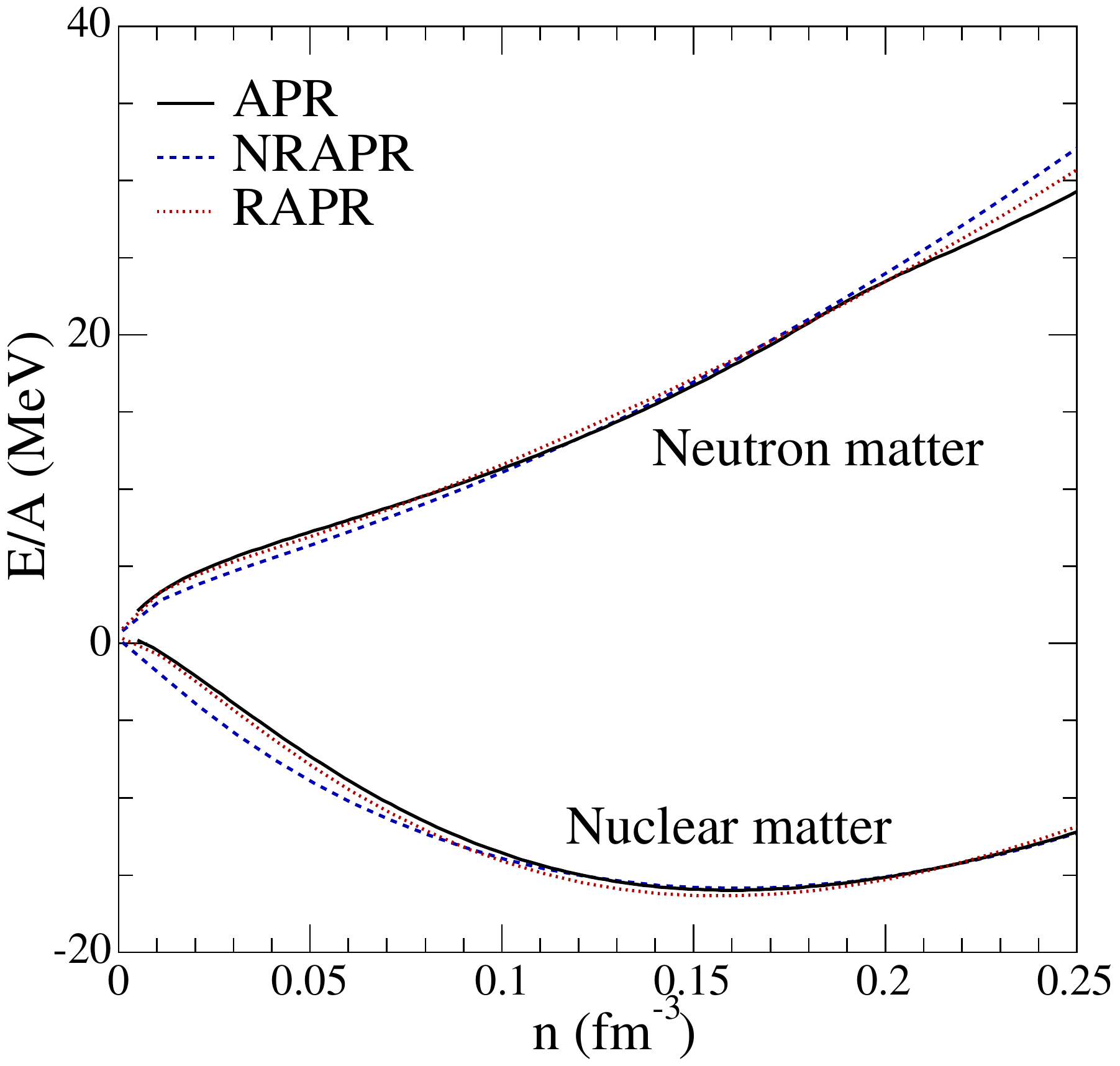}
   \vspace*{-.1in}
   \caption{Binding energies of nuclear and neutron
   matter from Akmal et al.\ for several equations of state~\cite{Akmal:1998cf}.}
   \label{fig:matter}
\end{center}
\end{figure}

\begin{figure}[t!]
\begin{center}
   \includegraphics[width=2.7in]{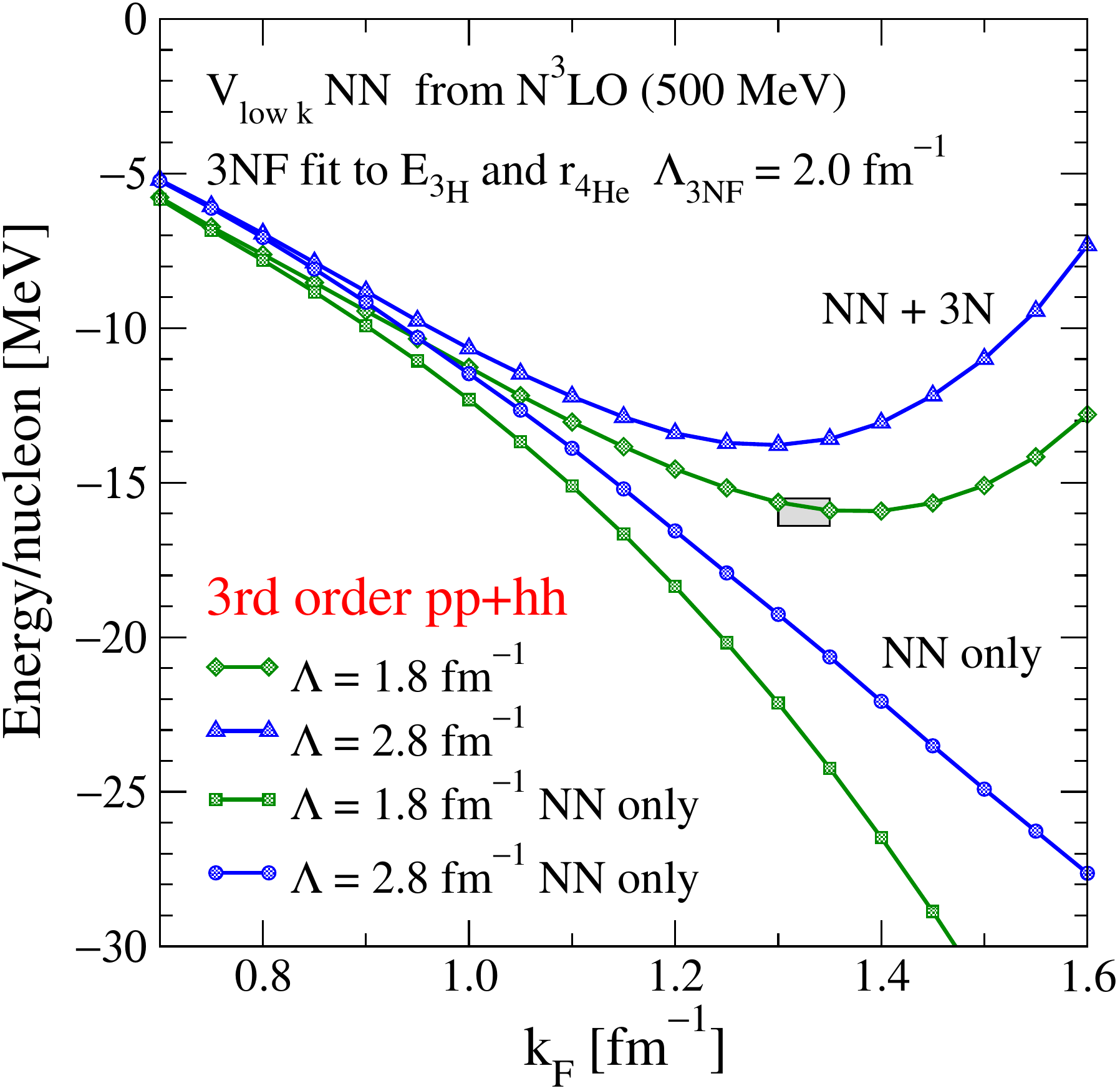}
   \vspace*{-.1in}
   \caption{Nuclear matter at third order in perturbation theory
   in the particle-particle channel (this appears sufficient for
   convergence) using a $\vlowk$ RG-evolved 
   potential~\cite{Hebeler:2010xb}.  The ``NN only'' curves include
   no three-body interactions while the ``NN + 3N'' curves include
   a 3NF fit to the triton binding energy and the alpha particle radius.}
   \label{fig:nm_vlowk}
\end{center}
\end{figure}

In Fig.~\ref{fig:threebodyenergies}, the convergence of the
triton ground-state energy with harmonic oscillator basis
size ($\Nmax\hbar\omega$ excitations)
is shown for two chiral EFT NN potentials and for the corresponding
SRG-evolved potentials at $\lambda$'s from $4\fmi$ to $1\fmi$.
In accord with our previous discussion, we see 
increasingly rapid convergence as $\lambda$ decreases.
(Note that there is softening already at $\lambda = 3\fmi$ for
the N$^3$LO EFT with $\Lambda = 600\,$MeV, which corresponds to
$3\fmi$.  The moral is that it is not sufficient to simply compare the cutoff numerically
to the decoupling scale $\lambda$.)
However, we also see that the converged (or extrapolated) binding energies are different for each $\lambda$!
How could this be, if
the SRG is supposed to generate unitary transformations?

There are further signs of trouble, such as nuclear matter failing
to saturate after the two-body potential is softened by
RG evolution.
Let's review the facts about uniform nuclear matter.
Figure~\ref{fig:matter} shows the binding energy per particle
for pure neutron matter ($Z=0$) and symmetric nuclear matter ($N=Z=A/2$) from calculations
adjusted to be consistent with extrapolations from nuclei.
Neutron matter has positive pressure, while symmetric
nuclear matter \emph{saturates}; that is, there is
a minimum at a density of about $0.16\,\mbox{fm}^3$ with a binding energy
of about $-16\,$MeV/$A$.
Reproducing this minimum with microscopic interactions fit only
to few-body data is one of the holy grails of nuclear
structure theory.
But if we evolve an NN potential with either the $\vlowk$ RG or
the SRG, we find that nuclear matter does not converge.  This is shown
for $\vlowk$ by the ``NN only'' curves in Fig.~\ref{fig:nm_vlowk} and
similar behavior is found for SRG.

The failure of softened low-momentum potentials to reproduce nuclear matter
saturation should sound familiar to anyone who knows the long
history of low-energy nuclear theory.
There were active attempts to transform away hard cores and soften
the tensor interaction in the late sixties and early seventies.
But the requiem for soft potentials was given by Bethe in 1971~\cite{Bethe:1971xm}:
  ``Very soft potentials must be excluded because they do not give
  saturation; they give too much binding and too high density.  In
  particular, a substantial tensor force is required.''
The next thirty-five years were spent 
struggling to solve accurately with such ``hard'' potentials. 
But the story is not complete: the three-nucleon forces (3NF)
were not properly accounted for!

Three-body forces between protons and neutrons have a classical analog
in tidal forces:  the gravitational force on the
Earth is not just the pairwise sum of Earth-Moon and Earth-Sun forces. 
Quantum mechanically, an analog is with the three-body force
between atoms and molecules, 
which is called the Axilrod-Teller term and dates from 1943~\cite{Axilrod:1943aa}.  
The origin is from triple-dipole mutual polarization.
It is a third-order perturbation correction, so the weakness of
the fine structure constant means that these forces are usually
negligible in metals and semiconductors.  However, in solids bound
by van der Waals potentials it can be significant; for example, it is
ten percent of the binding energy in solid xenon~\cite{Bell:1976aa}.
This is the same relative size as needed in the triton.

\begin{figure}[t!]
\begin{center}
   \includegraphics[width=2.4in]{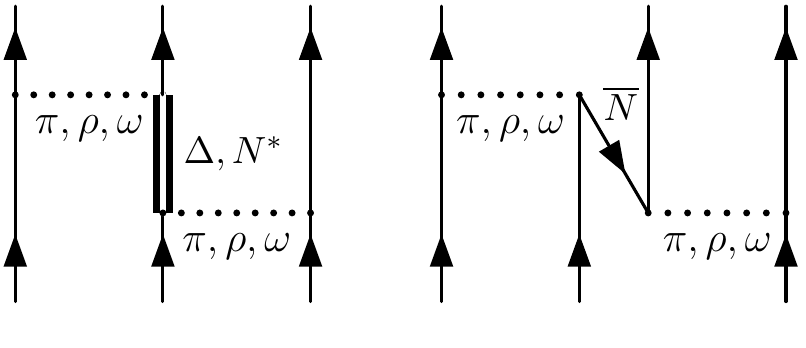}
   \vspace*{-0.05in}
   \caption{Sources of internucleon three-body forces.  On the
   left, the intermediate state includes an excitation of the
   nucleon ($\Delta$ or $N^\ast$).  On the right, the intermediate
   state includes a virtual anti-nucleon ($\overline N$).  When
   these degrees of freedom are integrated out, the resulting 
   nucleons-only vertex is a 3NF.}
   \label{fig:3N_diags_left}
\end{center}
\end{figure}

In general, three-body forces arise from eliminating degrees of
freedom.  In the nuclear case, this can mean eliminating excited
states of the nucleon ($N^\ast$ or $\Delta$) or 
from relativistic effects;
see Fig.~\ref{fig:3N_diags_left} for diagrammatic representations.
If the intermediate states are not included in the low-energy degrees
of freedom, we have irreducible vertices with three nucleons in and
three nucleons out.  
But 3NF's also result from
the decoupling of high-momentum intermediate states, whether
they are eliminated explicitly by a cutoff (as with the $\vlowk$ RG)
or the degree of coupling modified (as with the SRG).
Omitting three-body forces leads to model dependence: observables
will depend on the decoupling scale, whether it is the $\vlowk$ $\Lambda$
or the SRG $\lambda$.
This dependence also becomes a tool, because it is a diagnostic
for errors (more on this later).
To eliminate this dependence, 
the 3NF at different $\Lambda$ or $\lambda$ must be either fit or
evolved.  

It is easy to see that RG flow equations lead to many-body operators.
Consider the SRG operator flow equation written with
second-quantized $a$'s and $a^\dagger$'s:
   \bea
     \frac{dV_s}{ds} \ampseq
     \Bigl[ \Bigl[\sum\underbrace{a^\dagger a}_{G_s}, 
     \sum \underbrace{a^\dagger a^\dagger a a}_{\mbox{2-body}} \Bigr],
    \sum \underbrace{a^\dagger a^\dagger a a}_{\mbox{2-body}} \Bigr]
     \nonumber \\
     \ampseq \cdots + {\sum \underbrace{a^\dagger a^\dagger a^\dagger a a a}_{\mbox{3-body!}}}
     + \cdots
     \label{eq:2ndquant}
   \eea
where the second equality reflects that even if the initial
Hamiltonian is two-body, the commutators give rise to
three-body terms.  (For future reference, recall 
that the creation and destruction operators are always defined
with respect to a single-particle basis and a reference state, which
in this case is the vacuum.)

\begin{figure}[t!]
\begin{center}
   \includegraphics[width=2.4in]{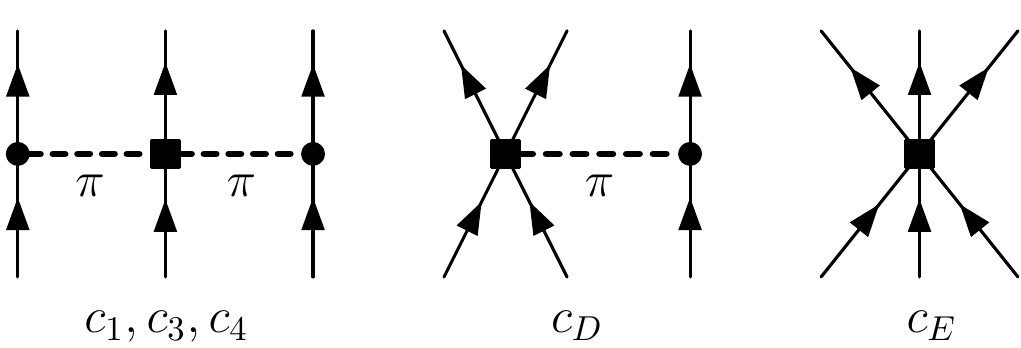}
   \vspace*{-0.05in}
   \caption{Leading three-body forces from chiral EFT. These contributions
   represent three different ranges:  long-range 2-pion exchange, short-range
   contact with one-pion exchange, and pure contact interaction.}
   \label{fig:3N_diags_right}
\end{center}
\end{figure}

%\begin{figure*}[bh!]
%\begin{center}
%   \includegraphics[width=5.in]{fig_srg_diagrams}
%   \caption{Diagrams for SRG: disconnected diagrams cancel.}
%   \label{fig:srg_diagrams}
%\end{center}
%\end{figure*}

As the evolution continues,   
there inevitably will be $A$-body forces (and operators) generated.
Is this a problem?
Not if these ``induced'' many-body forces are the same size
as those that naturally occur.	   	   
Indeed, nuclear three-body forces are already needed in 
almost all potentials in common use to get even the triton binding
energy correct.
In fact, low-energy effective theory
tell us generalized diagrams such as those in Fig.~\ref{fig:3N_diags_left}
with four or more legs imply that
there are $A$-body forces (and operators) initially!

However, there is a natural hierarchy predicted from chiral EFT, whose leading
contributions are given in Fig.~\ref{fig:3N_diags_right}
(we'll return to this in Section~\ref{subsec:chiralEFT} and supply additional details).
If we stop the flow equations before induced $A$-body forces are unnaturally
large 
or if we can
tailor the SRG $G_s$ to suppress their growth, we will be ok.
(Another option is to choose a non-vacuum reference state,
which is what is done with in-medium SRG, to be discussed later.)
Note that analytic bounds on $A$-body growth have not been derived,
so we need to explicitly monitor the contribution in different
systems.
But the bottom line that makes the SRG attractive as a method
to soften nuclear Hamlitonians is that
it is a tractable method to evolve many-body operators.

To include the 3NF using SRG with normal-ordering in the vacuum,
we start with the SRG flow equation
$dH_s/ds = [[G_s,H_s],H_s]$ (e.g., with $G_s = \Hzero$).
The right side is evaluated without 
solving bound-state or scattering equations, unlike the
situation with $\vlowk$, so the SRG
can be applied directly in the three-particle space.
The key observation is that for normal-ordering in the vacuum,
$A$-body operators are completely fixed in the $A$-particle subspace. 
Thus we can first solve for the evolution of the two-body potential
in the $A=2$ space, with no mention of the 3NF (either initial
or induced), and then use this NN potential in the equations
applied to $A=3$. 

What about spectator nucleons?
There is a decoupling of the 3NF part.
We can see this from the first-quantized version of the SRG
flow equation,
\bea
    \frac{dV_s}{ds} \ampseq 
    \frac{d\Vtwo12}{ds} + \frac{d\Vtwo13}{ds} + \frac{d\Vtwo23}{ds}
    + \frac{d\Vthree}{ds} 
    \nonumber \\
    \ampseq [[\Trel, V_s], H_s] \,,    
\eea
where we isolate the contributions from each pair and the 3NF.
Using each SRG equation for the two-body derivatives,
we can cancel them against terms on the right side.
The result is~\cite{Bogner:2006pc}:
\bea
   \frac{d\Vthree}{ds} \ampseq
   [[\Ttwo12,\Vtwo12], (T_3 + \Vtwo13 + \Vtwo23 + \Vthree)]
   \nonumber \\
   && \null
   + \{123\rightarrow132\} 
   + \{123\rightarrow231\} \nonumber \\
   && +  [[\Trel,\Vthree],H_s] \;. 
\eea	
The key is that there are no ``multi-valued'' two-body interactions
remaining
(i.e., dependence on the excitation energy of unlinked spectators);
all the terms are connected.
An implementation of these equations in a momentum basis would
be very useful and has very recently been achieved by Hebeler~\cite{Kai}.
But an alternative approach has also succeeded:
a direct solution in a discrete  basis~\cite{Jurgenson:2008jp,Jurgenson:2009qs,Jurgenson:2010wy}.

The idea is that the SRG flow equation is an operator equation,
and thus we can choose to evolve in any basis.  If one chooses
a discrete basis, than a separate evolution of the three-body
part is not needed.  This was first done for nuclei by Jurgenson and
collaborators in 2009 using an anti-symmetrized Jacobi harmonic
oscillator (HO) basis~\cite{Jurgenson:2009qs}.  
The technology for working with such a basis had already been
well established for applications to the no-core shell model (NCSM)~\cite{Navratil:1999pw}.
This approach leads to SRG-evolved matrix elements of the potential
directly in the HO basis, which is just what is needed for many-body 
applications such as NCFC or coupled cluster.

\begin{figure}[t!]
\begin{center}
   \includegraphics[width=2.5in]{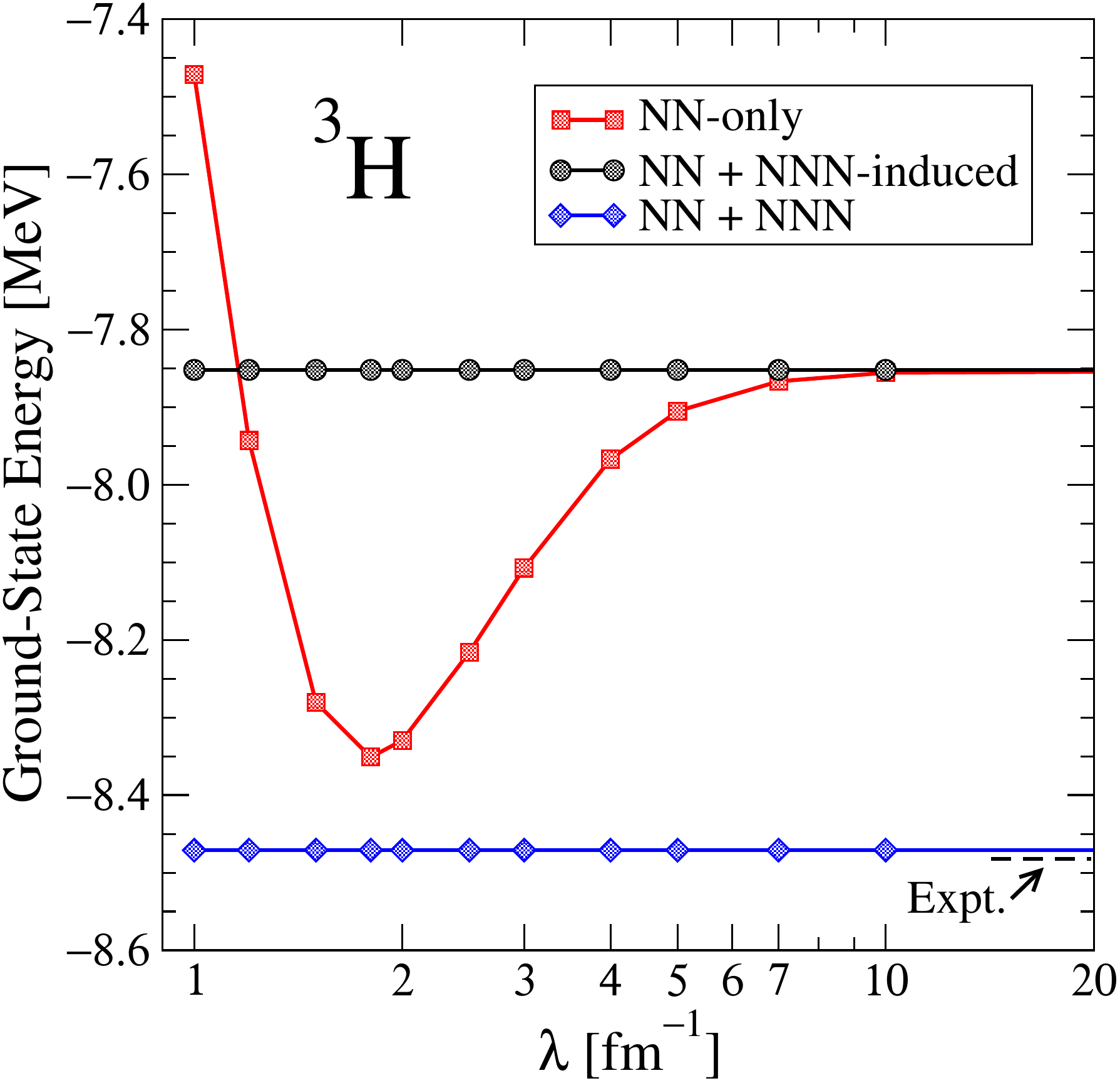}
   \caption{Triton binding energy during SRG evolution.  The three curves
   are for an initial potential with only NN components where the
   induced 3NF is not kept (``NN-only''), for the same initial NN
   potential but keeping the induced 3NF (``NN + NNN-induced''), and
   with an initial NNN included as well (``NN + NNN'').}
   \label{fig:H3_Ebind_lambda}
\end{center}
\end{figure}

\begin{figure}[tb!]
\begin{center}
   \includegraphics[width=2.5in]{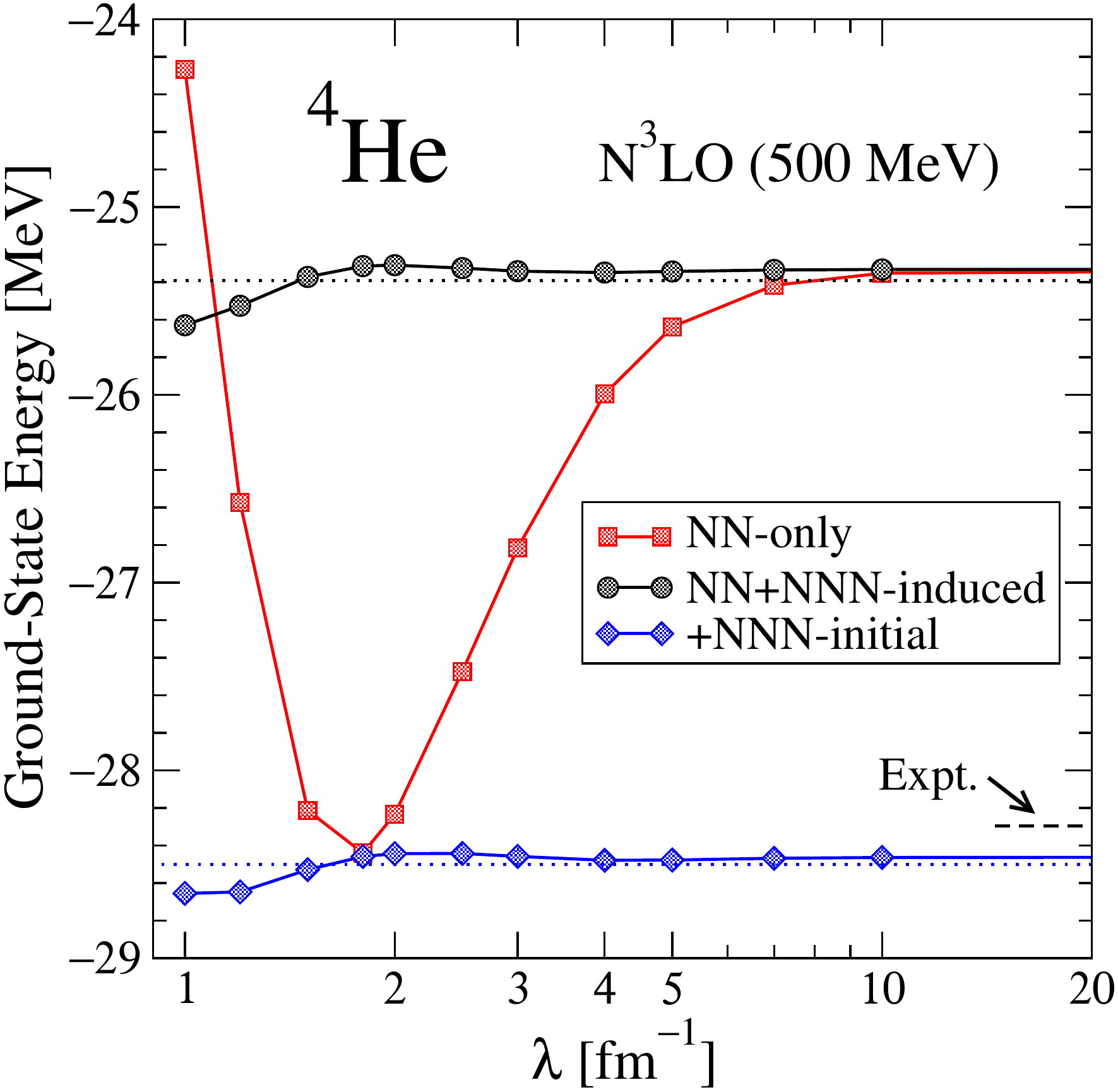}
   \caption{Alpha particle binding energy during SRG evolution.
   The curves correspond to those in Fig.~\ref{fig:H3_Ebind_lambda}.}
   \label{fig:He4_Ebind_lambda}
\end{center}
\end{figure}

In Fig.~\ref{fig:H3_Ebind_lambda}, the comparison of two-body-only
to full two-plus-three-body evolution is shown for the
triton ($^3$H).  The NN-only curve
uses the evolved two-body potential.  The change in energy with
$\lambda$ reflects the violation of unitarity by omission of the
induced three-body force.  When this induced 3NF is included
(``NN + NNN--induced''), the energy is independent of $\lambda$
for $A=3$.  If we now turn to the alpha particle ($^4$He) in
Fig.~\ref{fig:He4_Ebind_lambda}, we see similar behavior, except
now the inclusion of the induced
3NF does not lead to a completely flat
curve at the lowest $\lambda$ values. If there is sufficient 
convergence, this is a signal of missing induced 4NF.  

\begin{figure}[t!]
\begin{center}
   \includegraphics[width=2.5in]{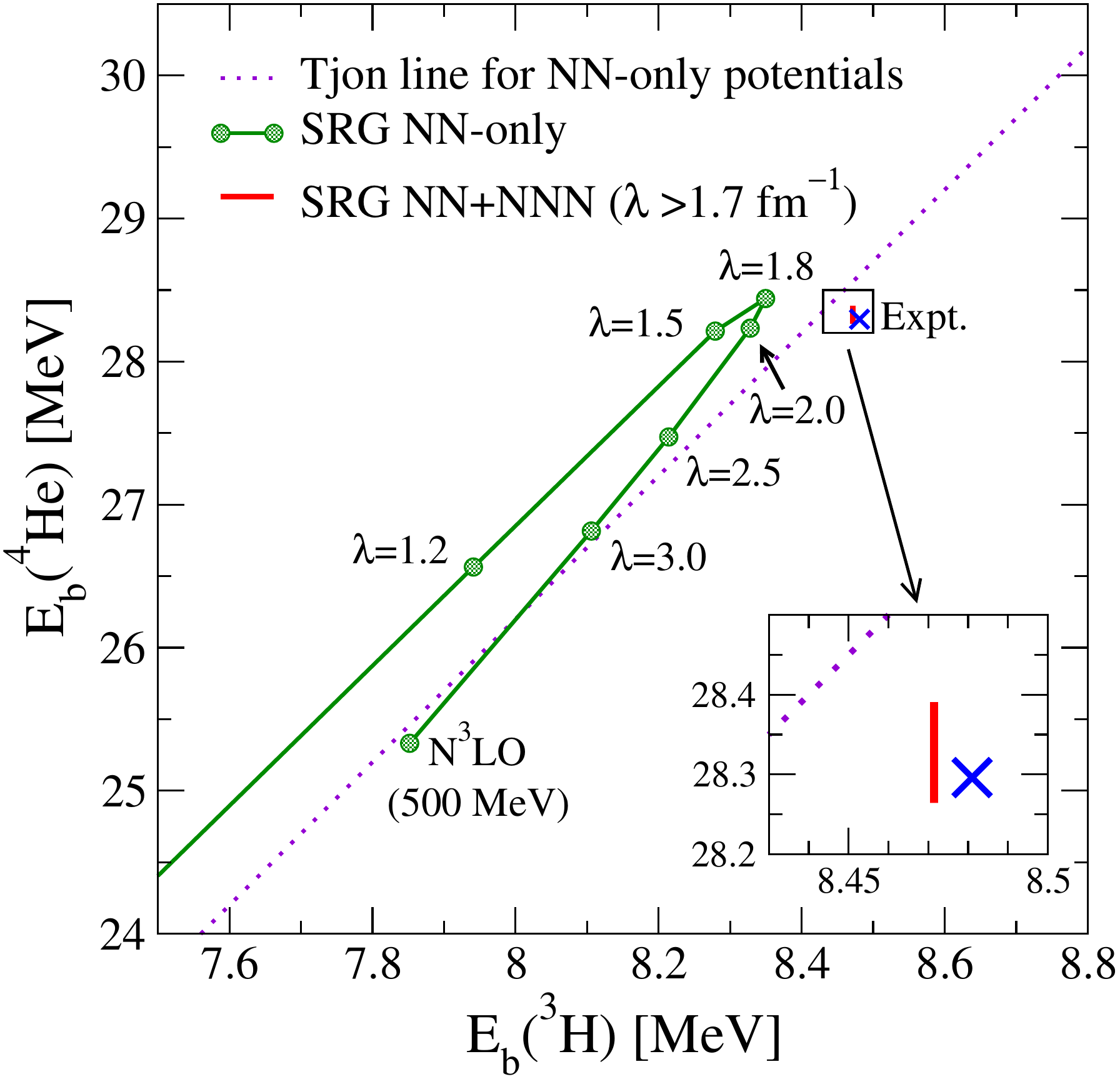}
   \caption{Correlation plot of the binding energies of the alpha
   particle and the triton.  The dotted line connects (approximately) 
   the locus of
   points found for phenomenological potentials, and is
   known as the Tjon line~\cite{Nogga:2000uu}.}
   \label{fig:tjon_line}
\end{center}
\end{figure}

In both cases, it is evident that starting with an initial NN-only
interaction (in this case, an \nthreelo(500\,MeV) interaction~\cite{Entem:2003ft}), does
not reproduce experiment.  The third line in each plot 
of Figs.~\ref{fig:H3_Ebind_lambda} and \ref{fig:He4_Ebind_lambda} shows
that an initial 3NF (labeled NNN) contribution leads to a good
reproduction of experiment.  The triton energy is part of the
fit of this initial force, but the alpha particle energy
is a prediction.  
Note that the magnitude of the NN-only variation is comparable
to the initial 3NF needed.  This is an example of the natural
size of the 3NF being manifested by the running of the potential
(which is, in effect, the beta function).

The nature of the evolution is illustrated in Fig.~\ref{fig:tjon_line},
which is a correlation plot of the binding energies in each nucleus.
The dotted line is known as the Tjon line for NN-only phenomenological
potentials.  It was found that different potentials that fit NN
scattering data gave different binding energies, but that they
clustered around this line.  With the SRG evolution starting with
just an NN potential, the path follows
the line, passing fairly close to the experimental point.
With an initial NNN force and keeping the induced 3-body part,
the trajectory is greatly reduced (see inset), at least until
$\lambda$ is small.

\begin{figure}[t!]
\begin{center}
  \includegraphics[width=2.5in]{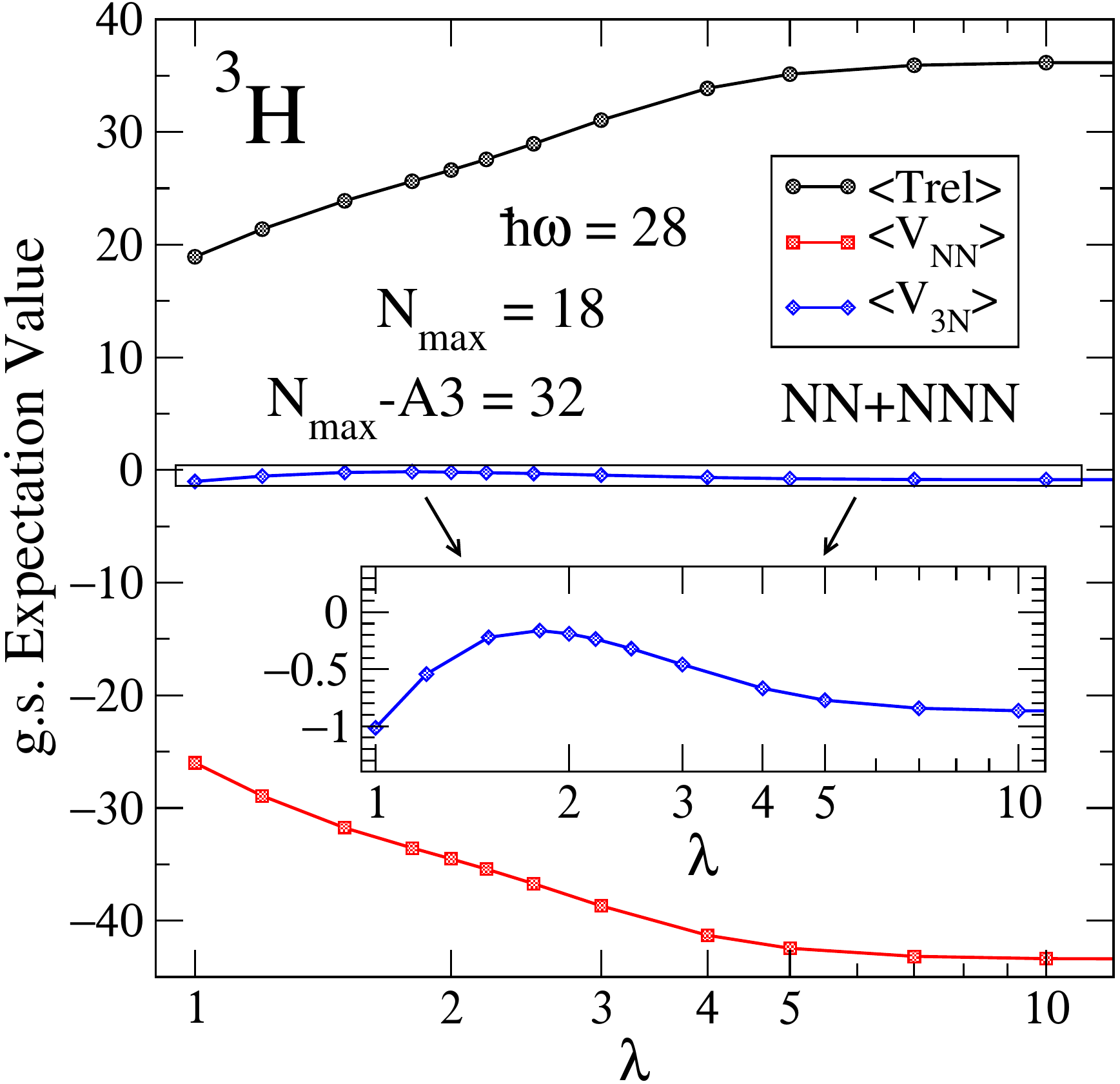}
   \caption{Contributions to the triton binding energy during SRG evolution.
   Plotted are the expectation values of the kinetic energy, the two-body
   potential, and the three-body potential~\cite{Jurgenson:2010wy}.}
   \label{fig:H3_Ebind_contributions}
\end{center}
\end{figure}

\begin{figure}[tbh!]
\begin{center}
 \includegraphics[width=2.5in]{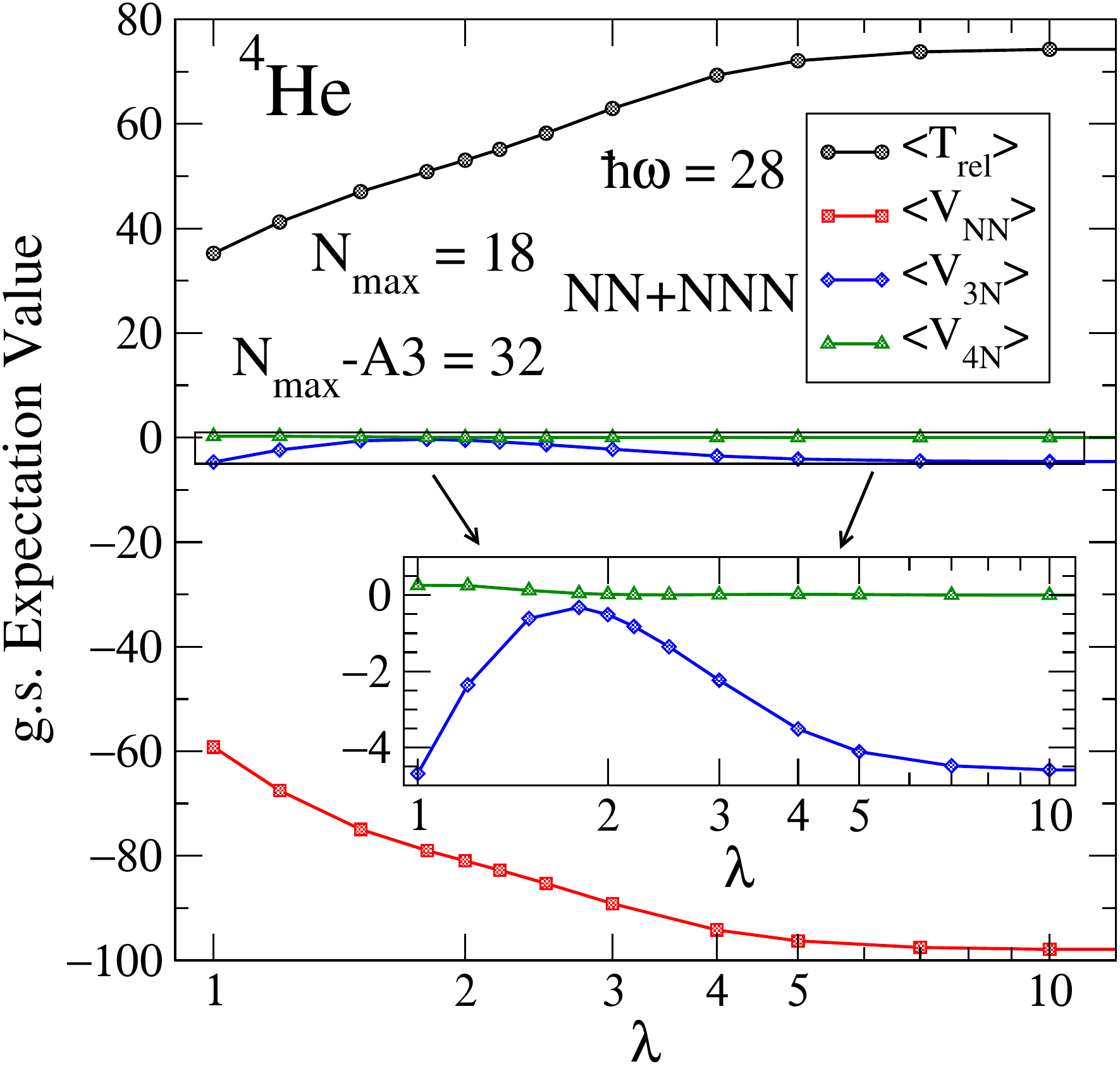}
   \caption{Contributions to the alpha particle binding energy during SRG evolution.  Plotted are the expectation values of the kinetic energy, the two-body
   potential, the three-body potential, and the four-body potential~\cite{Jurgenson:2010wy}.}
   \label{fig:He4_Ebind_contributions}
\end{center}
\end{figure}

Figures~\ref{fig:H3_Ebind_contributions} and \ref{fig:He4_Ebind_contributions}
show individual contributions to the energy in the form of ground-state
matrix elements of the kinetic energy, two-body, three-body, and (implied)
four-body potentials.  The hierarchy of contributions is quite clear
but the graphs also manifest the strong cancellations between the NN
and kinetic energy contributions.  These cancellations magnify the impact
of higher-body forces.
Even so, it appears that a truncation including the NNN but omitting
higher-body forces is workable, particularly with $\lambda > 1.5\fmi$.
But what about the $A$ dependence of the 4NF (and beyond)?

\begin{figure}[t]
\begin{center}
 \includegraphics[width=2.5in]{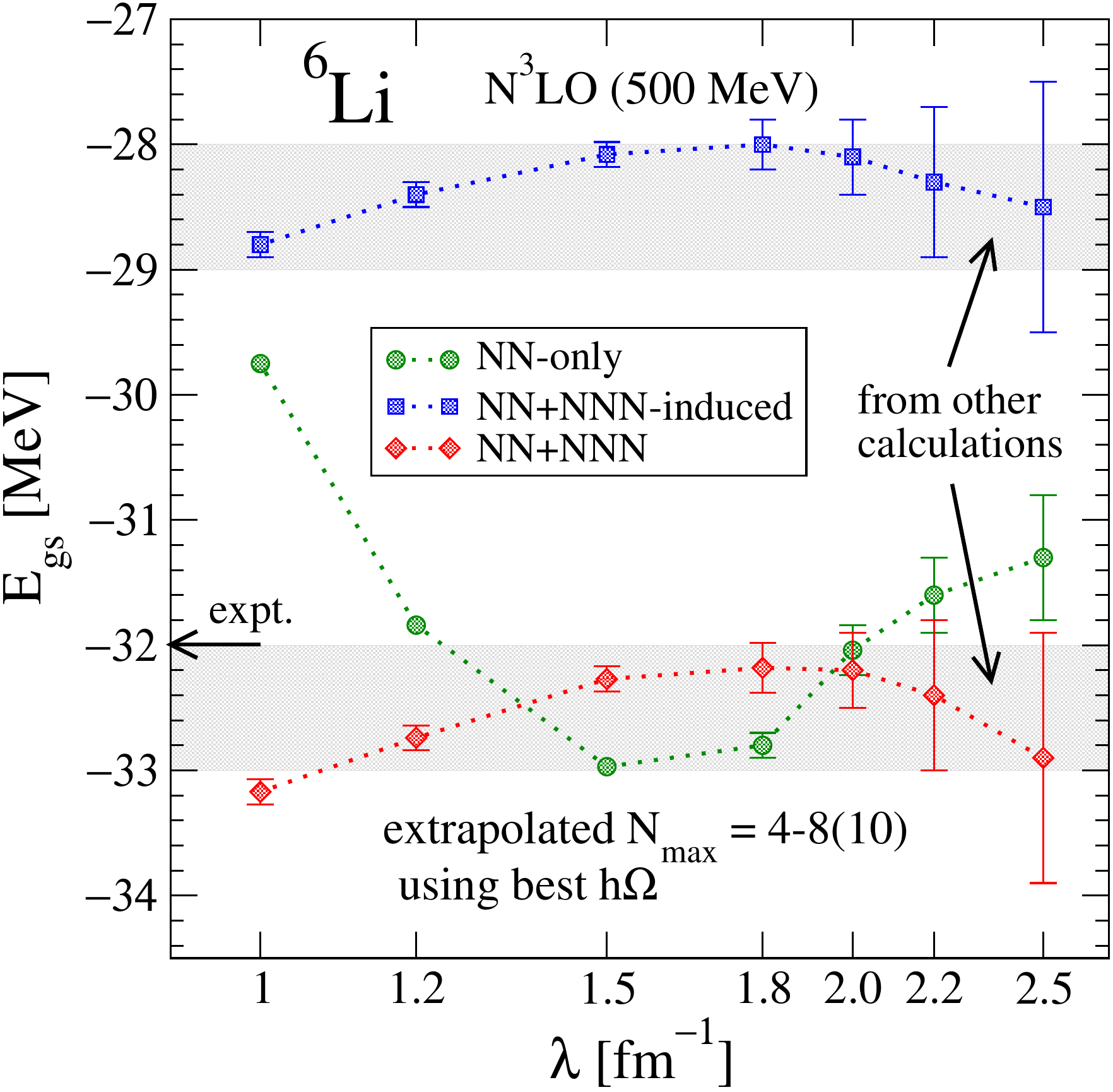}
   \caption{SRG running of the $^6$Li ground-state energy.  The error
   bars are estimates of the errors from extrapolating the results
   to $N_{\rm max} = \infty$.  The gray shaded region shows the
   uncertainty from NSCM calculations using a Lee-Suzuki effective
   interaction~\cite{Jurgenson:2010wy}.}
   \label{fig:Li6_Ebind_running}
\end{center}
\end{figure}

This $A$ dependence is the topic of current research.
In Fig.~\ref{fig:Li6_Ebind_running}, results for $^6$Li 
are shown~\cite{Jurgenson:2010wy}.
Assessing these results is made difficult because of insufficient
convergence of comparison calculations (the shaded areas) and of calculations
at the large $\lambda$ values.  Note, however, that the variations are
with the 1\,MeV level; nevertheless we expect to do better.

Roth and collaborators have since used importance truncated NCSM
(IT-NCSM) to extend these calculations to much higher $\Nmax$ and
all the way to $^{16}$O~\cite{Roth:2011ar}.  
They find that the SRG with initial NN-only
but including the induced NNN shows only small running with $\lambda$.
On the other hand, with increasing $A$ they find significant deviations.
This has been traced to the influence of the initial long-range NNN
interaction.
If they lower the cutoff of this part of the interaction, then approximate
SRG unitarity is restored and with coupled-cluster methods
they find reasonable results even
for medium-size nuclei (although not yet with fully consistent Hamiltonians)~\cite{Roth:2011vt}. 

\subsection{Summary points}

Renormalization group flow equations dramatically reduce correlation
in many-body wave functions, leading to faster convergence of
many-body calculations.
Flow equations (SRG) achieve this lower resolution by decoupling
via a series of unitary transformations, which leave observables
invariant (if no approximations are made) but alter the physics
interpretation.
Few-body forces are inevitable, but the flow-equation approach allows
the evolution of vacuum interactions.  

%%%%%%%%%%%%%%%%%%%%%%%%%%%%%%%%%%%%%%%%%%%%%%%%%%%%%%%%%%%%%%%%%%%%%%%%
%%%%%%%%%%%%%%%%%%%%%%%%%%%%%%%%%%%%%%%%%%%%%%%%%%%%%%%%%%%%%%%%%%%%%%%%

\section{Features of SRG applied to nuclear problems}
\label{sec:features}

\subsection{Chiral EFT, many-body forces, and the SRG}
\label{subsec:chiralEFT}

Before continuing with the SRG for nuclear systems, let's say a bit
more about chiral effective field theory (EFT).  In the SRG flow
equations, the input interaction is merely an initial condition;
the equations are the same whether we start with an EFT potential
or a more phenomenological potential such as Argonne $v_{18}$.
However, increasingly nuclear theorists are moving toward using
EFT interactions because they promise a more systematic construction
of many-body forces and consistent operators.

There are three fundamental ingredients of an effective field theory
(e.g., see Ref.~\cite{Beane:2000fx}).  The first is to use the most
general lagrangian with low-energy degrees of freedom consistent with
the global and local symmetries of the underlying theory.  For nuclei,
the underlying theory is quantum chromodynamics (QCD).
We can identify a hierarchy of nuclear QCD scales:
      \bi
        \I $M_{\rm QCD} \sim 1\,$GeV [$M_{\rm hadrons}$, $4\pi f_\pi$]
	\I $M_{\rm nuc} \sim 100\,$MeV [$\kf$, $f_\pi$, $m_\pi$,
	$\delta_{\Delta N}$]
	\I $M_{\rm nuc}^2/M_{\rm QCD} \sim 10\,$MeV [nuclear binding energy
	per nucleon]
      \ei
For all but the lightest nuclei, the EFT of choice at present draws
a line between the first two levels to define the high-energy and
low-energy scales.  This is the chiral EFT, with degrees of freedom
consisting of the nucleon (proton and neutron) and the pion~\cite{Epelbaum:2008ga}.  In the
near future, the $\Delta$ resonance will be included as well because
of the small mass difference with the nucleon, $\delta_{\Delta N}$.
Besides the usual space-time symmetries,
terms in the chiral EFT lagrangian are constrained by the
requirements of spontaneously broken (as well as explicitly broken)
chiral symmetry~\cite{Epelbaum:2008ga}.

The other two ingredients are the declaration of a regularization and
renormalization scheme, and the identification of a well-defined
power counting based on well-defined expansion parameters.  
The separation of scales provides the expansion parameter as
ratios of $Q/M_{\rm QCD}$, where $Q$ is one of the quantities
lumped together above as $M_{\rm nuc}$.
The chiral EFT potentials used here are derived
using a momentum cutoff and what is called ``Weinberg counting'',
in which the counting is done at the level of an irreducible
potential that is summed non-perturbatively with the Lippmann-Schwinger
equation.  
This scheme has been criticized because it does not allow
systematic renormalization (meaning, in this context, the removal of
cutoff dependence at each order), which limits the range of cutoffs used.
This in turn hinders the validation of the EFT, because
the sensitivity to the cutoff is used as a measure of uncertainties,
See Ref.~\cite{Epelbaum:2008ga} for a thorough overview of
chiral EFT for nuclei and the status of the power counting and renormalization
controversies.

With any scheme, however, the power counting implies a hierarchy
of many-body contributions.  Weinberg counting associates a power
$\nu$ of $Q/M_{\rm QCD}$ with diagrams for the potential,
where
\beq
  \nu = -4 + 2N + 2L + \sum_i (d_i + n_i - 2) \;.
\eeq
The definitions and details of its implementation can be found in 
Ref.~\cite{Epelbaum:2008ga}.  For our purposes, the relevant
term is ``$2N$'', which says that adding a nucleon to go from
an $A$--body potential to an $A+1$--body potential generally suppresses
the contribution by $Q^2/M_{\rm QCD}$.
In the theory without $\Delta$'s, the suppression of the leading 3NF
compared to the leading NN interaction is actually $(Q/M_{\rm QCD})^3$,
and a four-body force first appears at order $(Q/M_{\rm QCD})^4$
\cite{Epelbaum:2008ga}.
It is this hierarchy that we want to preserve as we run our SRG
flow equations.

As noted, the flow equation technology discussed here does not rely
on a particular implementation of chiral EFT, except that the SRG
is inherently non-perturbative.  This apparently
excludes alternative renormalization
and power counting schemes that require a perturbative treatment
beyond leading order~\cite{Beane:2001bc,Bedaque:2002mn,Nogga:2005hy}.  The current belief is that the two approaches
should give comparable results as long as the EFT cutoff is taken
to be of order $M_{\rm QCD}$, but the issue is far from settled~\cite{Epelbaum:2008ga}.

Another consideration is whether one could bypass the SRG by
simply applying chiral EFT with a lower cutoff.
Indeed, there exists low-cutoff implementations on the market 
that display similar characteristics to low-momentum RG-evolved
interactions.  However, the lower cutoff also means the effective
expansion parameter is smaller, and therefore the truncation error
is reduced.  With the RG, one preserves the truncation error from
the cutoff of order $M_{\rm QCD}$.  An additional advantage of
the RG is the controlled variation of the decoupling scale, which
provides a tool for assessing errors from the Hamiltonian truncation
and many-body approximations.  But this is also not a settled issue
and further study would be welcome.

\subsection{More perturbative nuclear systems flow equations}

Earlier we mentioned the role of RG in high-energy physics in improving
perturbation theory.
Much of low-energy nuclear physics is intrinsically non-perturbative
because of large scattering lengths and bound states, so how do
we quantify ``perturbativeness''?  
We study the convergence of the Born series for scattering
to see how this can be done.

Consider whether the Born series for the T-matrix operator
at a given (complex) $z$,  
 \beq
   T(z) = V + V \frac{1}{z-H_0}V + V\frac{1}{z-H_0}V\frac{1}{z-H_0}V +
   \cdots
 \eeq
converges.
This is something like a geometric series, for which we know clearly
the convergence criterion:  $1 + w + w^2 + \cdots$ diverges if
$|w| \geq 1$.
We get a clue for how to use this if we consider a bound state
$|b\rangle$ and the special value $z = E_b$, which is the bound-state
energy.  Then we can rearrange the Schr\"odinger equation
  \beq 
    (H_0 +V)|b\rangle = E_b | b\rangle 
  \eeq
to the form
  \beq
     \frac{1}{E_b-H_0}V|b\rangle = |b\rangle
     \label{eq:bswein}
  \eeq
and look at $T(E_b)|b\rangle$.  Using Eq.~\eqref{eq:bswein} repeatedly,
the divergence is manifest, i.e., we get $V(1+1+1+\cdots)|b\rangle$.  

\begin{figure}[t!]
\begin{center}
 \includegraphics[width=2.5in]{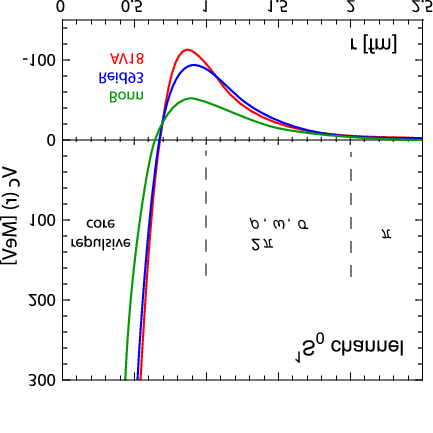}
   \vspace*{-0.25in}
   \caption{The potentials of Fig.~\ref{fig:phenpots} inverted as
   part of the Weinberg eigenvalue analysis (see text).}
   \label{fig:weinberg1}
\end{center}
  \vspace*{-0.1in}
\end{figure}

\begin{figure}[b!]
\begin{center}
 \includegraphics[width=2.8in]{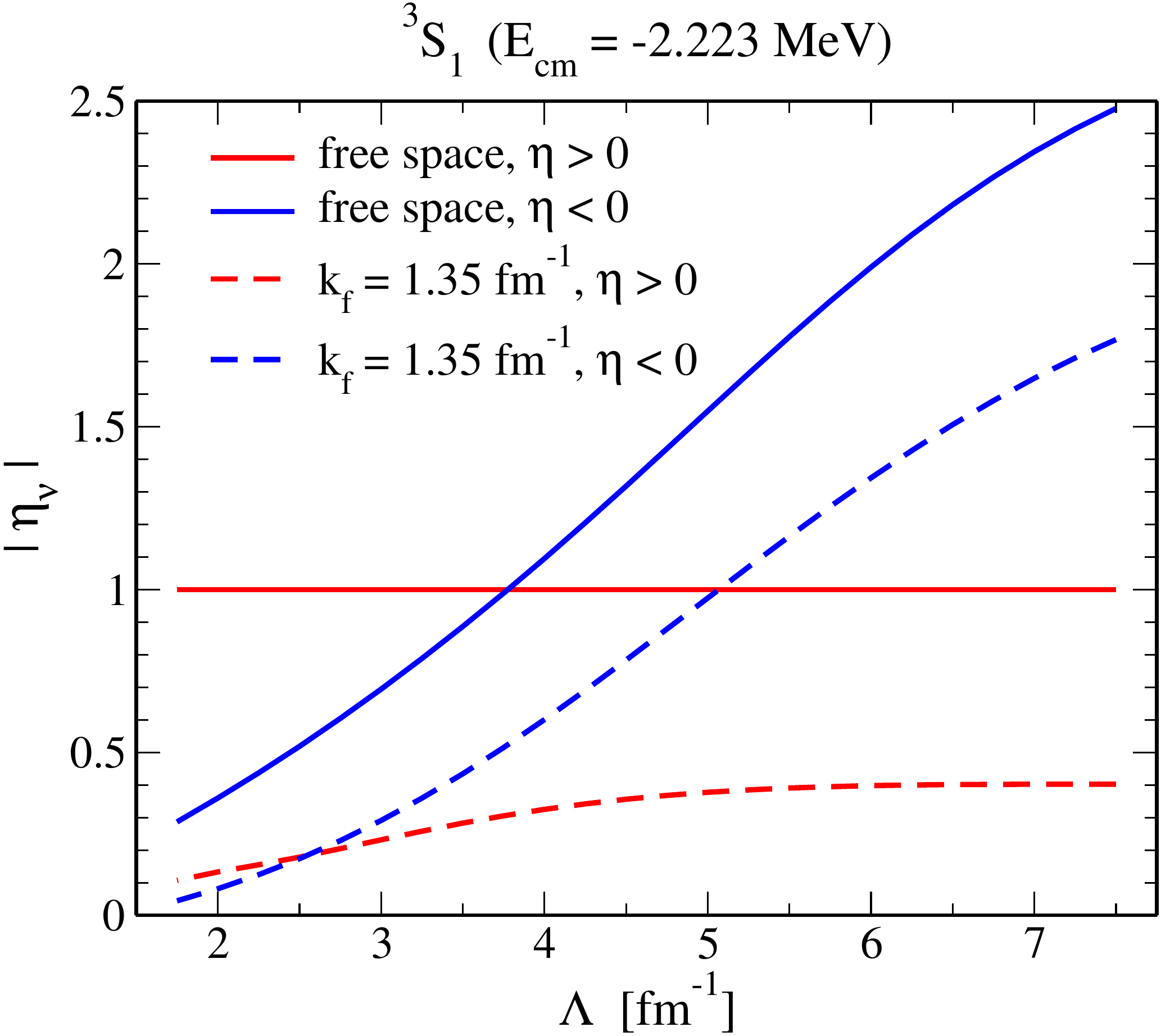}
   \vspace*{-0.05in}
   \caption{$\vlowk$ RG evolution of the largest positive and negative Weinberg
   eigenvalue at fixed energy (corresponding to the deuteron binding
   energy) for AV18 in the $^3$S$_1$ channel~\cite{Bogner:2006tw}.  }
   \label{fig:weinberg2}
\end{center}
\end{figure}
  
Now we see that we can generalize Eq.~\eqref{eq:bswein} for fixed $E$
by looking for eigenstates of $(E-H_0)^{-1}V$ with 
eigenvalue $\eta_\nu(E)$,
  \beq
     \frac{1}{E-H_0}V|\Gamma_\nu\rangle
       = \eta_\nu(E) |\Gamma_\nu\rangle  \;.
       \label{eq:eigwein}
  \eeq
Then with $T$ applied to these eigenstates, there is manifestly a divergence
     for $|\eta_\nu(E)| \geq 1$: 
  \beq
    T(E)|\Gamma_\nu\rangle 
      = V|\Gamma_\nu\rangle
        (1 + \eta_\nu + \eta_\nu^2 + \cdots)
	\;.
  \eeq
So we characterize the perturbativeness of a potential at energy
$E$ by the $\eta_\nu(E)$ with the largest magnitude, which dictates
convergence of the T~matrix.  
This analysis follows work in the early 1960's by Weinberg~\cite{Weinberg:1963zz}, 
so we call $\eta_\nu$ a ``Weinberg eigenvalue'',
although others have made similar treatments of the convergence of
the Lippmann-Schwinger series.

\begin{figure}[t!]
\begin{center}
 \includegraphics[width=2.4in]{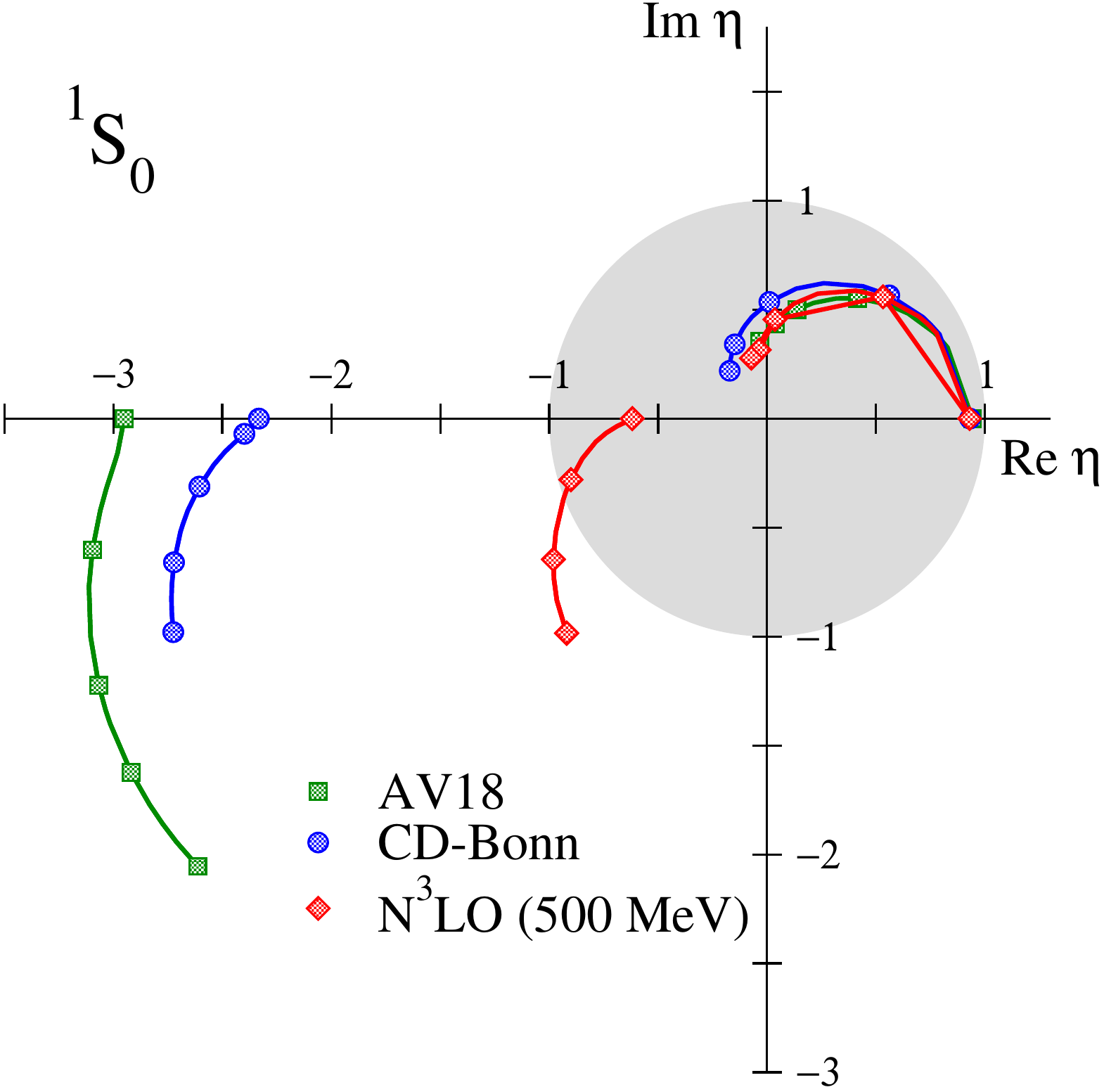}
 \includegraphics[width=2.4in]{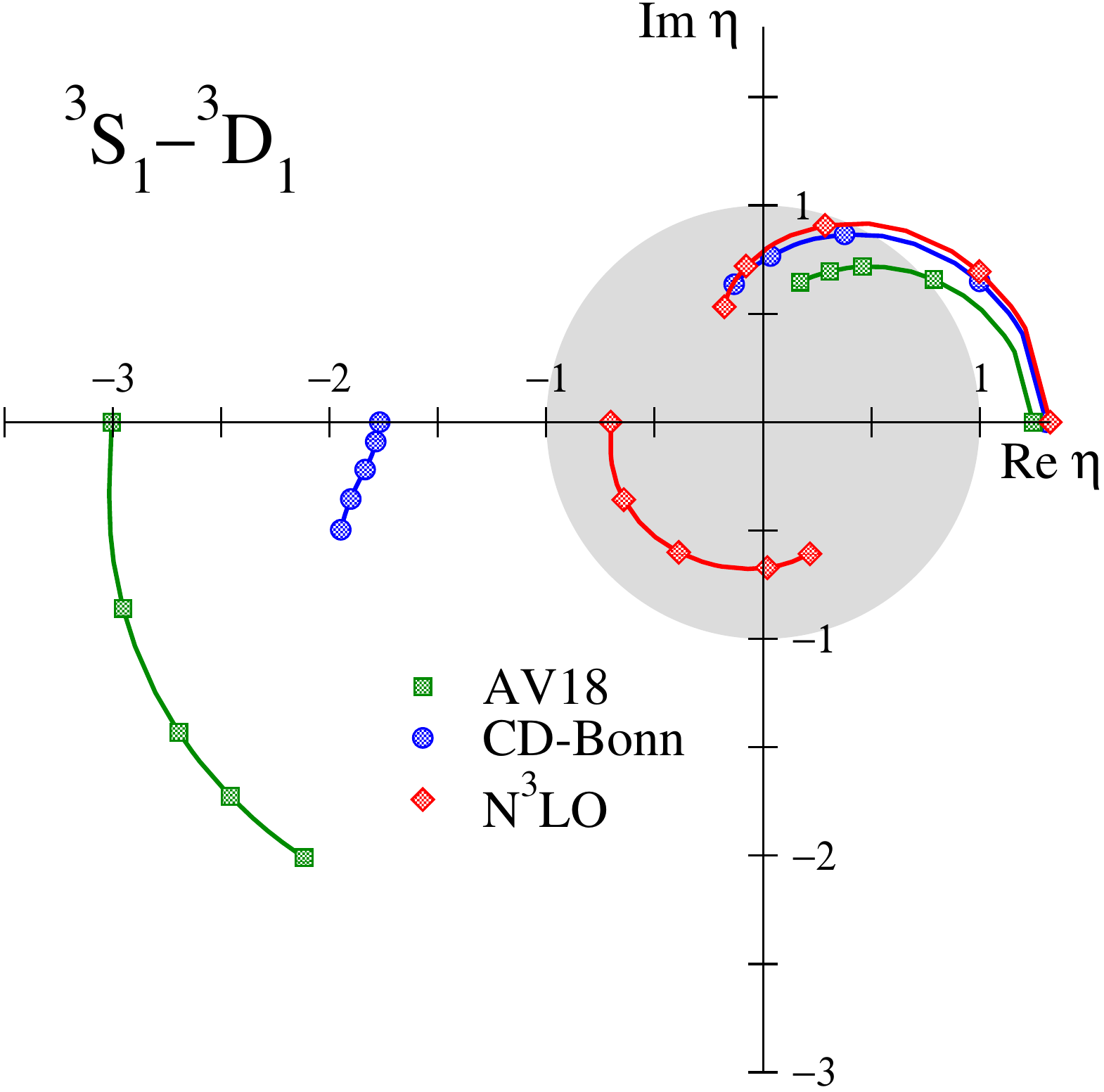}
 \vspace*{-.05in}
   \caption{Trajectories of Weinberg eigenvalues in the complex
   plane for several realistic NN potentials in two channels.  The symbols
   go from 0\,MeV on the axis to 25, 66, 100, and 150\,MeV~\cite{Bogner:2006tw}.}
   \label{fig:weinberg3}
\end{center}
 \vspace*{-.2in}
\end{figure}

If we compare Eq.~\eqref{eq:eigwein} to Eq.~\eqref{eq:bswein},
we can express the convergence criterion for $E < 0$ as saying that 
$T(E)$ diverges if there exists a bound state at $E$ for $V/\eta_\nu$
with $|\eta_\nu| \geq 1$.  Or, in other words, what is the
largest $\eta_\nu$ for which $V/\eta_\nu$ supports a bound state at $E$?
We have convergence only for $\eta_\nu < 1$.
This means we'll have two types of eigenvalues, because $\eta_\nu$ could
be negative (``repulsive'') as well as positive (``attractive'').
The negative eigenvalue corresponds to looking for bound states with
a scaled ``flipped'' potential, as in Fig.~\ref{fig:weinberg1}.
We see that the repulsive core becomes a deep attractive well, 
implying that we will have a large negative eigenvalue in these
cases.  But then we expect that RG evolution to a softened form,
which eliminates the core, should result in decreased eigenvalues.

\begin{figure*}[t!]
%
%\vfill
%
\begin{center}
 \includegraphics[width=2.5in]{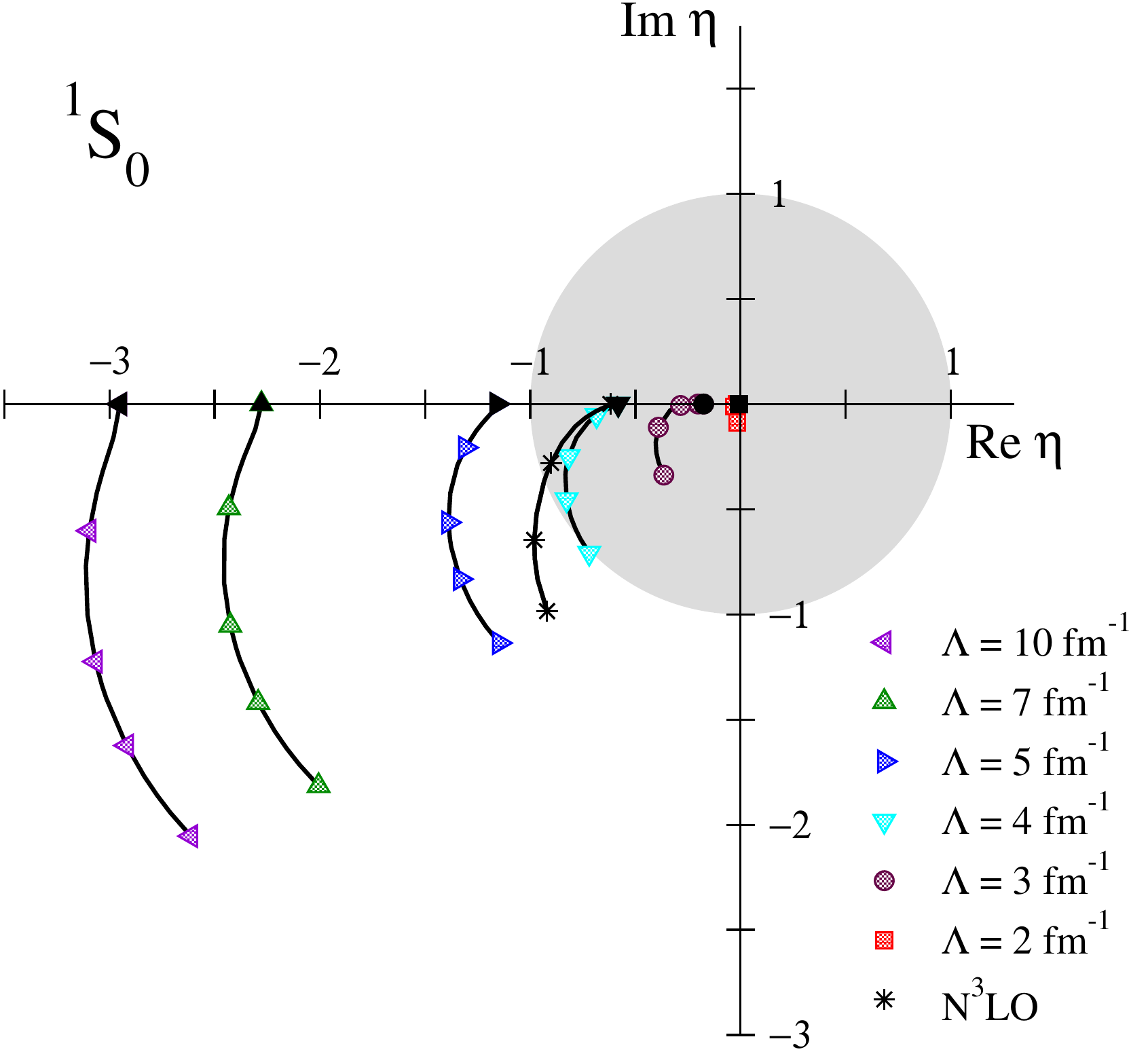}
 \hspace*{.2in}
 \includegraphics[width=2.5in]{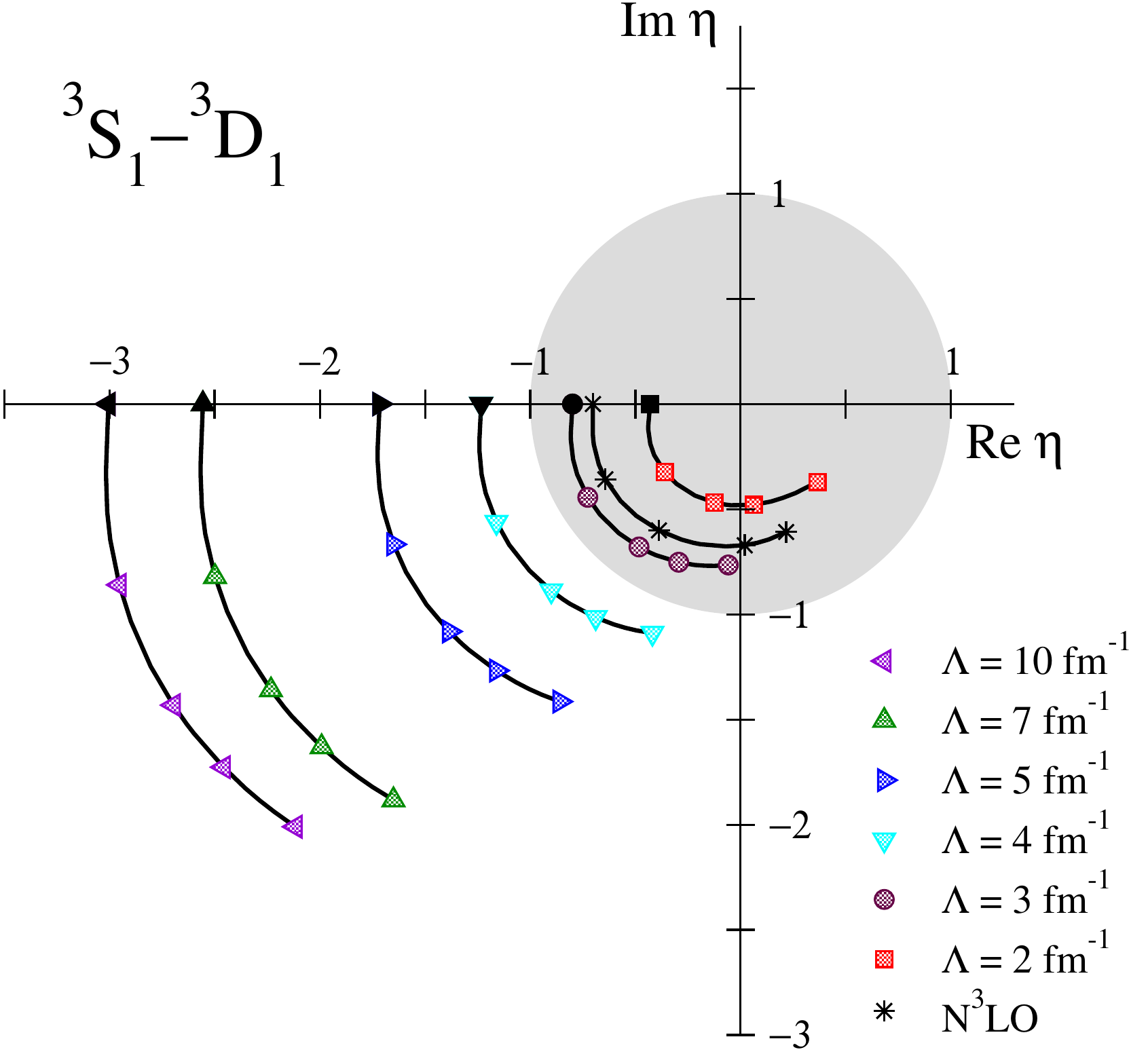}
 \vspace*{-.05in}
   \caption{Trajectories of Weinberg eigenvalues in the complex
   plane for the AV18 NN potential in two channels at various
   states in a $\vlowk$ RG evolution (labeled by $\Lambda$).  The symbols
   go from 0\,MeV on the axis to 25, 66, 100, and 150\,MeV~\cite{Bogner:2006tw}.}
   \label{fig:weinberg4}
\end{center}
%
%\vfill
%
\begin{center}
 \includegraphics[width=2.5in]{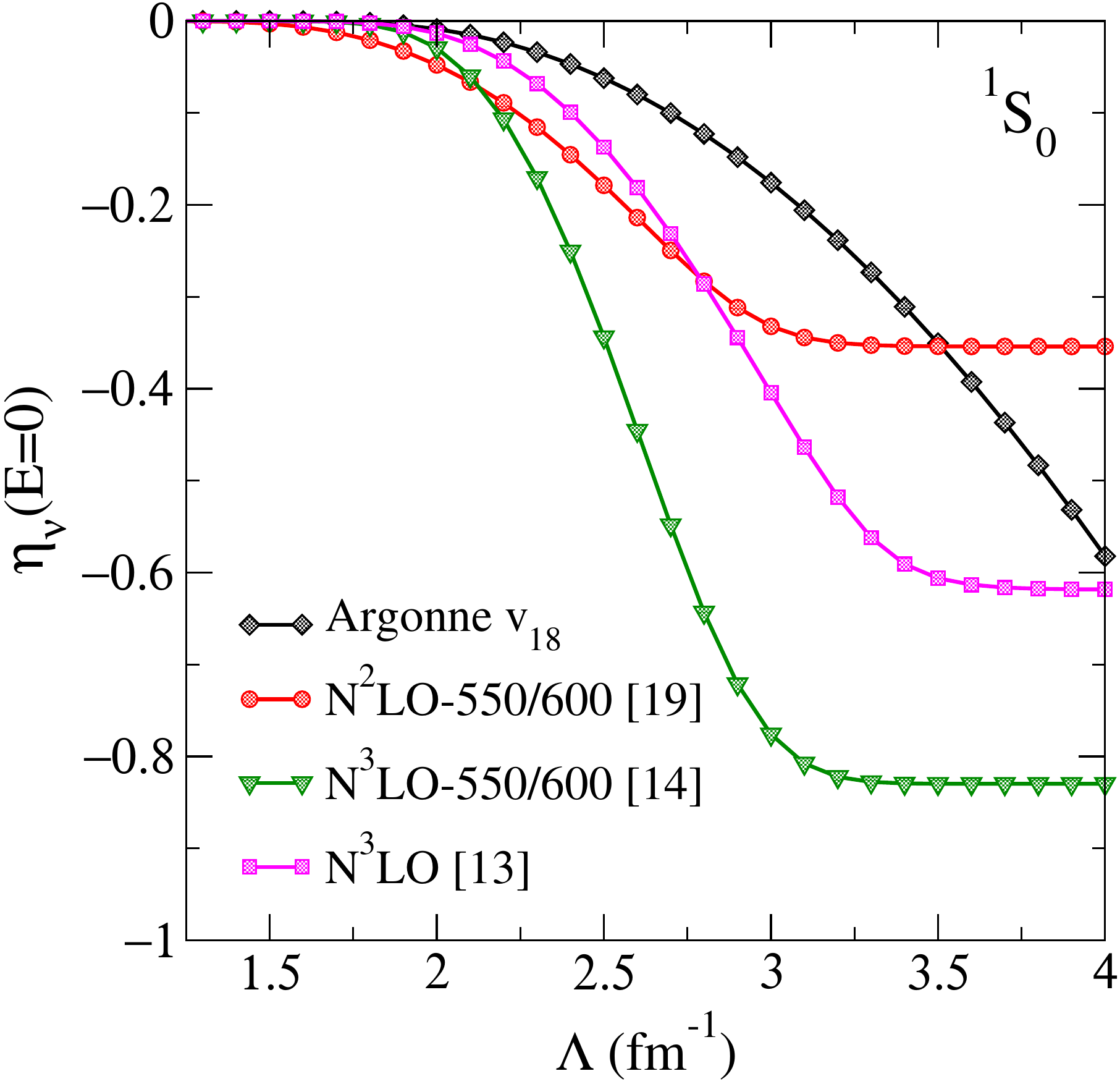}
 \hspace*{.2in}
 \includegraphics[width=2.5in]{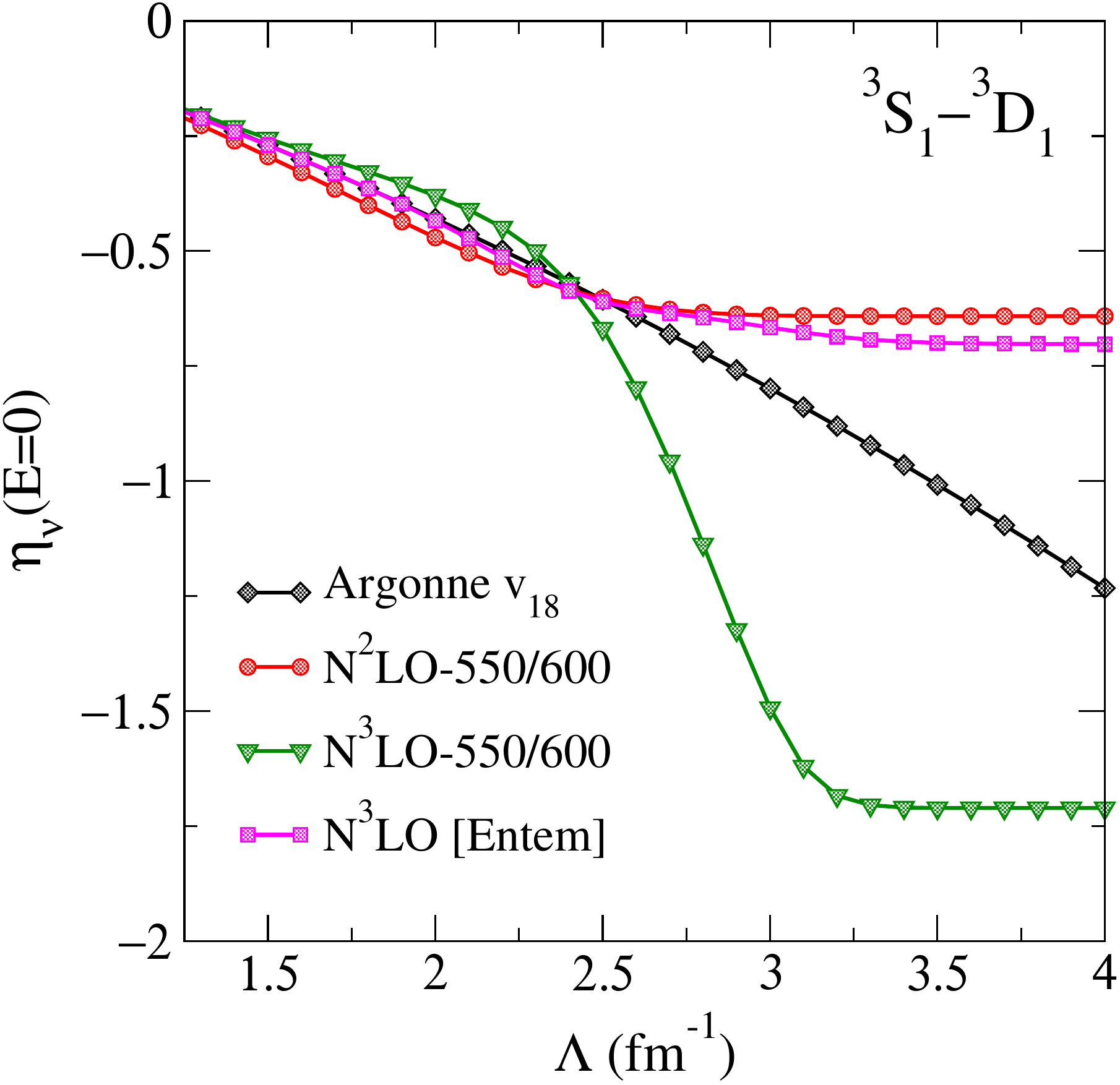}
 \vspace*{-.05in}
   \caption{Repulsive (negative) Weinberg eigenvalues at $E=0$
   for several N$^3$LO chiral EFT potentials as a function
   of $\vlowk$ $\Lambda$~\cite{Bogner:2006tw}.}
   \label{fig:weinberg5}
\end{center}
\end{figure*}

These expectations hold in practice, as illustrated in 
Fig.~\ref{fig:weinberg2}, which shows the evolution of the largest
positive and negative Weinberg eigenvalues as a function of the
$\vlowk$ cutoff $\Lambda$.  (Very similar behavior is observed for
 the SRG with $\lambda$ replacing $\Lambda$.)
In free space, the largest attractive eigenvalue is unity at all
$\Lambda$'s, corresponding to the deuteron bound state.  But the
negative eigenvalue starts very large (because of the repulsive
core) and drops dramatically as the potential is evolved, ending
up well less than unity, indicating it is perturbative (but the
positive eigenvalue still makes this channel nonperturbative).
The situation is even more dramatic in the medium.  The other
curves in the figure show the result from considering the T~matrix
in the nuclear medium, where Pauli blocking effects are included
in the intermediate states.
We see a further reduction of the negative eigenvalue and now the
positive eigenvalue is perturbative as well: the deuteron has dissolved!

Starting from negative energies,
we can follow the evolution of
the largest eigenvalue into the complex plane as we increase $E$ to
positive values.  Figures~\ref{fig:weinberg3} and
\ref{fig:weinberg4} show paths in the
complex plane for Weinberg eigenvalues, with the symbols indicating
energies of 0, 25, 66, 100, and 150\,MeV.
The shaded region is the unit circle; the potential is perturbative
for energies where both eigenvalues lie inside. 
The softening effect of the RG evolution is manifest in
Fig.~\ref{fig:weinberg4} and the eigenvalues provide a quantitative
measure of the perturbativeness.
It's also clear from Fig.~\ref{fig:weinberg3} that at least
the particular chiral potential
used there is already quite soft.
However,
Fig.~\ref{fig:weinberg5} shows that significant \emph{additional} softening
is possible with RG evolution.  In all of these plots,
the differences between
the $^1$S$_0$ channel and the $^3$S$_0$--$^3$D$_0$ coupled channel
stem from the latter having an additional source of nonperturbative
behavior: the short-range tensor force.

\begin{figure}[t!]
\begin{center}
 \includegraphics[width=2.8in]{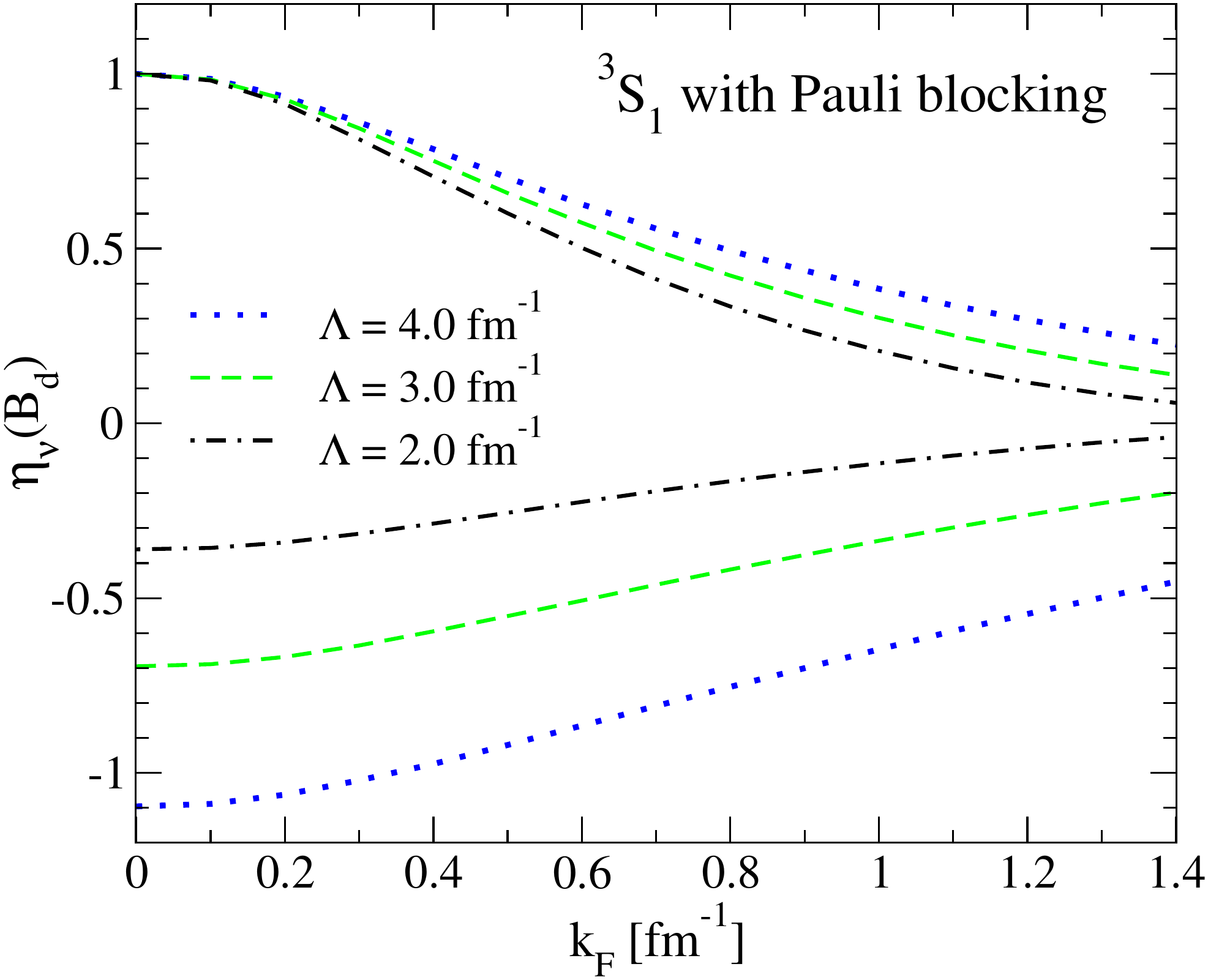}
   \caption{Dependence of the largest Weinberg eigenvalues on
   density for several evolved $\vlowk$ potentials~\cite{Bogner:2006tw}.}
   \label{fig:weinberg6}
\end{center}
\end{figure}

\begin{figure}[th!]
\begin{center}
 \includegraphics[width=1.9in]{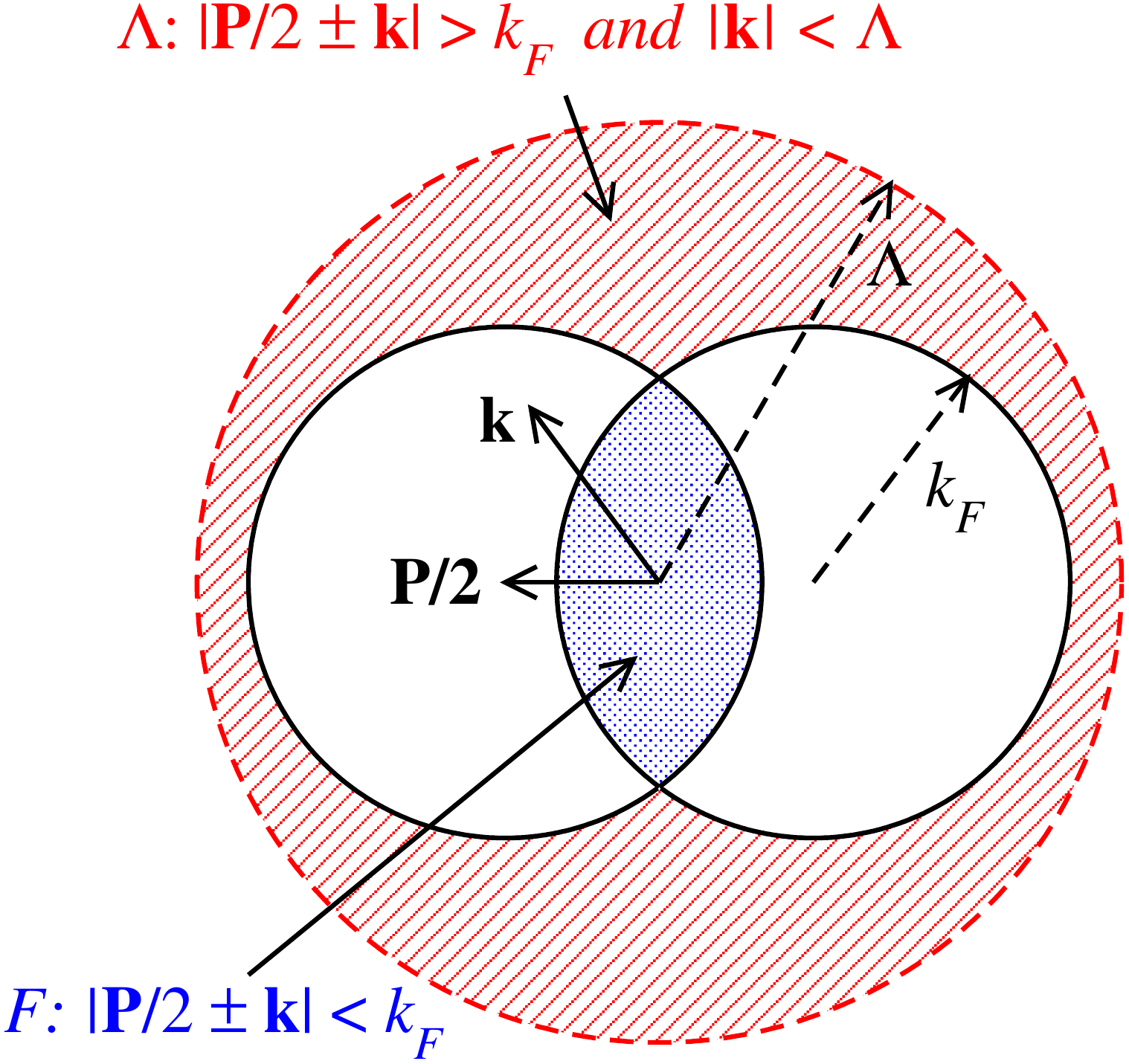}
   \caption{Overlapping Fermi spheres showing available phase space
   for two nucleons excited above the Fermi surface~\cite{Bogner:2005sn}.}
   \label{fig:weinberg7}
\end{center}
\end{figure}

The increasing ``perturbativeness'' at finite density is documented again
in Fig.~\ref{fig:weinberg6}. At typical nuclear densities with
$1\fmi \leq \kf \leq 1.3\fmi$, both positive and negative eigenvalues
are small at the lowest $\Lambda$'s.  This implies that nuclear
matter may actually be perturbative!  
(Note that at the Fermi surface, pairing as a nonperturbative
phenomenon is revealed by $|\eta_\nu|>1$~\cite{Ramanan:2007bb}.)
We can understand how this
happens from Fig.~\ref{fig:weinberg7}, which shows the phase space
available to two nucleons that scatter in the medium.  Pauli blocking
means they must go outside the two Fermi spheres, but the volume
is increasingly restricted with decreasing $\Lambda$.  In addition,
the magnitudes of the matrix elements that scatter such particles
decrease as well~\cite{Bogner:2005sn}.

\begin{figure}[t!]
\begin{center}
 \includegraphics[width=2.8in]{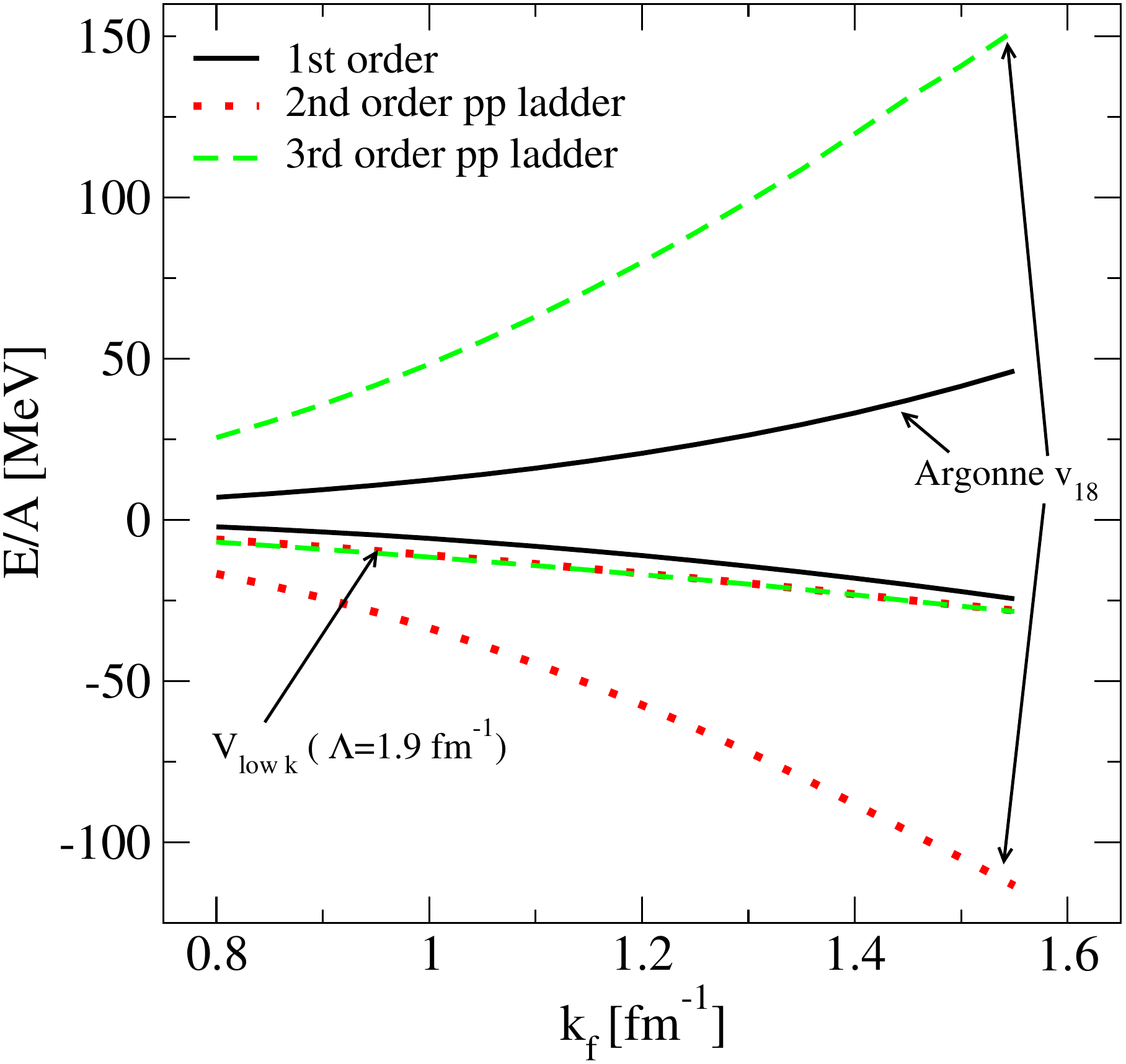}
   \caption{Many-body perturbation theory for symmetric nuclear
   matter up to third order in the particle-particle channel~\cite{Hebeler:2010xb}.}
   \label{fig:weinberg8}
\end{center}
\end{figure}

\begin{figure*}[th!]
\begin{center}
 \includegraphics[width=5.2in]{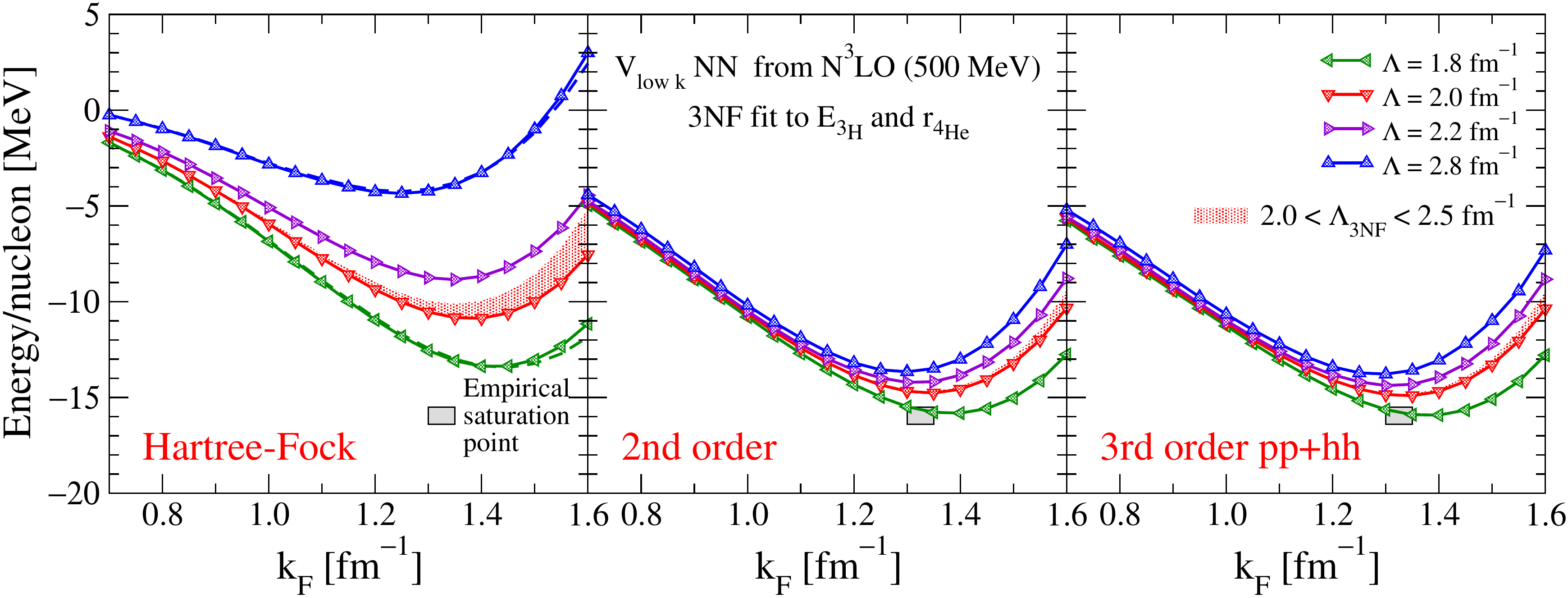}
   \caption{Many-body perturbation theory calculations of symmetric
   nuclear matter with $\vlowk$ NN potentials at four different
   cutoffs and the leading-order 3NF from chiral EFT~\cite{Hebeler:2010xb}.  
   The two
   free parameters in the 3NF are fit at each density to the triton binding
   energy and the alpha particle radius.}
   \label{fig:weinberg9}
\end{center}
\end{figure*}

Perturbation theory in the particle-particle channel is shown
in Fig.~\ref{fig:weinberg8} for the high-resolution Argonne $v_{18}$ potential
initially and after evolution by the $\vlowk$ RG to low resolution ($\Lambda=1.9\fmi$).
Whereas many-body perturbation theory (MBPT) manifestly diverges for the 
original potential, it converges for the low-momentum interaction
at second order
(at least in this channel; more complete many-body approximations
must be studied to be more definite). 
It is also evident that there is no saturation.  But adding a 
3NF fit only to few-body properties, as in Fig.~\ref{fig:weinberg9}, shows that the empirical saturation
point can be reproduced with an uncertainty of about 2--3\,MeV/particle.
On-going work to improve this result includes the development of SRG
evolution for the 3NF in momentum space~\cite{Kai} 
and of coupled cluster methods
for infinite matter with 3NF's to provide a high-order resummation
of perturbation theory to test convergence.

\subsection{Universality from flow equations}

Another general aspect of RG flows known from the study of critical
phenomena is the appearance of universal behavior.  In the application
of RG to nuclear interactions, the universality we observe is that
distinct initial NN potentials that reproduce the experimental low-energy
scattering phase shifts, are found to collapse to a single universal
potential. We've already seen an indication of this, but here we
document it in more detail.

\begin{figure*}[p!]
\begin{center}
 \includegraphics[width=5.3in]{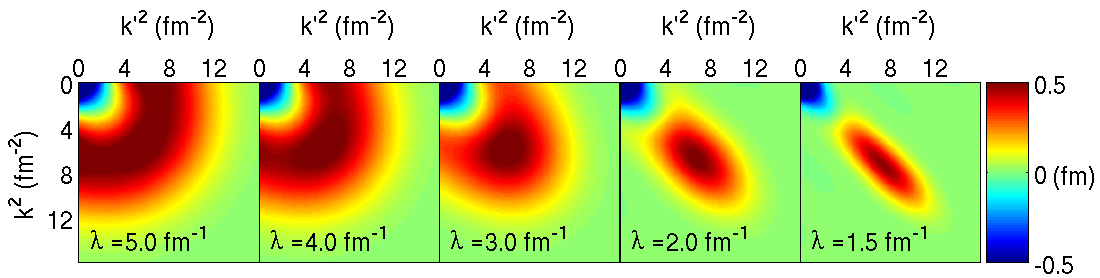}
 
 \vspace*{0in}
 
 \includegraphics[width=5.3in]{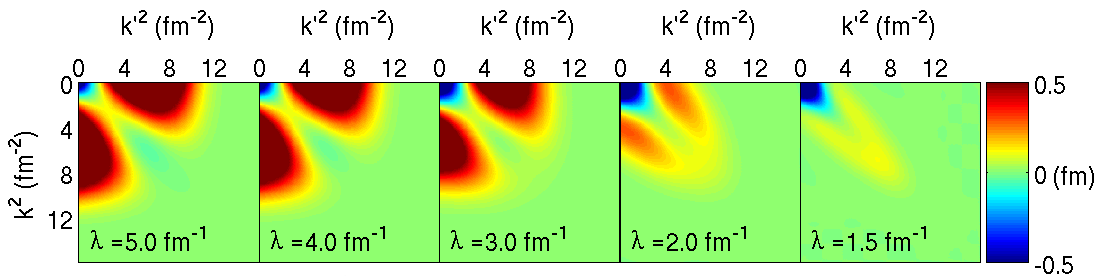}
 \vspace*{-.1in}
 \caption{SRG evolution of two chiral EFT potentials in
 the $^1$S$_0$ channel~\cite{Bogner:2009bt}.}
 \label{fig:srguniv1}
 
 \vspace*{.1in}
 
 \includegraphics[width=5.3in]{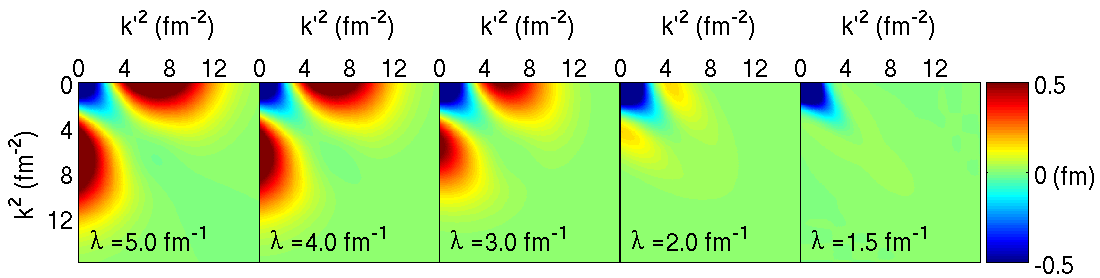}
 
 \vspace*{0in}
 
 \includegraphics[width=5.3in]{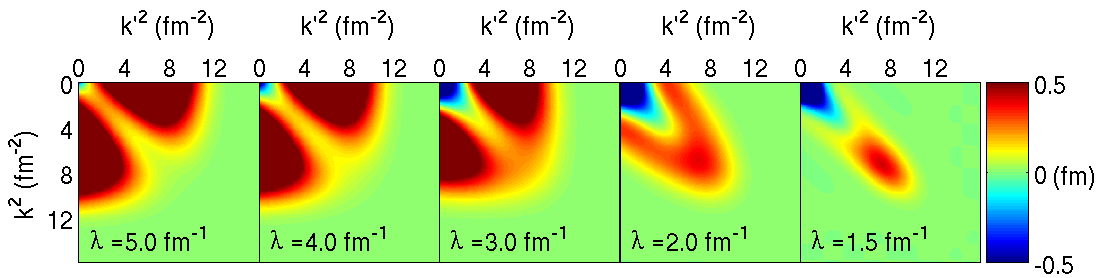}
 \vspace*{-.1in}
   \caption{SRG evolution of two chiral EFT potentials
    in the $^3$S$_1$ channel~\cite{Bogner:2009bt}.}
   \label{fig:srguniv2}
 
 \vspace*{.14in}

    \includegraphics[width=2.3in]{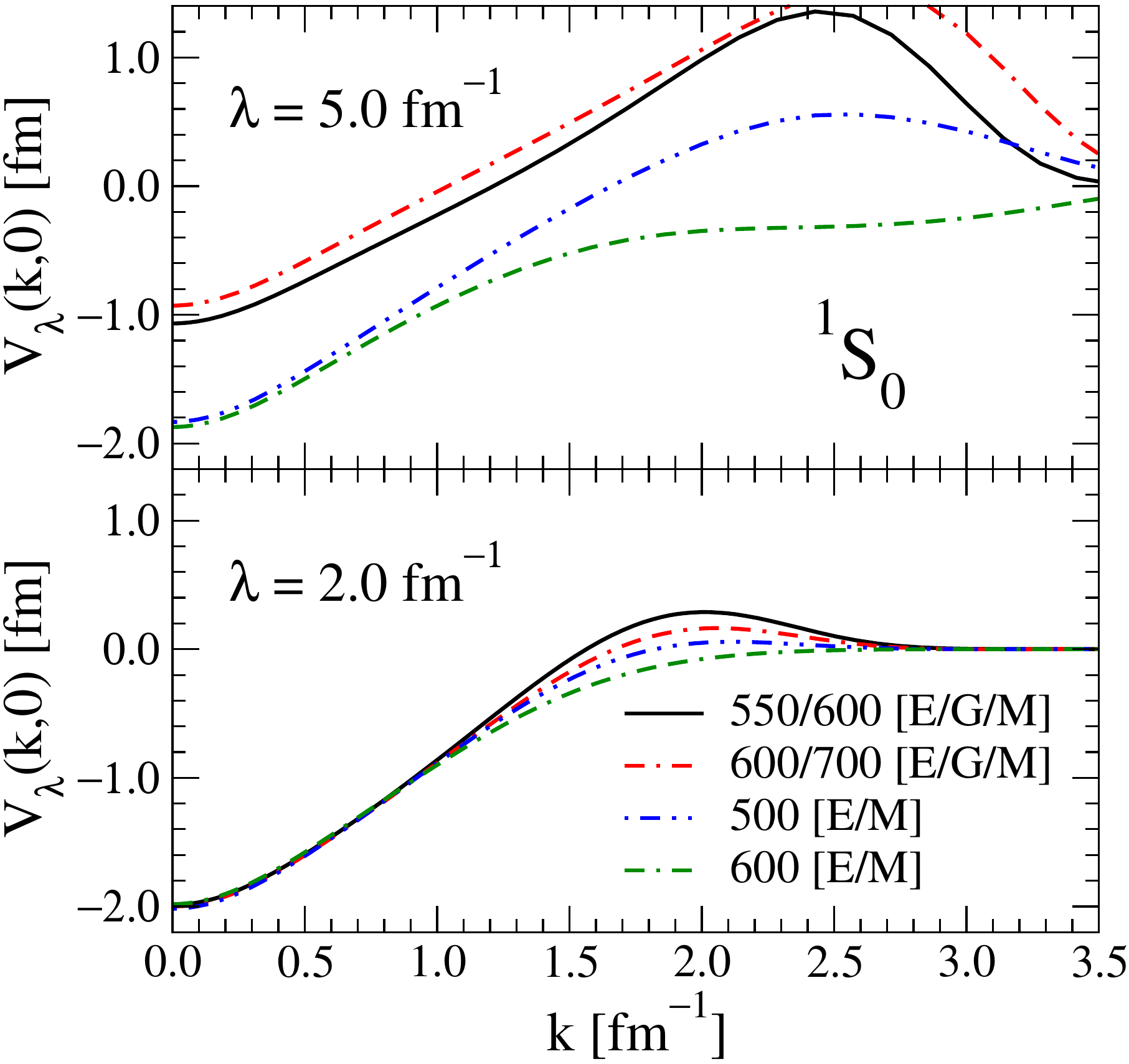}
    \hspace*{.2in}
   \includegraphics[width=2.3in]{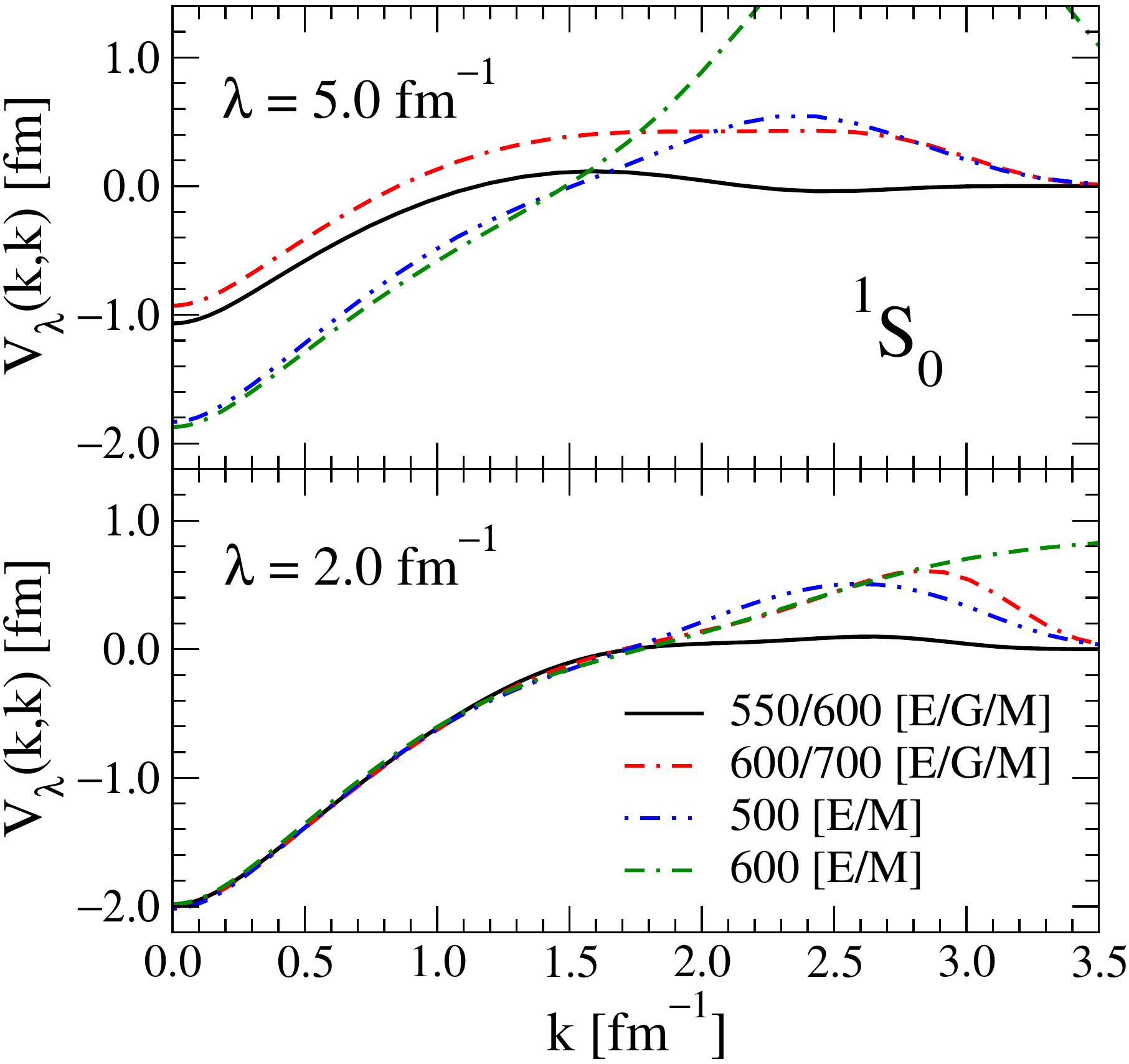}
   \vspace*{-.1in}
   \caption{SRG flow toward universality for several chiral EFT 
   potentials~\cite{Bogner:2009bt}.}
   \label{fig:srguniv3}

\end{center}
\end{figure*}

We focus on the SRG, but very similar conclusions are found
for $\vlowk$ evolution.  In Figs.~\ref{fig:srguniv1} and \ref{fig:srguniv2},
S-wave N$^3$LO chiral EFT potentials from Refs.~\cite{Entem:2003ft} and ~\cite{Epelbaum:2004fk} are evolved with the SRG.
Although the level of truncation is the same and the cutoffs
approximately equal, the methods of regulating the potential
differ (particularly for the two-pion exchange part).
The result is very different looking initial interactions.  This is not
a concern, because the NN potential is not an observable.
We also observe there is significant off-diagonal strength coupling
low and medium momenta in the initial potentials.

As the potentials are evolved, we see the characteristic driving
toward the diagonal, with the diagonal width in $k^2$ given roughly by
$\lambda^2$.
At the end of the evolution shown, the interactions still look quite
different at first glance.  However, if we focus on the low-momentum
region, where $k^2 < 2\,\mbox{fm}^{-2}$, they appear much more
similar.  We can quantify this by taking a slice along the edge
(i.e., $V(k,0)$ and along the diagonal
(i.e., $V(k,k)$) and plot these quantities for these potentials and two
additional ones.
This is done in Fig.~\ref{fig:srguniv3}.
We see a dramatic collapse of the interaction between $\lambda = 5\fmi$
and $\lambda = 2\fmi$ for the region of $k$ below $\lambda$
(or maybe 3/4 $\lambda$).
An open question under active investigation is whether evolved
3NF interactions will be universal. 

\subsection{Operator evolution via flow equations}

We have focused almost exclusively on the evolution of the
Hamiltonian, but an RG transformation will also change the
operators associated with measurable quantities.
It is essential to be able to start with operators consistent
with the Hamiltonian
and then to evolve them maintaining this consistency.
The first step is a prime motivation for using EFT; we will
assume that we have consistent initial operators in hand.
The second step can be technically difficult, especially
since we will inevitably induce many-body operators as we
evolve (for the same arguments as we made for the Hamiltonian).
This is where the SRG is particularly advantageous, because
it is technically feasible to evolve operators along with the
Hamiltonian.  

The SRG evolution with $\flow$ (recall $s = 1/\lambda^4$) 
of \emph{any} operator $O$ 
is given by:
\beq
  O_\flow = U_\flow O U^\dagger_\flow \;,
  \label{eq:Oflow}
\eeq
so $O_\flow$ evolves via
 \beq
   \frac{dO_\flow}{d\flow}
   = [ [G_\flow,H_\flow], O_\flow]  \;,
  %  = [\eta(\flow),O_\flow] = [ [\Hzero,V_\flow], O_\flow] 
 \eeq
where we use the same $G_\flow$ to evolve the Hamiltonian
and all other operators. 
While we can directly evolve any operator like this in parallel
to the evolution of the Hamiltonian, in practice it is more
efficient and numerically robust to either evolve the unitary
transformation $U_s$ itself:
\beq
  \frac{dU_\flow}{d\flow}
    = \eta_s U_s = [G_s,H_s] U_s
    \;,
\eeq
with initial value $U_{s=0} = 1$, or calculate it directly
from the eigenvectors of $H_{s=0}$ and $H_s$:
\beq
  U_s = \sum_i |\psi_i(s)\ra \la \psi_i(0)| \;.
\eeq
Then any operator is directly evolved to the desired $\flow$
by applying Eq.~\eqref{eq:Oflow} as a matrix multiplication.
The second method works well in practice.

\begin{figure}[t!]
\begin{center}
 \includegraphics[width=2.5in]{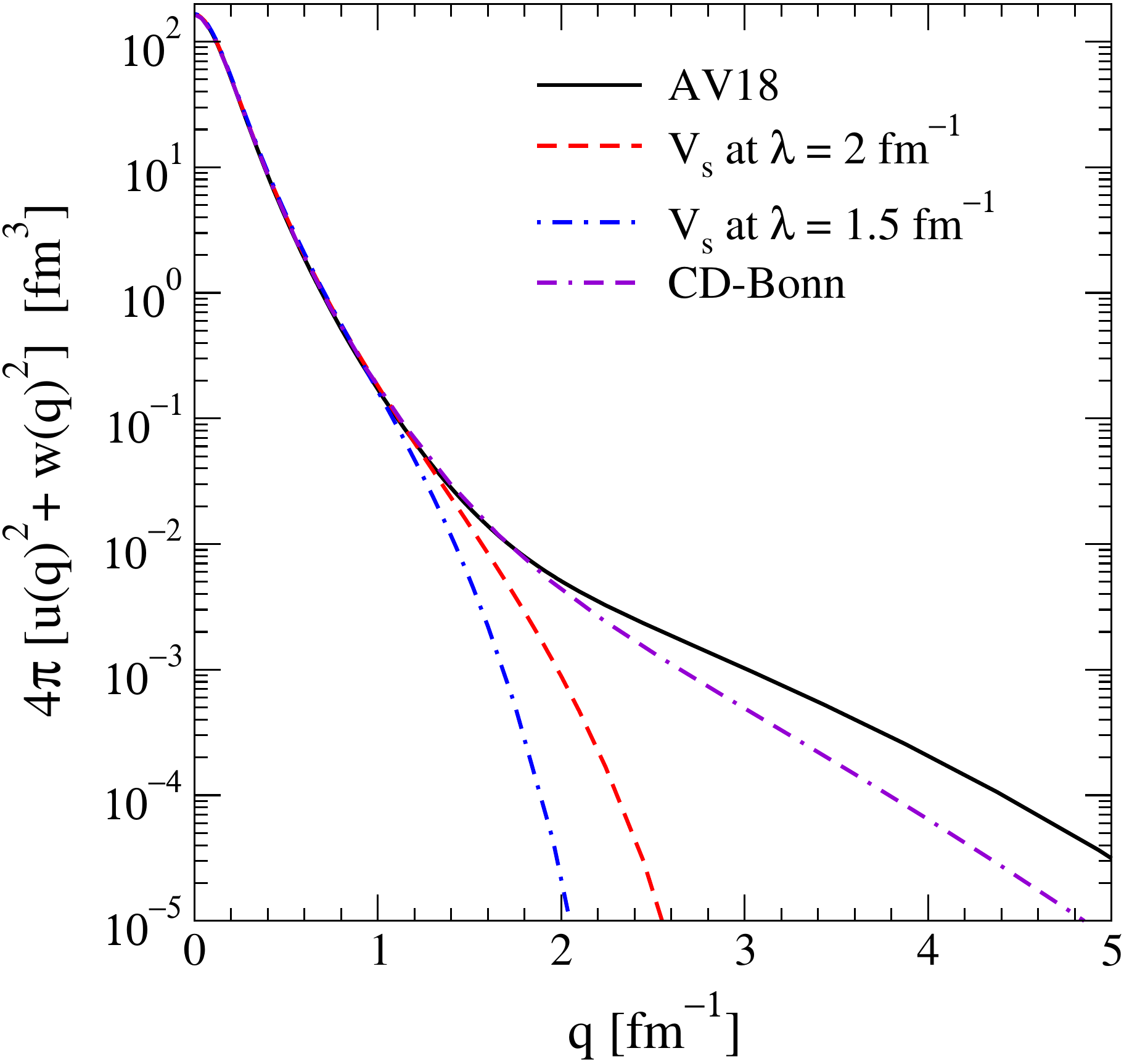}
   \caption{Momentum distribution of the deuteron for the
   AV18 potential and CD-Bonn potentials, compared to
   results from SRG potentials evolved to $\lambda=2.0\fmi$
   and $1.5\fmi$.}
   \label{fig:deuteron_md_1}
\end{center}
\end{figure}

\begin{figure}[t!]
\begin{center}
 \includegraphics[width=2.5in]{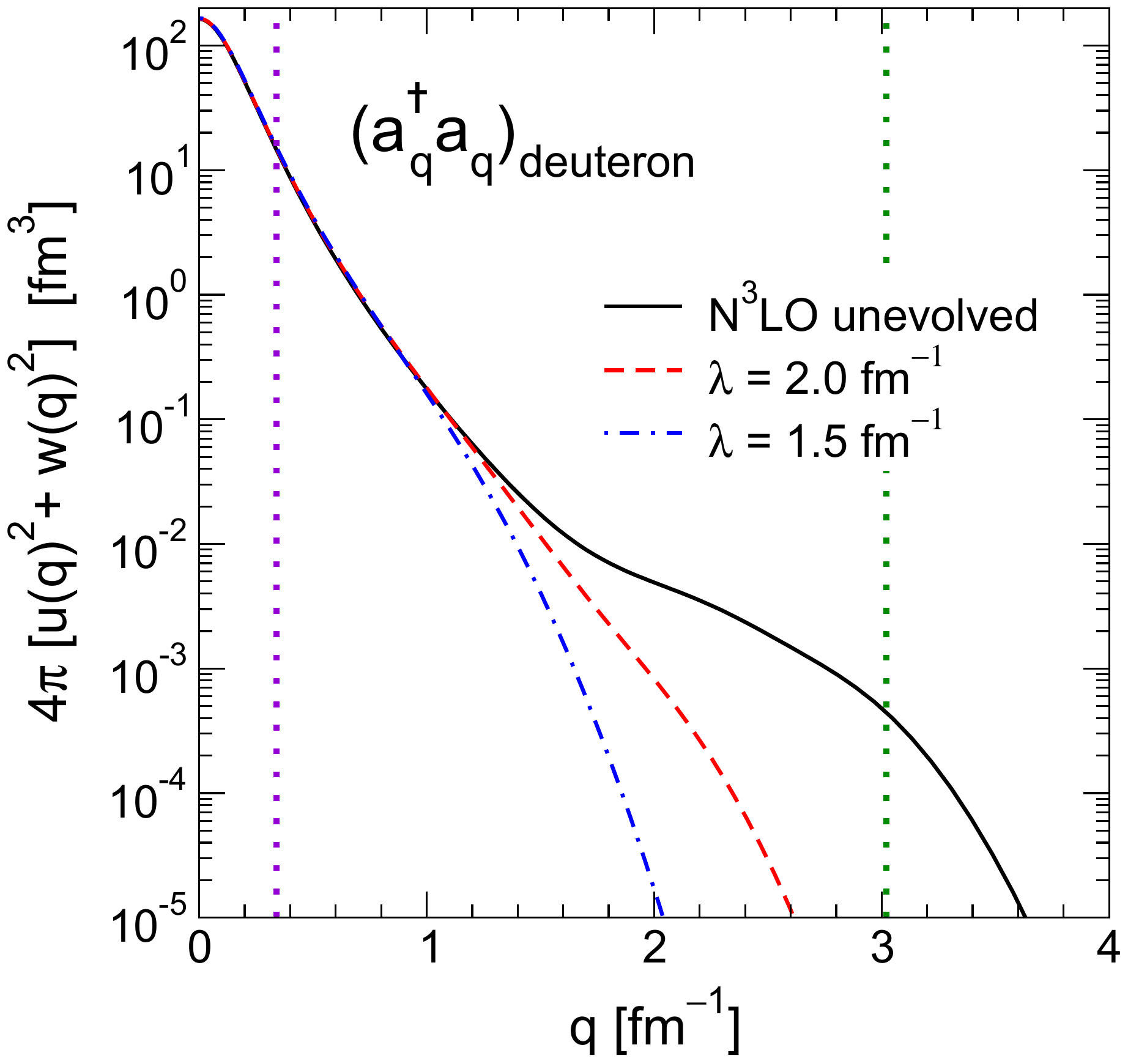}
   \caption{Momentum distribution for an unevolved chiral N$^3$LO 
   potential compared to results from SRG potentials evolved to $\lambda=2.0\fmi$
   and $1.5\fmi$~\cite{Anderson:2010aq}.}
   \label{fig:deuteron_md_2}
\end{center}
\end{figure}

To simplify our study of operator evolution, we consider the simplest
possible operator: the momentum number operator $a^\dagger_q a_q$.
In Fig.~\ref{fig:deuteron_md_1}, we show the momentum distribution
in the deuteron, i.e., $<\psi_d| a^\dagger_q a_q |\psi_d> $
for two different realistic potentials, AV18 and CD-Bonn.
As implied by the $y$-axis label, the momentum distribution is just the
square of the deuteron wave function in momentum space.
The results for the two potentials agree up to about $2\fmi$ and then
are different.
If we evolve the AV18 potential \emph{and} the momentum operator, then
the matrix element
$<\psi_d^\lambda| O_\lambda |\psi_d^\lambda> $
will be the same for any $\lambda$ and the curve will exactly reproduce
the AV18 deuteron momentum distribution.
However, if we
calculate $<\psi_d^\lambda| a^\dagger_q a_q |\psi_d^\lambda> $ for
$\lambda = 2\fmi$ and $1.5\fmi$, that is, we use the evolved wave function
but the unevolved operator, then we get the other curves.
Besides directly illustrating that the high-$q$ part of the momentum
distribution is not an observable, since we can change it at will
by unitary transformations, this manifests that the high-momentum
part of the wave function is removed.  

\begin{figure*}[p!]
\begin{center}
\includegraphics[width=5.2in]{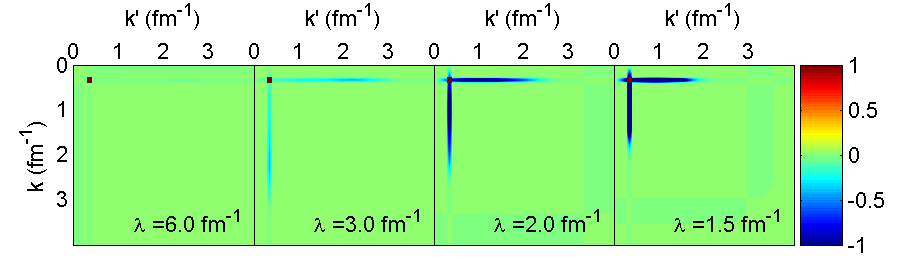}
\includegraphics[width=5.2in]{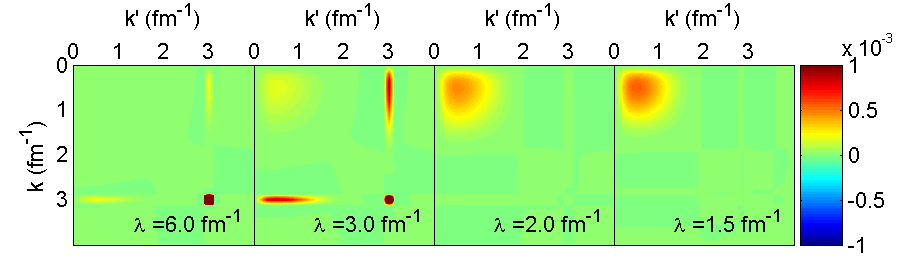}
   \caption{Integrand of momentum distribution operator~\cite{Anderson:2010aq}.}
   \label{fig:contourmd}
%\end{center}
%\end{figure*}
%
  \vspace*{.3in}
%
%\begin{figure*}[t!]
%\begin{center}
   \includegraphics[width=5.8in]{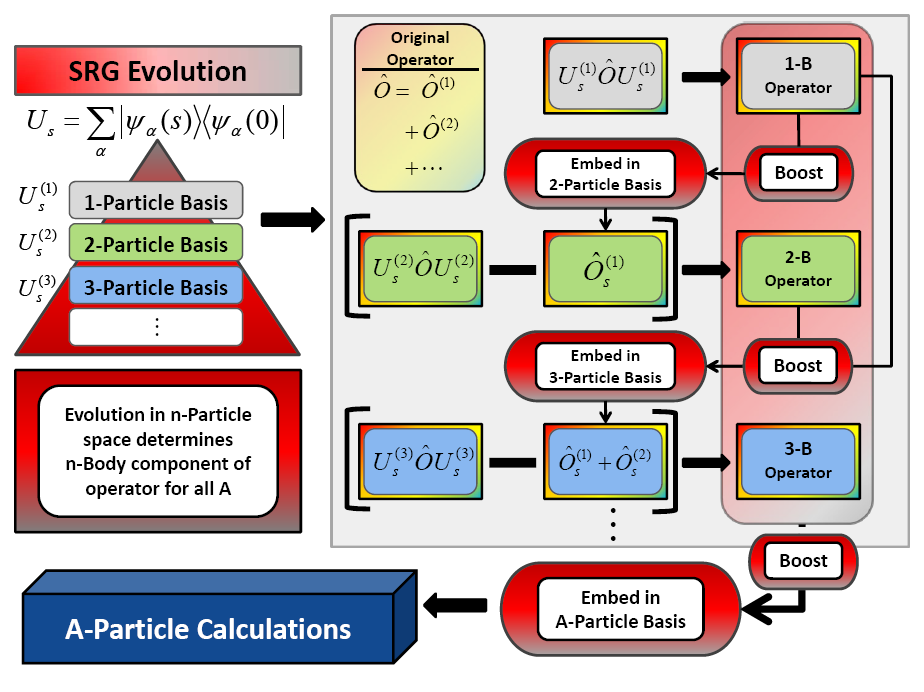}
   \caption{Schematic of the SRG operator evolution and embedding 
   process~\cite{Anderson:2011private}.}
   \label{fig:operator_embed}
\end{center}
\end{figure*}

The latter is potentially disturbing, because if the evolved operator
is supposed to reproduce a high momentum result when the evolved
wave function has a vanishingly small component at that momentum, this
may be because the operator is becoming pathological. To explore
this further, we consider the momentum distribution in Fig.~\ref{fig:deuteron_md_2} at low ($q=0.34\fmi$) and moderately
high ($q=3.0\fmi$) values of the momentum~\cite{Anderson:2010aq}.
In Fig.~\ref{fig:contourmd} we plot the \emph{integrand} of
\beq
 \left\langle\psi_{d}^\lambda \right \vert   (Ua^{\dagger}_{q}a_{q}U^{\dagger})\left\vert \psi_d^\lambda \right\rangle 
 \;,
\eeq
at each of these two $q$ values.  The full integral is the momentum
distribution at those $q$'s, so the plots tell us where the strength
of the operator lies.  
For the low-momentum operator, there is little renormalization, but
the nature of the high-momentum operator changes completely.
Originally, the integral comes entirely from the region of $q=3.0\fmi$,
but the evolution of the operator shifts its strength entirely
to low momentum.  This result is similar for other operators, such
as electromagnetic form factors~\cite{Anderson:2010aq}.

As we move to $A \geq 3$, the operator evolution and extraction process
becomes more involved.  A flowchart for the procedure is given in
Fig.~\ref{fig:operator_embed}.
Imagine we initially have a one-body operator  and we want to evolve 
and then evaluate
it in an $A$-particle nucleus.
The difficulty is that $n$-body components are induced as we evolve
and these must be separated out so we can correctly embed them
in the nucleus.  In particular, to embed an $n$-body operator in an anti-symmetrized $A$-particle nucleus, we need an embedding factor of $\binom{A}{n}$, so we need to isolate the components first.

With the usual SRG generators, there is no evolution of one-body
operators.  This is easiest to see in second quantization, where the
commutators on the right side yield operators that have at least
four creation/destruction operators, meaning it is two- and higher-body. 
To isolate the two-body part, we first evolve the Hamiltonian
in the two-particle basis to find the unitary operator $U_s^{(2)}$
and use it to evolve our operator.  Then if we subtract the embedded
one-body operator, we will have our two-body part.  The one- and two-body
parts are then embedded in the 3-particle basis and subtracted from
the evolved operator in that basis.  And so on until we reach the
desired level of truncation, at which point we embed in the $A$-particle
basis and perform the $A$-particle calculation.
More details and examples can be found about operator evolution in Ref.~\cite{Anderson:2010aq}.

\subsection{Computational aspects}

Before concluding this lecture, we'd like to make some brief comments
on the computational aspects of the calculations behind the figures
we've seen.
As noted earlier, the continuous momentum is discretized into a finite
number of momentum points.  
The subsequent discretization of integrals leads directly to matrices,
and most of the manipulations are efficiently cast in this language.
For example, momentum-space flow equations have integrals like:
    \beq
      I(p,q) \equiv \int\!dk\, k^2\, V(p,k) V(k,q) \;.
    \eeq 
The usual choice of discretization is to use gaussian quadrature to
accurately evaluate integrals with a minimum of points.  (The size
is not an issue for two-body operators, but becomes critical for
higher-body operators.)    
We introduce gaussian nodes and weights $\{k_n,w_n\}$ with ($n = 1,N$)
to reduce integrals to finite sums:
       \beq
          \quad \int\!dk\, f(k) \approx
	     \sum_{n} w_n\, f(k_n) \;.
       \eeq
(Note: these sets of nodes and weights are generally a combination
of separate smaller rules over adjacent intervals with a \emph{total} of $N$ points.)       
Then $I(p,q) \rightarrow I_{ij}$, where $p = k_i$
      and $q = k_j$, and
    \beq 
    I_{ij} = \sum_n k_n^2 w_n\, V_{in} V_{nj}
      \rightarrow \sum_n \wt V_{in} \wt V_{nj}
    \eeq
    where  
    \beq
       \wt V_{ij} \equiv \sqrt{w_i}k_i\, V_{ij}\, k_j \sqrt{w_j} \;. 
    \eeq
This allows us to solve SRG equations and integral equations
    for phase shifts,
    Schr\"odinger equation in momentum representation.
In practice, $N\approx 100$ gauss points is adequate
        for accurate nucleon-nucleon partial waves.

A computer code that carries out SRG evolution can be remarkable
simple.
Here is a possible pseudocode that is suitable:
  \be
    \I Set up basis (e.g., momentum grid with gaussian quadrature
         or HO wave functions with $N_{\rm max}$)
    \I Calculate (or input) the initial Hamiltonian and $G_s$
         matrix elements (including any weight factors)
    \I Reshape the right side $[[G_s,H_s],H_s]$ to a vector
      and pass it to a coupled differential equation solver
    \I Integrate $V_s$ to desired $s$ (or $\lambda = s^{-1/4}$)
    \I Diagonalize $H_s$ with standard symmetric eigensolver
     \Lra energies and eigenvectors
    \I Form $U = \sum_i |\psi^{(i)}_s\rangle
             \langle \psi^{(i)}_{s=0} |$
	 from the eigenvectors
    \I Output or plot or calculate observables
  \ee
Such a code has been implemented in
MATLAB, Mathematica, Python, C++, and Fortran-90.
While any basis can be used, so far only discretized momentum
and harmonic oscillator bases have been implemented.
Note that the same procedure (and even the same code in some cases)
is relevant for a many-particle basis, but the number of differential
equations will grow rapidly.  For accurate results in the two-body
evolution, $100^2 = 10,000$ matrix elements are needed, so there are
that many differential equations.  For an accurate 3NF evolution
in a harmonic oscillator basis, at least 20~million coupled differential
equations need to be solved.  This sounds intimidating, but is well
within reach of MATLAB (for example) on a machine with a moderate
amount of memory.

\subsection{Summary points} 

Chiral EFT establishes a hierarchy of many-body forces.
Using flow equations to run to low resolution makes many-body
calculations more perturbative and interactions flow
to universal form (at least for NN; this is not yet
established for the 3NF).
Operators can be evolved consistently with interaction.
Long-distance operators change very little, while
short-distance operators renormalize significantly,
accompanied in some cases by a change in physics interpretation.
The basic SRG flow equations in a partial wave momentum basis
can be cast in a form that
involves 
just matrix manipulations and the solution of
ordinary first-order, coupled differential equations.

%%%%%%%%%%%%%%%%%%%%%%%%%%%%%%%%%%%%%%%%%%%%%%%%%%%%%%%%%%%%%%%%%%%%%%%%
%%%%%%%%%%%%%%%%%%%%%%%%%%%%%%%%%%%%%%%%%%%%%%%%%%%%%%%%%%%%%%%%%%%%%%%%

\section{Nuclear applications and open questions}
\label{sec:applications}

In this final lecture, we take a look at the in-medium
similarity renormalization group (IM-SRG) and make a broad
survey of nuclear applications.  We conclude with a summary
of open problems and new challenges.

\subsection{In-medium similarity renormalization group}
 \label{subsec:imsrg}

\begin{figure}[th!]
\begin{center}
 \includegraphics[width=2.8in]{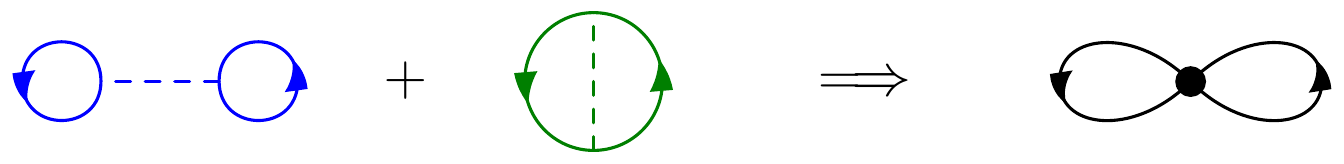}
   \caption{Feynman and Hugenholtz diagrams for Hartree-Fock.}
   \label{fig:HartreeFock}
\end{center}
\end{figure}

We start with a review of Hartree-Fock.
The Hartree-Fock wave function is the
best single Slater determinant
   \beq
     | \Psi_{\rm HF} \rangle = \det\{\phi_i(\xvec), i=1\cdots A  \}
     \,, \quad \xvec = ({\bf r}, \sigma, \tau)
   \eeq
 in the variational sense.   
The $\phi_i(\xvec)$ single-particle wave functions 
satisfy \emph{non-local} Schr\"odinger equations:
   \beq
      -\frac{\bm{\nabla}^2}{2M}\phi_i(\xvec) 
      + V_{\rm H}(\xvec)
      \phi_i(\xvec)
       + \int\! d\yvec\,  V_{\rm E}(\xvec,\yvec) \phi_i(\yvec)
       = \epsilon_i \phi_i(\xvec) 
   \eeq
with direct
  \beq
    V_{\rm H}(\xvec) = \int\! d\yvec \sum_{j=1}^A |\phi_j(\bf y)|^2
      v(\xvec,\yvec)
  \eeq
and exchange 
  \beq    
   V_{\rm E}(\xvec,\yvec) = - v(\xvec,\yvec) \sum_{j=1}^A
        \phi_j(\xvec) \phi_j^\ast(\yvec) 
  \eeq
potentials~\cite{FETTER71,Ring:2005}.  
The direct and exchange potentials are shown diagrammatically 
in Fig.~\ref{fig:HartreeFock}, with the rightmost diagram an
abbreviated form called a Hugenholtz diagram~\cite{Negele:1988vy}.   
We must solve self-consistently using occupied orbitals for $V_H$ and $V_E$.
Then Slater determinants from \emph{all} orbitals form an $A$-body basis.

The in-medium SRG (IM-SRG) 
for nuclei, developed recently by Tsukiyama, Bogner, and
Schwenk~\cite{Tsukiyama:2010rj}, applies the flow equations in an $A$--body system using
a different reference state than the vacuum.
For example, we can choose the Hartree-Fock ground state as a
reference state.
The appealing consequence is that,
unlike the free-space SRG evolution, the in-medium
SRG can approximately evolve
$3,...,A$-body operators using only two-body machinery.
However, also in contrast to the free-space SRG,
the in-medium evolution must be repeated for each nucleus or
density.   

The key to
the IM-SRG
simplification is the use of \emph{normal-ordering} with respect to a
finite-density reference state. 
That is, starting from the
second-quantized Hamiltonian with two- and three-body interactions,
\bea
  H \ampseq \sum_{12} T_{12} \ad_1 a_2 + \frac{1}{(2!)^2} \sum_{1234}
  \, \langle12|V|34\rangle \ad_1\ad_2a_4a_3
  \nonumber \\
  \amps{} 
  \hspace*{-.2in}
  \null + \frac{1}{(3!)^2} \, \sum_{123456} \langle123|V^{(3)}|456\rangle
  \ad_1\ad_2\ad_3a_6a_5a_4 \,,
  \label{eq:Ham}
\eea
all operators are normal-ordered with respect to a finite-density 
Fermi vacuum $|\Phi\rangle$ (for example, the Hartree-Fock ground
state or the non-interacting Fermi sea in nuclear matter), as 
opposed to the zero-particle vacuum. Wick's theorem can then be
used to {\it exactly} write $H$ as
\bea
  H \ampseq E_0 + \sum_{12} f_{12} \{\ad_1a_2\} 
  \nonumber \\
  \amps{} 
  \hspace*{-.25in}
  \null
  + \frac{1}{(2!)^2} \sum_{1234} \, \langle12|\Gamma|34\rangle 
  \{\ad_1\ad_2a_4a_3\}
  \nonumber \\
  \amps{} 
  \hspace*{-.25in}
  \null
  + \frac{1}{(3!)^2} \sum_{123456} \, \langle123|\Gamma^{(3)}|456\rangle
  \{\ad_1\ad_2\ad_3a_6a_5a_4\} \,,
  \label{eq:NorderedHam}
\eea
where the zero-, one-, and two-body normal-ordered terms are given by
\bea
  E_0 \ampseq \langle \Phi | H | \Phi \rangle = \sum_1 T_{11} n_1 
  \nonumber \\ \amps{} \null
+ \frac{1}{2} \sum_{12} \, \langle12|V|12\rangle n_1 n_2 
  \nonumber \\ \amps{} \null
+ \frac{1}{3!} \sum_{123} \, \langle123|V^{(3)}|123\rangle n_1 n_2 n_3
\,, \label{eq:NorderedCoeff1} \\[10pt]
f_{12} \ampseq T_{12} + \sum_i \langle1i|V|2i\rangle n_i 
  \nonumber \\ \amps{} \null
+ \frac{1}{2} \, \sum_{ij} \langle1ij|V^{(3)}|2ij\rangle n_i n_j \,,
\label{eq:NorderedCoeff2} \\[10pt]
\langle12|\Gamma|34\rangle \ampseq \langle12|V|34\rangle  
+ \sum_i \langle12i|V^{(3)}|34i\rangle n_i \,,
  \label{eq:NorderedCoeff3}
\eea
and $n_i = \theta(\varepsilon_{\rm F} - \varepsilon_i)$ denotes the
sharp occupation numbers in the reference state, with Fermi level or
Fermi energy $\varepsilon_{\rm F}$. 

By construction, the
normal-ordered strings of creation and annihilation operators obey
$\langle\Phi|\{a^{\dagger}_1\cdots a_n\}|\Phi\rangle = 0$.  It is
evident from Eqs.~\eqref{eq:NorderedCoeff1}--\eqref{eq:NorderedCoeff3}
that the coefficients of the normal-ordered zero-, one-, and two-body
terms include contributions from the three-body interaction $V^{(3)}$
through sums over the occupied single-particle states in the reference
state $|\Phi\rangle$. Therefore, truncating the in-medium SRG
equations to two-body {\it normal-ordered} operators will
(approximately) evolve induced three- and higher-body interactions
through the density-dependent coefficients of the zero-, one-, and
two-body operators in Eq.~(\ref{eq:NorderedHam}).

The in-medium SRG flow equations at the normal-ordered two-body level
are obtained by evaluating $dH/ds = [\eta,H]$ with the normal-ordered
Hamiltonian $H = E_0 + f + \Gamma$ and the SRG generator $\eta =
\eta^{1b} + \eta^{2b}$ (with one- and two-body terms) and neglecting
three- and higher-body normal-ordered terms. For infinite matter,
a natural generator choice is $\eta = [f,\Gamma]$ in analogy
with the free-space SRG. In this case, the explicit form of the SRG
equations simplifies because $\eta^{1b}=0$ and $f_{ij} = f_i \,
\delta_{ij}$. This leads to
\bea
  \frac{dE_0}{ds} \ampseq
  \frac{1}{2} \sum_{1234} \, (f_{12} - f_{34})
  |\Gamma_{1234}|^2 \, n_1 n_2 \bar{n}_3 \bar{n}_4
   \,, \label{eq:NorderedSRG0} \\[10pt]
   \frac{df_1}{ds} \ampseq \sum_{234} \, (f_{41} - f_{23})
  |\Gamma_{4123}|^2 
  \nonumber \\ \amps{} \qquad \null \times
  (\bar{n}_2\bar{n}_3n_4 + n_2n_3\bar{n}_4)
  \,, \label{eq:NorderedSRG1} \\[10pt]
  \frac{d\Gamma_{1234}}{ds} \ampseq -(f_{12} - f_{34})^2 \,
  \Gamma_{1234} 
  \nonumber \\ \amps{} \null
  + \frac{1}{2} \sum_{ab} \,
  (f_{12} + f_{34} - 2f_{ab})
  \Gamma_{12ab} \Gamma_{ab34} 
  \nonumber \\ \amps{} \qquad\qquad\null \times
  (1 - n_a - n_b) 
  \nonumber \\ \amps{} \null
  + \sum_{ab} \, (n_a - n_b) \nonumber \\
  \amps{} \hspace*{-.25in} \null\times \Bigl\{ \Gamma_{a1b3} \Gamma_{b2a4}
  \bigl[(f_{a1}-f_{b3}) - (f_{b2} - f_{a4})\bigr]
  \nonumber \\ \amps{} \hspace*{-.25in} \null
  - \Gamma_{a2b3}\Gamma_{b1a4}
  \bigl[(f_{a2}-f_{b3}) - (f_{b1} - f_{a4})\bigr] \Bigr\} \,,
  \label{eq:NorderedSRG2}
\eea
where the single-particle indices refer to momentum states and include
spin and isospin labels. 

While the in-medium SRG equations are of second
order in the interactions, the flow equations build up
non-perturbative physics via the successive interference between the
particle-particle and the two particle-hole channels in the SRG
equation for $\Gamma$, Eq.~(\ref{eq:NorderedSRG2}), and between the
two-particle--one-hole and two-hole--one-particle channels for $f$,
Eq.~(\ref{eq:NorderedSRG1}). In terms of diagrams, one can imagine
iterating the SRG equations in increments of $\delta s$. At each
additional increment $\delta s$, the interactions from the previous
step are inserted back into the right side of the SRG equations.
Iterating this procedure, one sees that the SRG accumulates
complicated particle-particle and particle-hole correlations to all orders.

\begin{figure}[th!]
\begin{center}
 \includegraphics[width=2.5in]{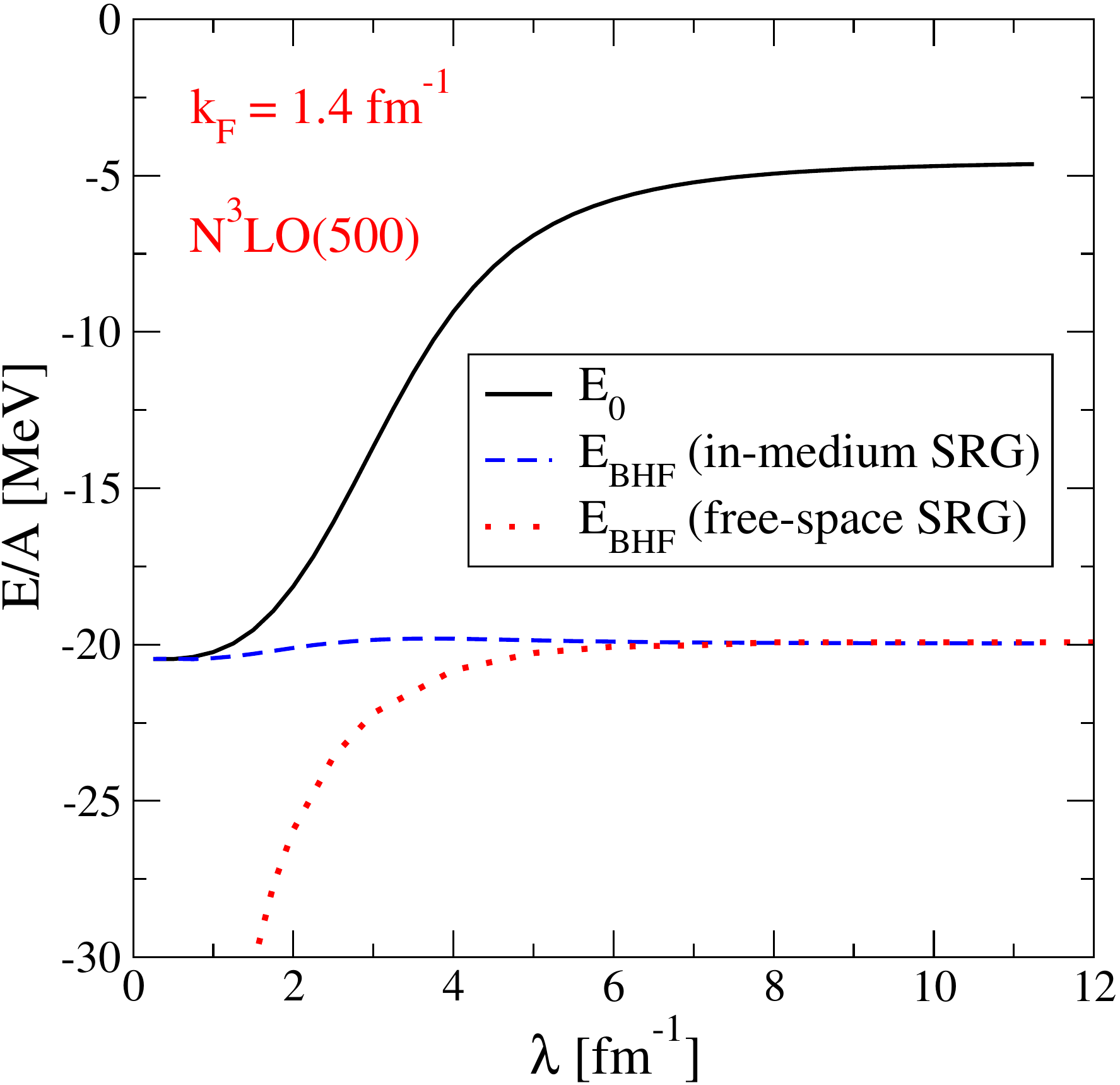}
   \caption{Symmetric nuclear matter energy/particle at $\kf = 1.4\fmi$
   as evolved in the IM-SRG~\cite{Tsukiyama:2010rj}.}
   \label{fig:SNMinmedium}
\end{center}
\end{figure}

\begin{figure}[th!]
\begin{center}
 \includegraphics[width=2.5in]{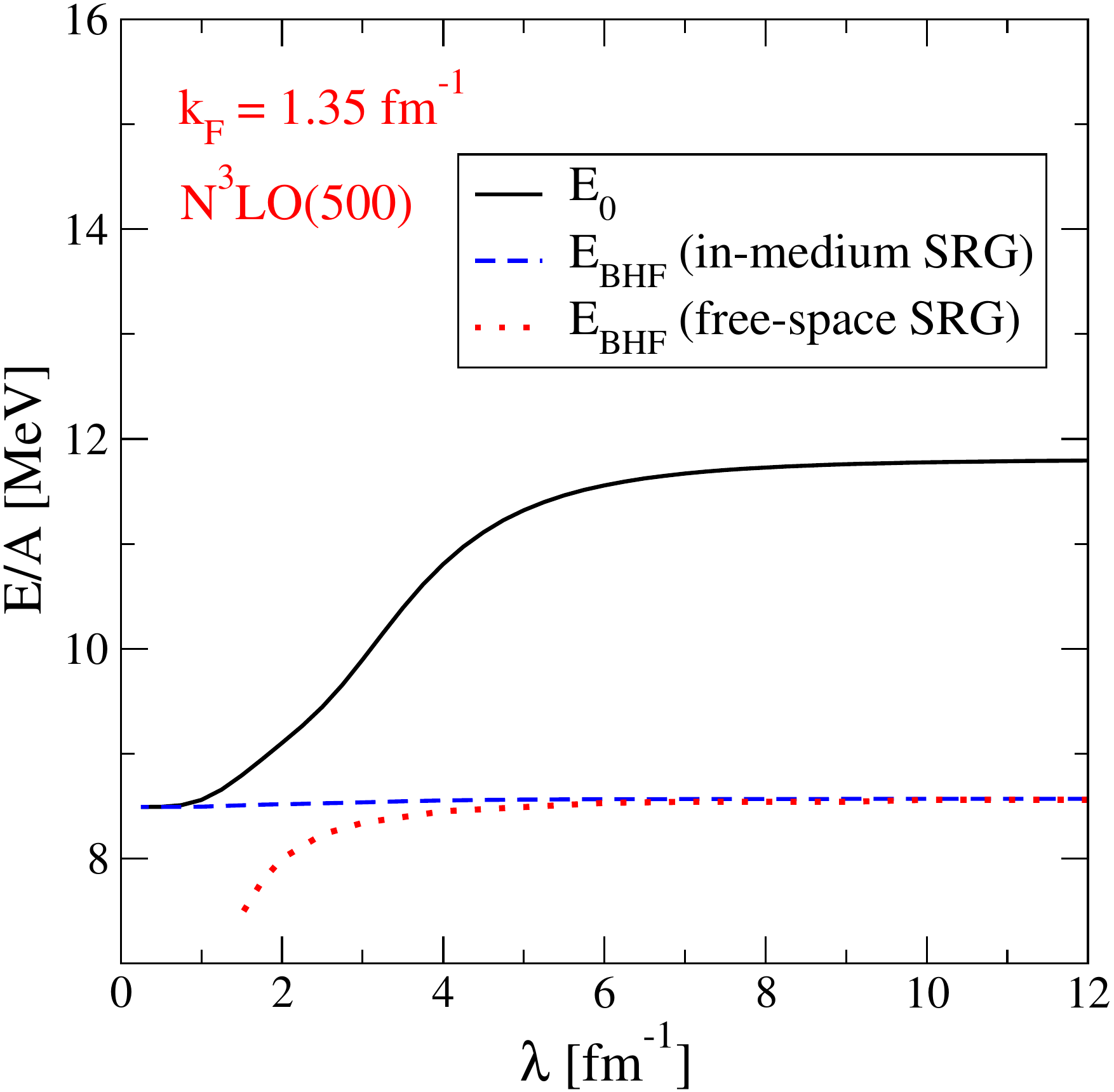}
   \caption{Neutron matter energy/particle at $\kf = 1.35\fmi$
   as evolved in the IM-SRG~\cite{Tsukiyama:2010rj}.}
   \label{fig:PNMinmedium}
\end{center}
\end{figure}

\begin{figure*}[p!]
\begin{center}
 \includegraphics[width=5.3in]{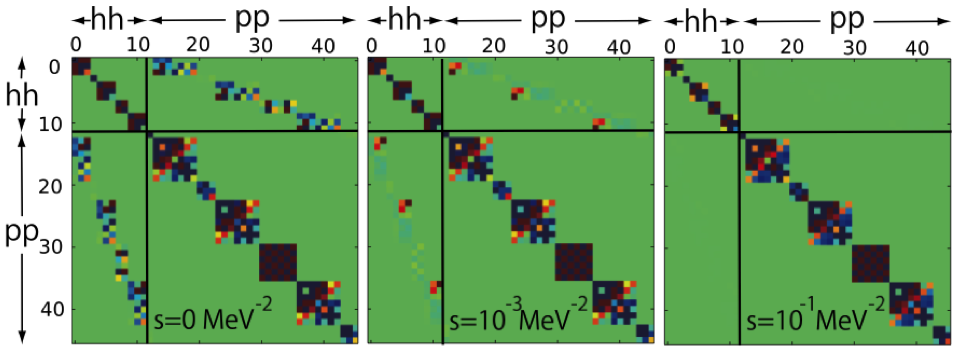}
   \caption{``Off-diagonal'' terms (e.g., $2p2h$ sectors) driven to zero
   as $s$ increases, 
 decoupling them from the Hartree-Fock reference state, which becomes exact as $s\rightarrow \infty$~\cite{Tsukiyama:2010rj}. }
   \label{fig:imsrg1}
 
 \vspace*{.1in}
 \vfill
  
 \includegraphics[width=4.in]{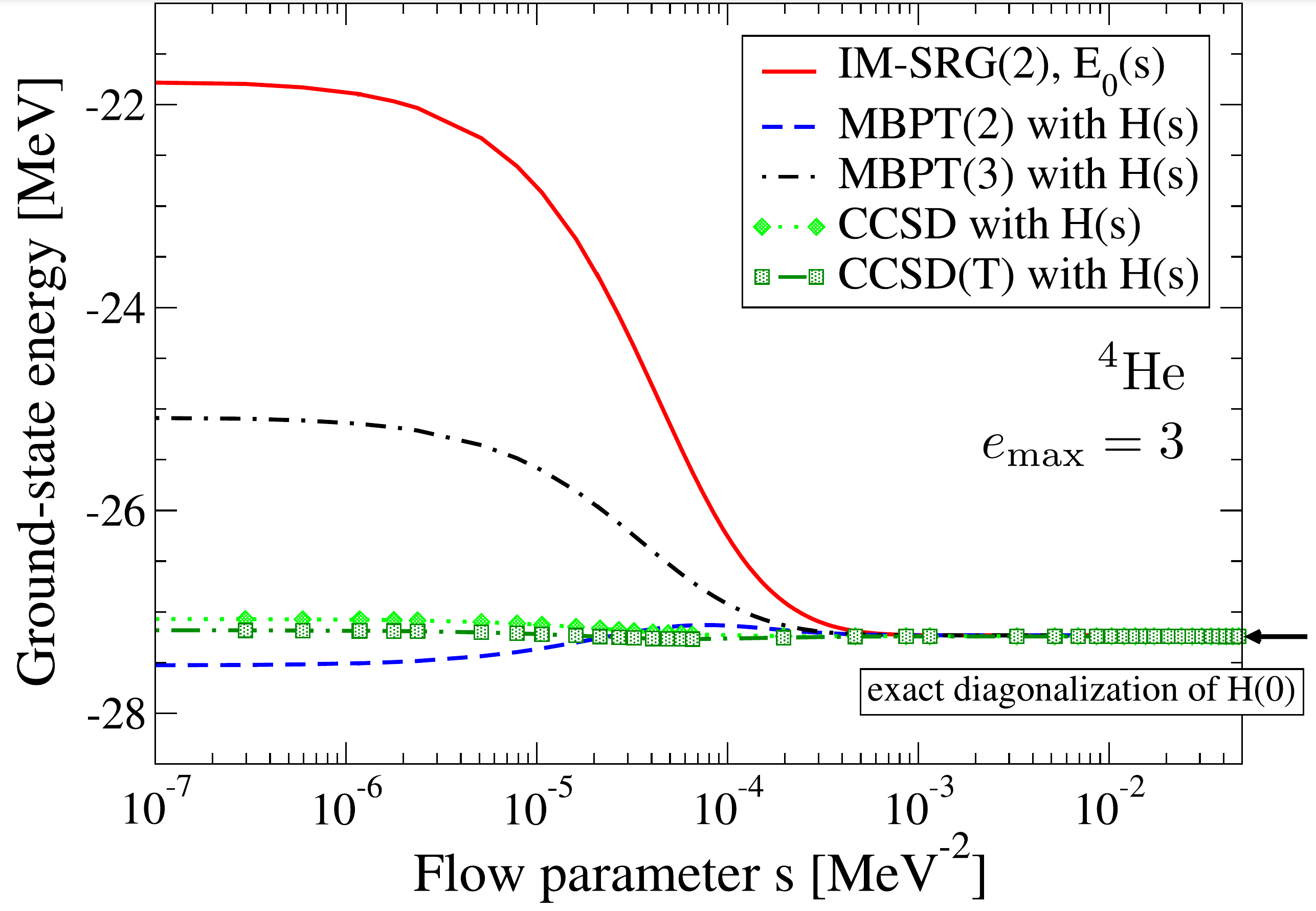}
 \vspace*{-.1in}
   \caption{IM-SRG flow for the energy of the alpha
   particle~\cite{Tsukiyama:2010rj}. }
   \label{fig:imsrg2}
 
 \vspace*{.1in}
 \vfill
 
 \includegraphics[width=5.3in]{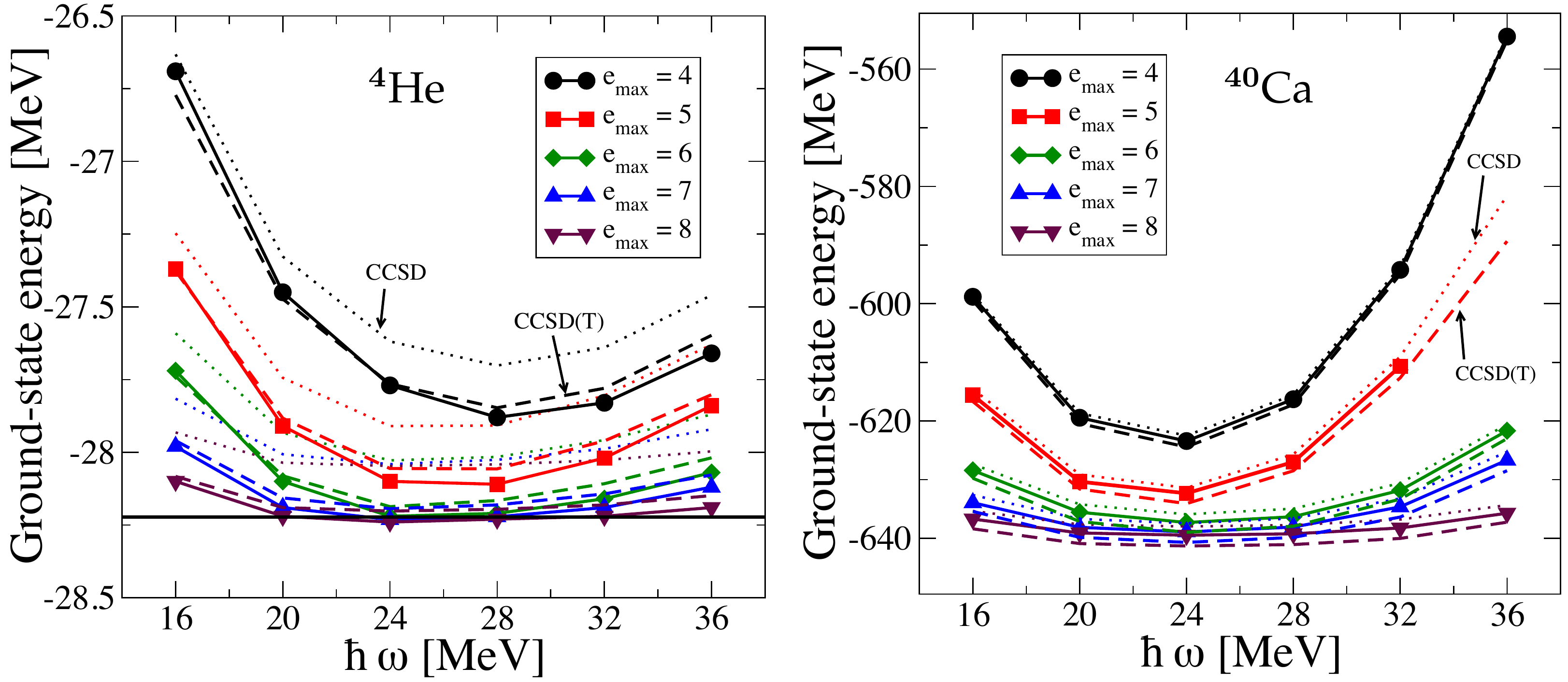}
 \vspace*{-.1in}
   \caption{IM-SRG convergence in finite nuclei compared
   to coupled cluster CCSD and CCSD(T) calculations~\cite{Tsukiyama:2010rj}. }
   \label{fig:imsrg3}

\end{center}
\end{figure*}

With the choice of generator $\eta = [f,\Gamma]$, the Hamiltonian is
driven towards the diagonal. This means that Hartree-Fock becomes
increasingly dominant with the off-diagonal $\Gamma$ matrix elements
being driven to zero. As with the free-space SRG, it is convenient for
momentum-space evolution to switch to the flow variable $\lambda
\equiv s^{-1/4}$, which is a measure of the resulting band-diagonal
width of $\Gamma$. In the limit $\lambda \rightarrow 0$,
Hartree-Fock becomes exact for the evolved Hamiltonian; the zero-body
term, $E_0$, approaches the interacting ground-state energy, $f$
approaches fully dressed single-particle energies, and the remaining
diagonal matrix elements of $\Gamma$ approach a generalization of the
quasiparticle interaction in Landau's theory of Fermi
liquids~\cite{Kehrein:2006}. 

An approximate 
solution of the $E_0$ flow equation for
symmetric nuclear matter and neutron matter as a function of $\lambda$
is shown in Figs.~\ref{fig:SNMinmedium} and \ref{fig:PNMinmedium} for two different Fermi momenta
$k_F$ (corresponding to different densities). As
expected, the in-medium SRG drives the Hamiltonian to a form where
Hartree-Fock becomes exact in the limit $\lambda \rightarrow 0$. In
contrast to the ladder approximation based on NN-only SRG interactions
evolved in free space, the same many-body calculation using
interactions evolved with the in-medium SRG at the two-body level
gives energies that are approximately independent of $\lambda$. This
indicates that truncations based on normal-ordering efficiently
include the dominant induced many-body interactions via the
density-dependent zero-, one-, and two-body normal-ordered terms.

In a similar manner, the in-medium SRG can be used as an ab initio
method for finite nuclei. Figure~\ref{fig:imsrg1} shows the off-diagonal normal-ordered two-body matrix elements $\Gamma$ being
driven to zero with the IM-SRG evolution. Figure~\ref{fig:imsrg2} shows
the evolution of the ground-state energy of
$^4$He~\cite{Tsukiyama:2010rj}. As the flow parameter $s$ increases,
the $E_0$ flow and
second-order (in $\Gamma$) many-body perturbation theory
contributions approach each other, as was the case for the infinite
matter results in Fig.~\ref{fig:SNMinmedium}.  In addition, the
convergence behavior with increasing harmonic-oscillator spaces in
Fig.~\ref{fig:imsrg3} for $^4$He and $^{40}$ Ca is very promising. 
Based on these
calculations, the in-medium SRG truncated at the
normal-ordered two-body level appears to give accuracies comparable to
coupled-cluster calculations truncated at the singles and doubles
(CCSD) level. Finally, we note that the
in-medium SRG is a promising method for non-perturbative calculations
of valence shell-model effective interactions and operators.

\subsection{Implications of RG for nuclear calculations}

There are many on-going and potential applications of low resolution
methods for calculations of nuclear structure and reactions.  In many
cases RG techniques are used explicitly but there are also examples
of low resolution being achieved by other means.  Here we'll survey
some of both types.
This will be far from a comprehensive list because it focuses largely
on developments
associated entirely or in part with a project called UNEDF, which stands for 
Universal Nuclear Energy Density Functional.  

\begin{figure*}[p!]
\begin{center}
 \includegraphics[width=3.9in]{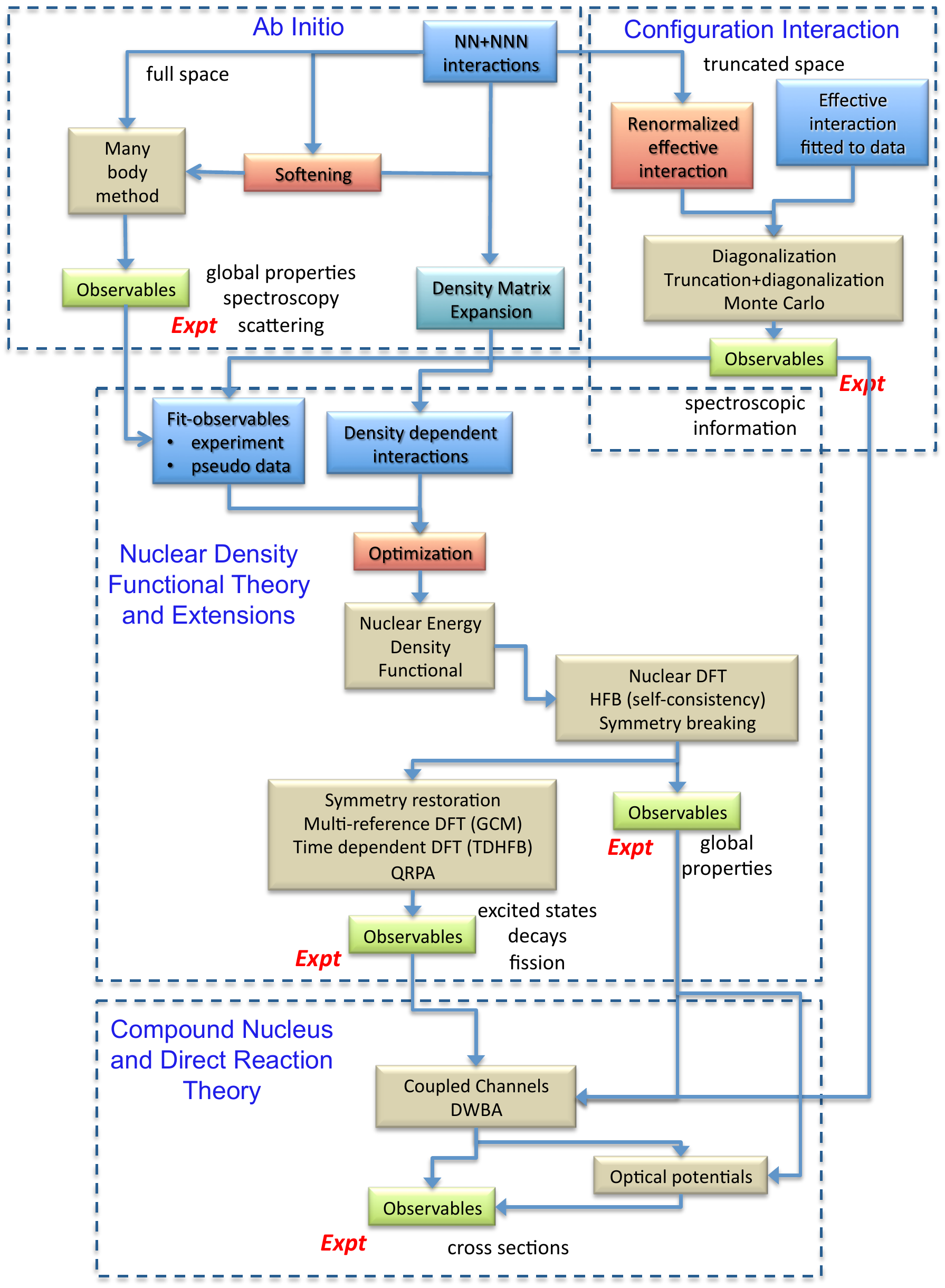}
   \vspace*{-.1in}
   \caption{Strategy diagram for UNEDF.}
   \label{fig:UNEDF}

 \includegraphics[width=2.6in]{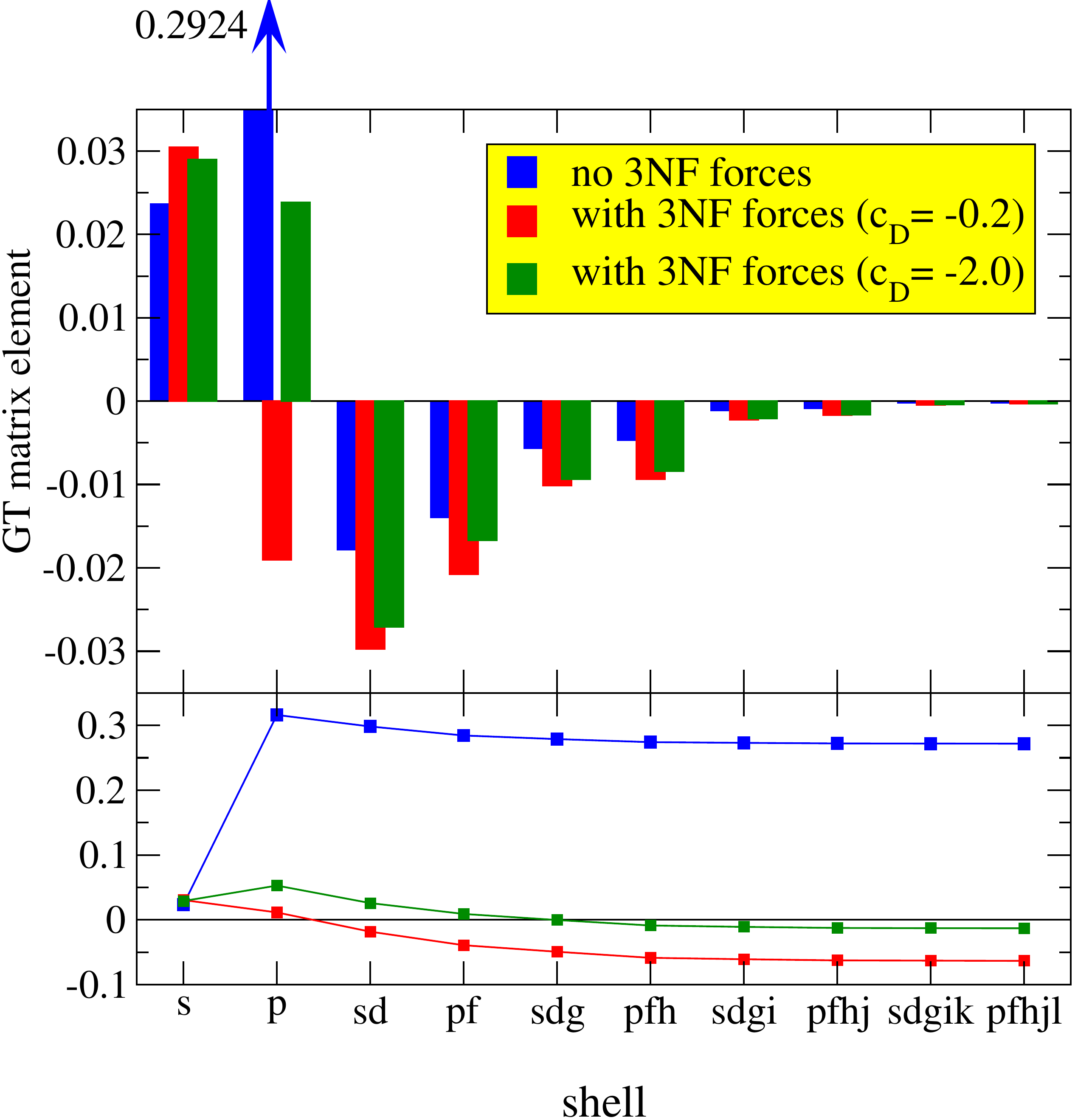}
 \hspace*{.3in}
 \raisebox{.5in}{\includegraphics[width=2.6in]{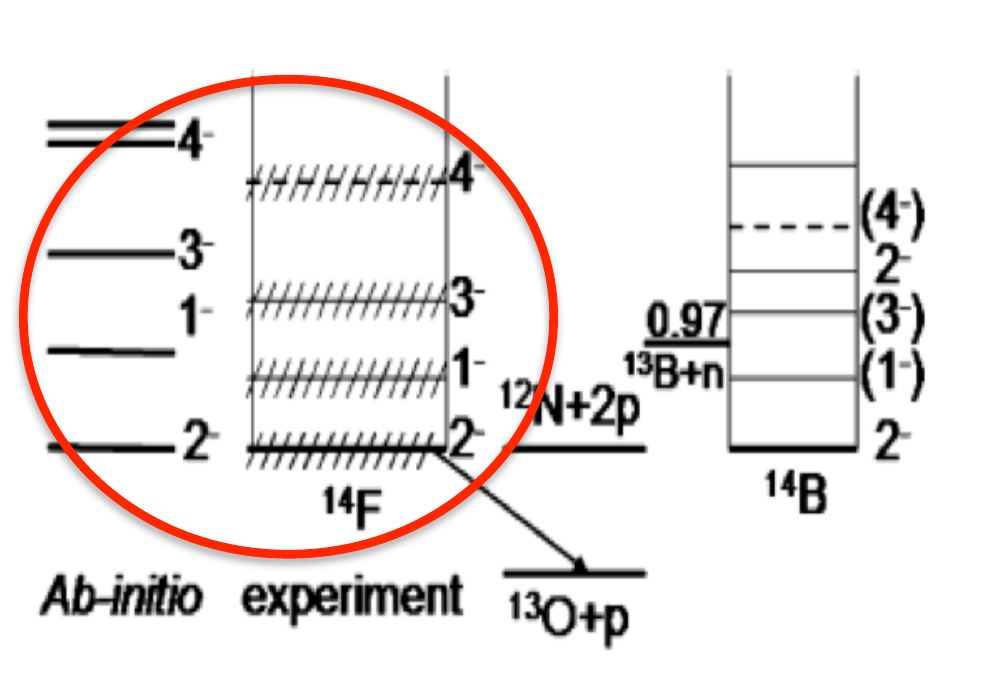}}
 \vspace*{-.1in}
 \caption{Left: Matrix elements from Carbon-14 lifetime calculation~\cite{Maris:2011as}.
 Right: Fluorine-14 spectrum predicted by NCSM and measured
   experimentally\cite{Maris:2009bx,Goldberg:2010zz}. }
   \label{fig:F14C14}
\end{center}
\end{figure*}

UNEDF is a collaboration
of more than fifty physicists, applied mathematicians, and computer
scientists in the United States plus many international
collaborators, funded through the U.S.\ Department of Energy's 
SciDAC program.
The long-term vision of the project is to arrive at a comprehensive and quantitative description of nuclei
and their reactions.  The focused mission is to construct, optimize, validate,
and apply energy density functionals for structure and reactions,
but to carry out this mission the team has developed many 
crosscutting physics collaborations where none existed previously between
the main physics areas: ab initio structure, ab initio
functionals, DFT applications, DFT extensions, reactions.
These interconnections are indicated schematically in the UNEDF
strategy diagram in Fig.~\ref{fig:UNEDF}.
This type of large-scale collaboration
represents a transformation in how low-energy nuclear theory
is done. 
UNEDF has been very productive, with over 200 publications to date, including 11 Physical Review Letters and a Science article in the 2011 calendar year
alone.  Further background, references, and scientific highlights can be found
at unedf.org, the project website.   

One of the important tools for nuclear structure enabled by low-resolution
interactions is the diagonalization of enormous but very sparse 
Hamiltonian matrices, usually in a harmonic
oscillator basis to permit center-of-mass effects to be excluded.  This
is referred to as no-core full configuration (NCFC) or no-core shell
model (NCSM), depending on the context.
Two recent examples of what is enabled are represented in Fig.~\ref{fig:F14C14}. On the left are Gamov-Teller matrix elements from a large-scale calculation
of Carbon-14 using a soft chiral EFT potential~\cite{Maris:2011as}.  
They highlight the critical role of the 3NF in suppressing
the beta decay rate, which explains the anomalously long lifetime
of $^{14}$C (which is used to great advantage for dating artifacts!).
On the right is the low-lying spectrum of Fluorine-14, which is unstable to 
proton decay. This spectrum was predicted in advance of the experimental measurements (not a common occurrence until now!), which meant solving a Hamiltonian matrix of dimension 2 billion using 30,000 cores with a soft interaction (derived from inverse scattering rather than RG, but with similar characteristics)~\cite{Maris:2009bx,Goldberg:2010zz}. The predictions and measurement agree within the combined experimental and theoretical uncertainties. 
These calculations would not be possible with the potentials of Fig.~\ref{fig:phenpots}.
RG-softened interactions will allow many more of these confrontations
of experiment with theory in the future.

%\begin{figure}[th!]
%\begin{center}
%\end{center}
%\end{figure}

\begin{figure}[t!]
\begin{center}
 \includegraphics[width=2.5in]{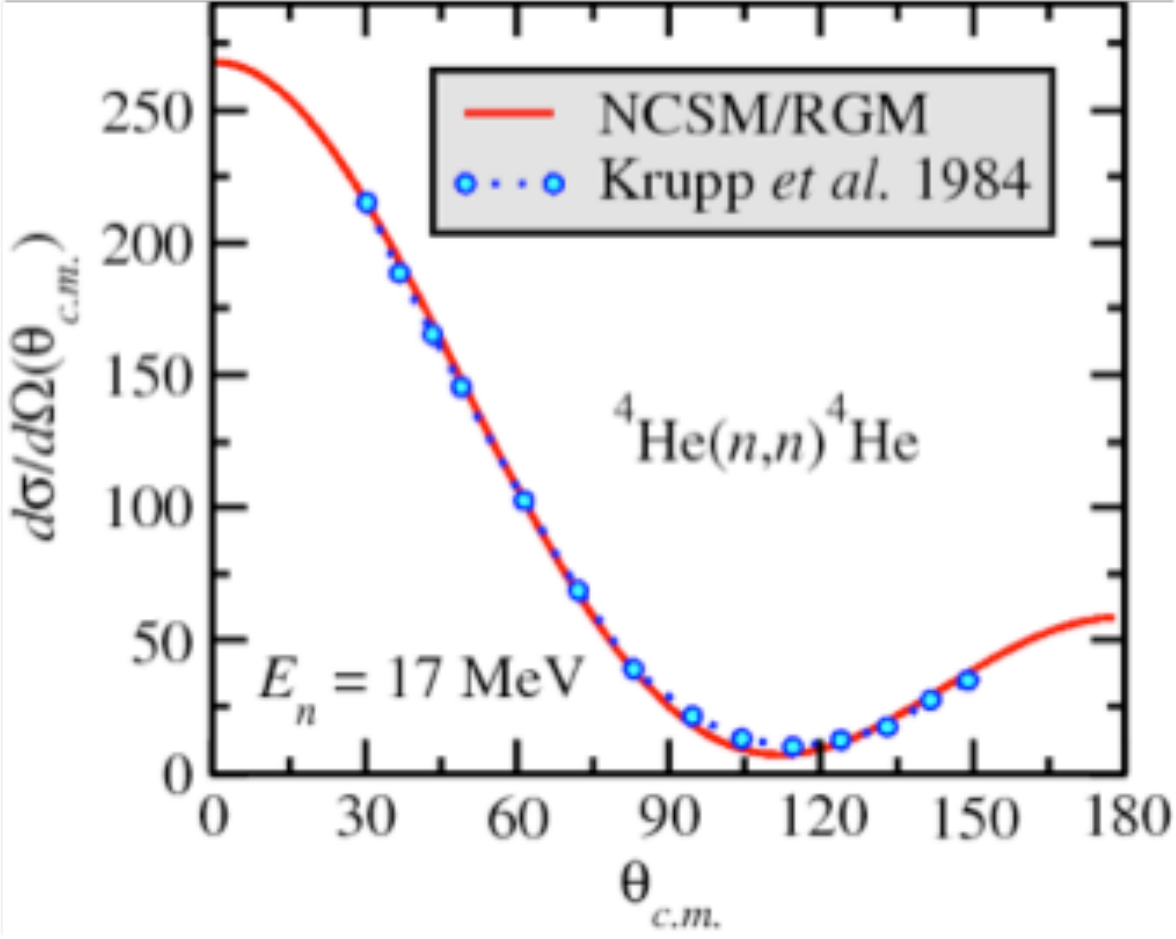}
   \caption{NCSM/RGM calculation of neutron scattering from the alpha particle using SRG-evolved interactions, compared with experimental measurements~\cite{Navratil:2010ey,Navratil:2010jn}.}
   \label{fig:ncsmrgm1}
\end{center}
\end{figure}

\begin{figure}[t!]
\begin{center}
 \includegraphics[width=2.8in]{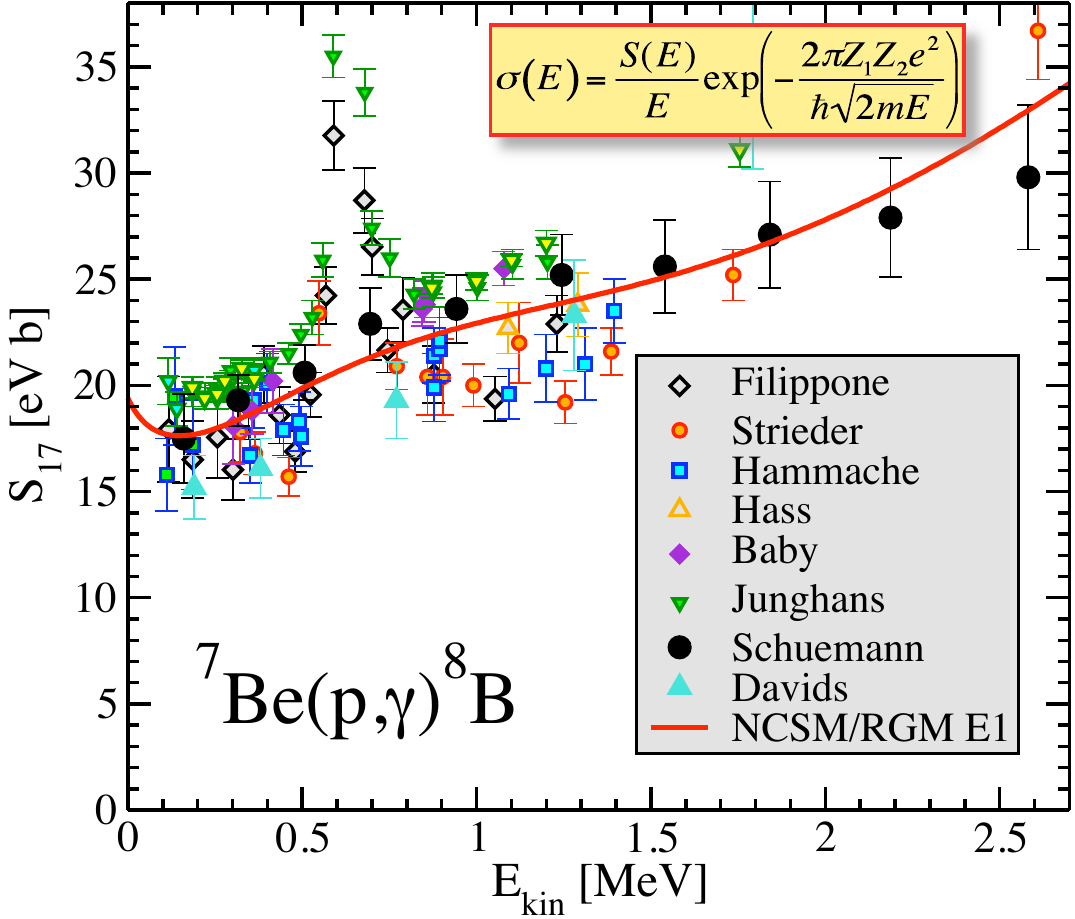}
   \caption{First ever ab initio calculations of $^7$Be$(p,\gamma){}^8$B
   astrophysical S-factor~\cite{Navratil:2011sa}.  Uses SRG-N3LO with $\lambda = 1.86\fmi$.}
   \label{fig:ncsmrgm2}
\end{center}
\end{figure}

One of the principal aims of the UNEDF project is to calculate reliable reaction cross sections for astrophysics, nuclear energy, and national security, for which extensions of standard phenomenology is insufficient. The interplay of structure and reactions is essential for a successful description of exotic nuclei as well. Such interplay is characteristic of the ab initio no-core shell model/resonating-group method (NCSM/RGM), which treats bound and scattering states within a unified framework using fundamental interactions between all nucleons. A quantitative proof-of-principle calculation of 
this approach is shown in Fig.~\ref{fig:ncsmrgm1}~\cite{Navratil:2010ey,Navratil:2010jn}. A wide range of applications is now possible including $^3$H$(d,n){}^4$He fusion and 
the $^7$Be$(p,\gamma){}^8$B reaction important for solar neutrino physics, and many more to come.
In Fig.~\ref{fig:ncsmrgm2}, the first-ever ab-initio calculation
of the $^7$Be$(p,\gamma){}^8$B
  astrophysical S-factor is shown~\cite{Navratil:2011sa}. 
This calculation uses NCSM/RGM with an N$^3$LO NN interaction evolved
by the SRG to a fine-tuned value of $\lambda = 1.86\fmi$.   
The ab initio theory predicts
both the normalization \emph{and} the shape of $S_{17}$.
This very promising techniques has many additional applications.
The inclusion of an SRG-evolved 3NF is planned for the near future.

\begin{figure}[tbh!]
\begin{center}
 \includegraphics[width=2.5in]{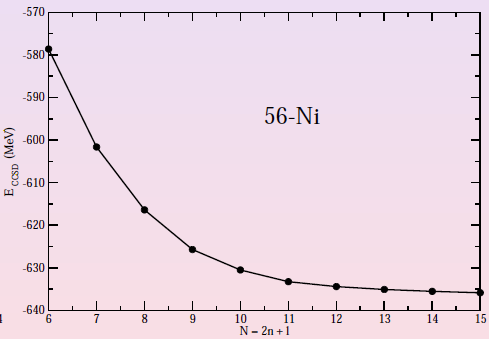}
   \caption{Convergence of CCSD for Ni-56 evolved N3LO to $\lambda = 2.5\fmi$.}
   \label{fig:cc_ni56}
\end{center}
\end{figure}

\begin{figure}[bh!]
\begin{center}
 \includegraphics[width=2.5in]{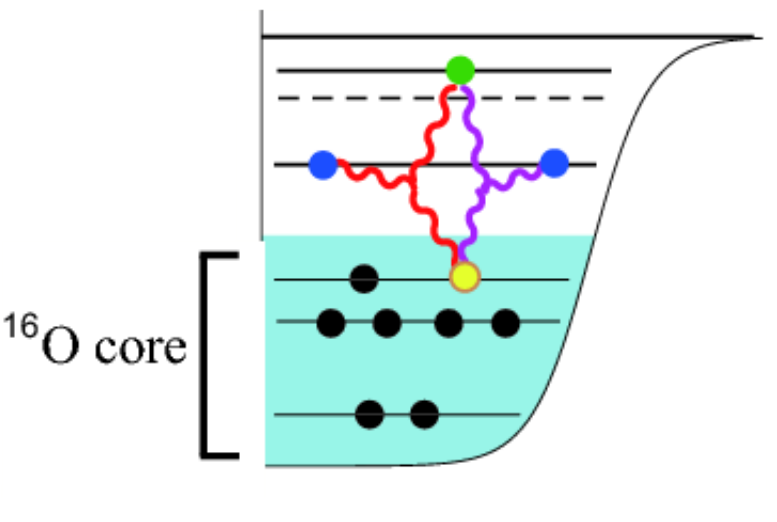}
   \caption{Influence of a three-body force on valence neutrons
   in oxygen isotopes.}
   \label{fig:oxygen_sm2}
\end{center}
\end{figure}

\begin{figure*}[tbh!]
\begin{center}
 \includegraphics[width=4.0in]{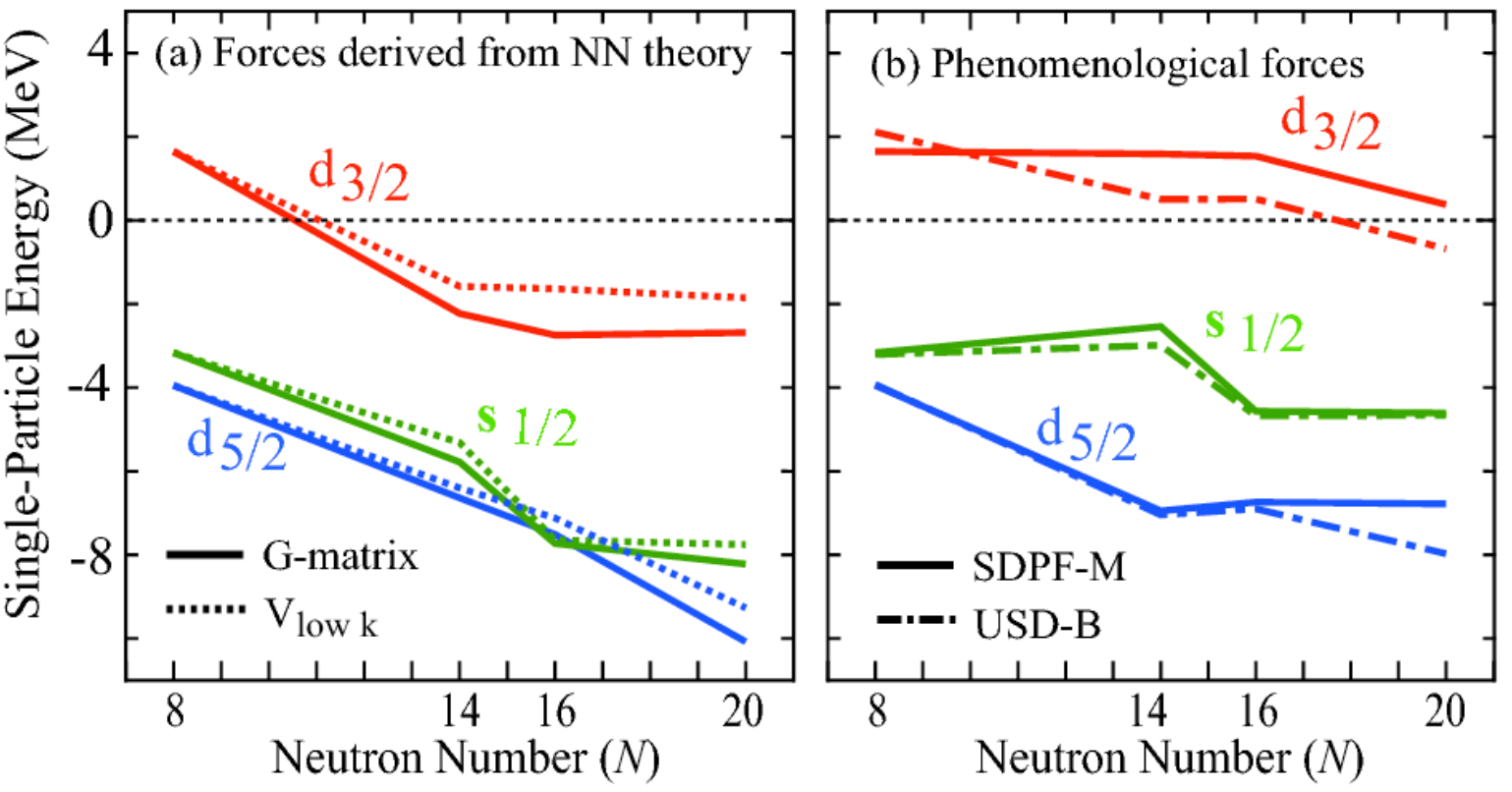}
 \hspace*{.1in}
 \includegraphics[width=2.0in]{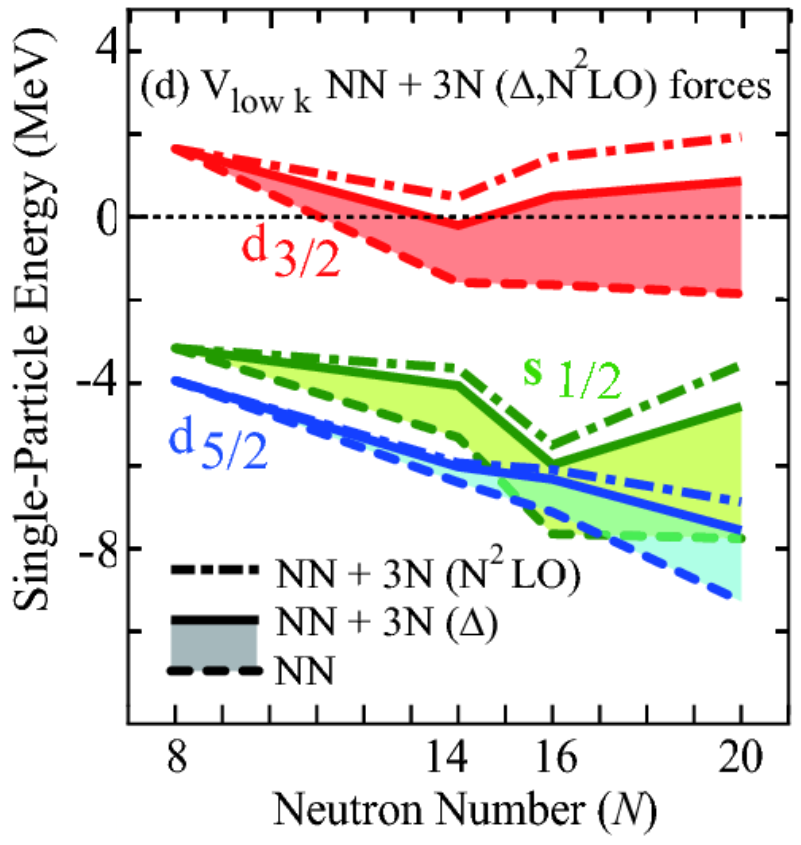}
   \caption{Three-body force impact on oxygen single-particle levels.
   Left: NN-only, middle: phenomenological forces, right: NN + 3NF.}
   \label{fig:oxygen_sm}
\end{center}
\end{figure*}

UNEDF members have demonstrated that the powerful coupled-cluster (CC) ab initio method, which is a workhorse in quantum chemistry, can be used to accurately calculate closed-shell medium-mass nuclei such as $^{40}$C, $^{48}$C, 
$^{56}$Ni with chiral EFT two-body interactions or the RG-softened versions~\cite{Hagen:2008iw,Hagen:2010gd}, as well as proton halo nuclei like $^{19}$F~\cite{Hagen:2010zz}. 
A proof-of-principle convergence curve (ground-state energy versus the
size of the orbital space) for $^{56}$Ni 
with an SRG-evolved potential is shown in
Fig.~\ref{fig:cc_ni56}.
The CC formalism has been extended to include NNN forces and their inclusion in calculations of the heavier nuclei will break new barriers. 

The in-medium SRG diagonalization of closed-shell nuclei such as
$^{40}$C~\cite{Tsukiyama:2010rj}, discussed in Section~\ref{subsec:imsrg} is a complementary approach to CC and is one of several advances
in our understanding of the phenomenological nuclear shell model enabled
by softened potentials.
Another is the direct use of MBPT to examine the effect of the 3NF
on the location of the neutron dripline: the limits of nuclear existence
where an added neutron is no longer bound---it ``drips'' away.
The new physics is indicated schematically
in Fig.~\ref{fig:oxygen_sm2}.  
It's been established experimentally that as you add neutrons to
stable oxygen-16, the neutrons stay bound until $^{24}$O.  But adding
one more proton to get fluorine extends the dripline all the
way to $^{31}$F.  This result is \emph{not} predicted by previous
microscopic calculations using NN interactions, because the single-particle
neutron energy levels that get filled are predicted to be bound
(leftmost panel in Fig.~\ref{fig:oxygen_sm}), leading to
$^{28}$O as the calculated dripline.
The phenomenological shell model, in which matrix elements of the
Hamiltonian are fit to nearby nuclei, has a very different pattern
(e.g., compare the $d3/2$ single-particle energies in the middle
and left panels of Fig.~\ref{fig:oxygen_sm}) 
and predicts the correct dripline.  However, recent calculations with a $\vlowk$ RG force and fitted 3NF yield the right panel of Fig.~\ref{fig:oxygen_sm}.  When the 3NF effect is added~\cite{Otsuka:2009cs}, the interaction of
valence neutrons with a core neutron, as in Fig.~\ref{fig:oxygen_sm2}, is repulsive, pushing up the
$d3/2$ level so that the dripline is at $^{24}$O.

\begin{figure}[b!]
\begin{center}
 \includegraphics[width=3.0in]{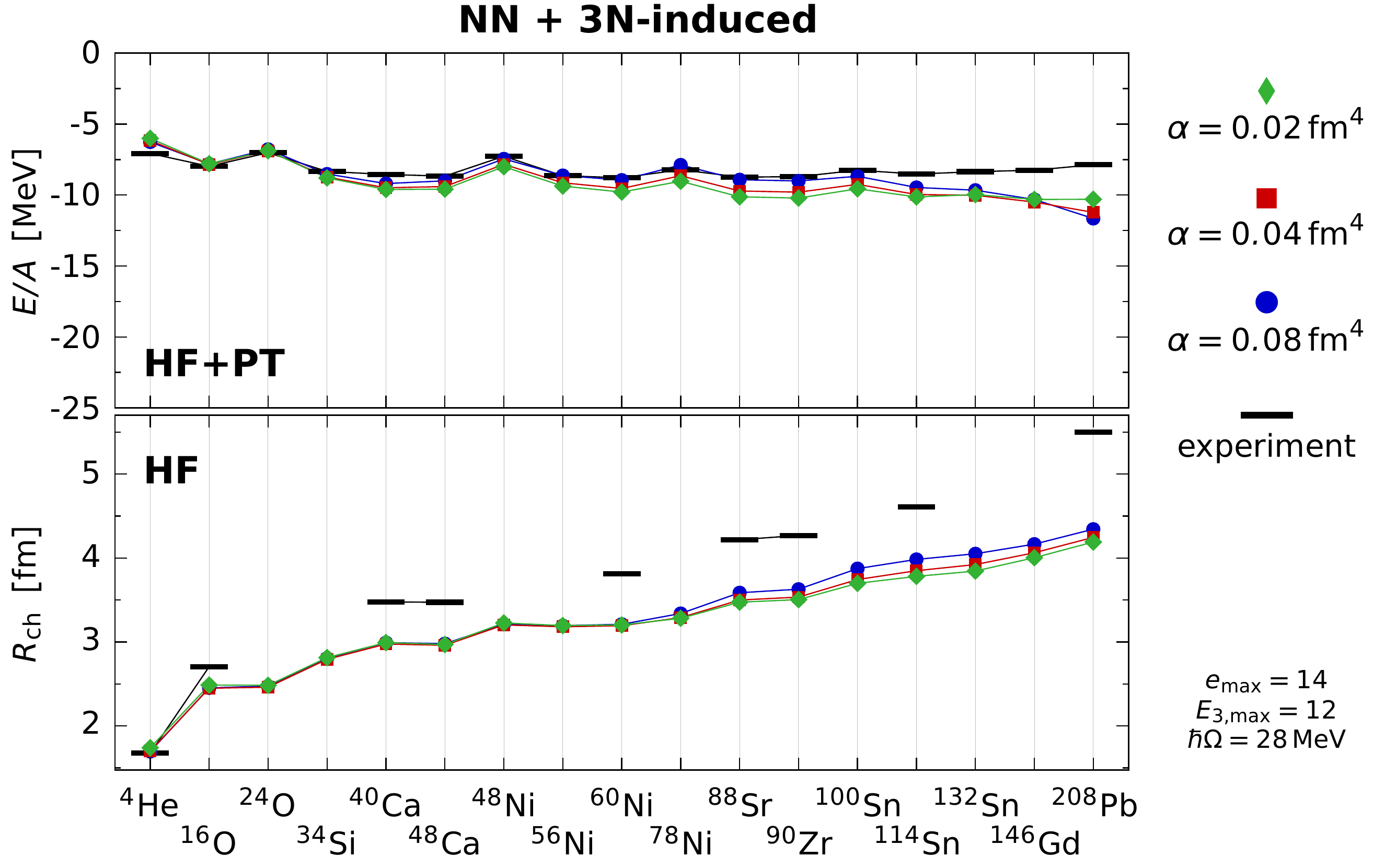}
   \caption{Second-order MBPT applied to closed shell nuclei
   with SRG-evolved NN interactions, including the induced 3NF~\cite{Roth:2009private}.  
   Several different values of the flow parameter are shown
   (note that $\alpha$ here is the same as $s=1/\lambda^4$).}
   \label{fig:rothmbpt}
\end{center}
\end{figure}

The apparent success of MBPT with low-momentum potentials has been
tested by Roth and collaborators~\cite{Roth:2009private}, who have 
done calculations of closed shell nuclei across the mass table
in second-order perturbation theory (first-order is Hartree-Fock).
Results for SRG evolved interactions from an initial NN chiral EFT
and including the induced 3NF show excellent independence of the
flow parameter $\lambda$ (see Fig.~\ref{fig:rothmbpt}) and both the energies
and radii are in good agreement with coupled cluster results~\cite{Hagen:2008iw,Hagen:2010gd}.
However, adding a 3NF leads to large $\lambda$ dependence.
Very recent results from Roth et al.\ that use importance truncated NCSM as well as coupled cluster calculations show that the long-range 3NF is the source  to apparent large 4NF contributions for oxygen and heavier nuclei, causing
a strong dependence on the flow parameter.  
However, by using a lower cutoff for the initial 3NF, remarkable agreement with experimental binding energies is achieved with fits only to
few-body properties~\cite{Roth:2011ar,Roth:2011vt}.
Work is in progress to identify SRG generators to better control the
RG evolution of the initial 3NF.

\begin{figure}[th!]
\begin{center}
 \includegraphics[width=2.5in]{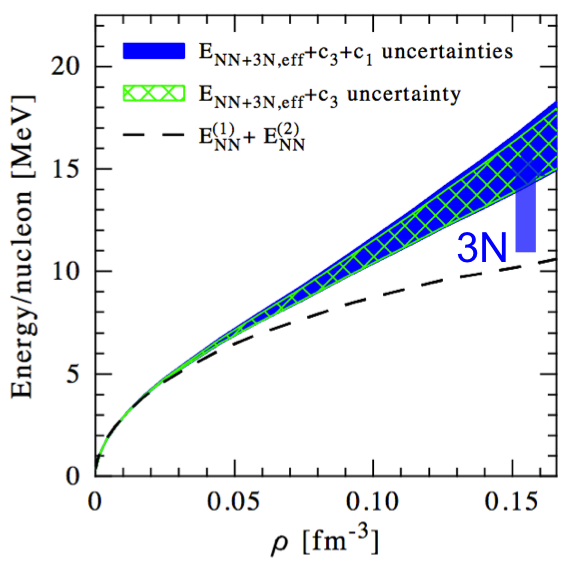}
  \vspace*{-.1in}
   \caption{Neutron matter energy/particle versus density calculated using a $\vlowk$ RG-evolved NN interaction plus fit 3NF.  The arrow shows the importance of the 3NF and the width of the band indicates the uncertainties due to 3NF couplings~\cite{Hebeler:2009iv}.}
   \label{fig:PNMachim1}
\end{center}
\end{figure}

\begin{figure}[th!]
\begin{center}
 \includegraphics[width=2.5in]{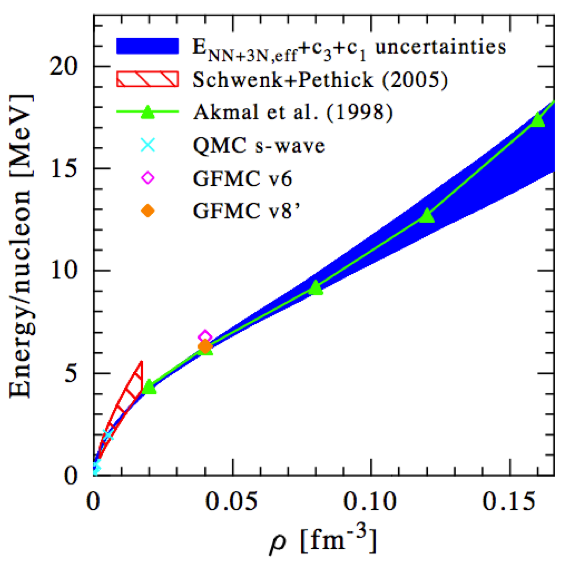}
  \vspace*{-.1in}
   \caption{Neutron matter energy/particle versus density as in Fig.~\ref{fig:PNMachim1} compared to other calculations~\cite{Hebeler:2009iv}.}
   \label{fig:PNMachim2}
\end{center}
\end{figure}

We have already seen the convergence of MBPT for symmetric
infinite nuclear matter.  Perturbation theory is even more
controlled for pure neutron matter, as illustrated in
Figs.~\ref{fig:PNMachim1} and \ref{fig:PNMachim2}, where 
$\vlowk$ RG-evolved interactions and fitted 3NF's are
used in calculations of the neutron matter energy per nucleon
as a function of the density~\cite{Hebeler:2009iv}.  
The cutoff dependence of the result is used to estimate the many-body uncertainty.
Figure~\ref{fig:PNMachim1}
shows that the 3NF contribution is important (compare to
the NN-only curves) but that the dominant theoretical uncertainty 
is the value of the coupling constants for the long-range part
of the N$^2$LO chiral EFT 3NF.
In Fig.~\ref{fig:PNMachim2}, comparisons with non-perturbative 
calculations demonstrate the consistency of the much easier
MBPT calculations.  The results for the neutron matter equation
of state have been used by Hebeler et al.\ to provide tight constraints
on neutron star masses and radii~\cite{Hebeler:2010jx}. 

The MBPT results for infinite matter are also valuable input for
work to develop a microscopic nuclear energy density functional (EDF). 
A key tool to incorporate microscopic input into nuclear
EDF's is the density matrix expansion (DME) originally proposed
by Negele and Vautherin, which has been revived
and improved in the UNEDF project.
The DME provides a route to an EDF based on microscopic nuclear
interactions through a quasi-local expansion of the energy in
terms of various local densities and currents, including resummations that can
treat long-range one- and two-pion exchange interactions given by chiral EFT.  
With sufficiently soft microscopic interactions,
many-body perturbation theory (MBPT) for nuclei is a quantitative
framework for implementing the DME.  
The formal development of DME with MBPT is on-going, 
but there are already hybrid formulations between purely ab initio and
phenomenological functionals~\cite{Stoitsov:2010ha}, 
which allow improvements to be made
while more systematic functionals are developed.  

\begin{figure}[t!]
\begin{center}
 \includegraphics[width=1.7in]{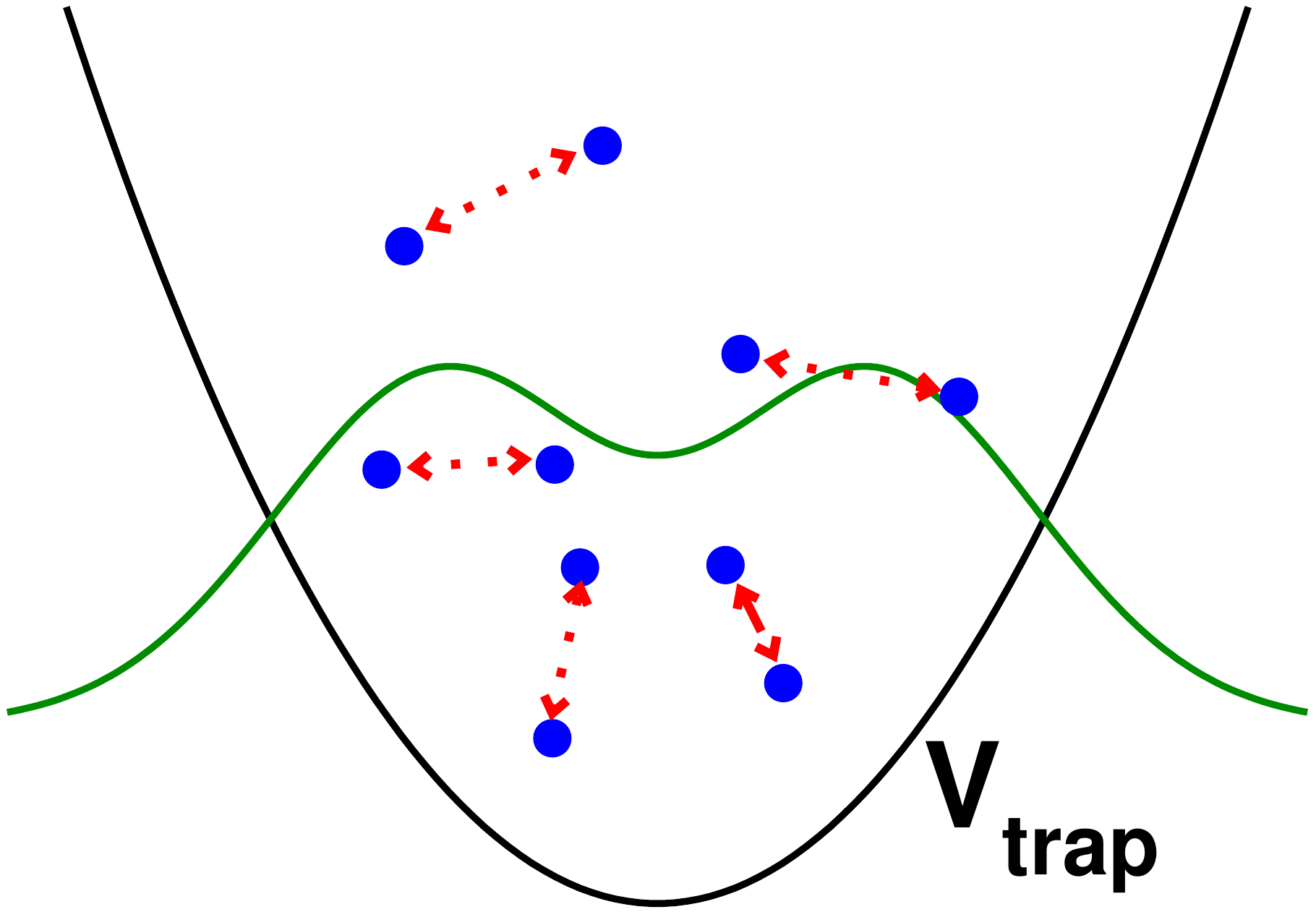}
   \caption{Schematic representation of interacting neutrons in a
   (theoretical) trapping potential.}
   \label{fig:KS}
\end{center}
\end{figure}

\begin{figure}[b!]
\begin{center}
\includegraphics[width=2.7in]{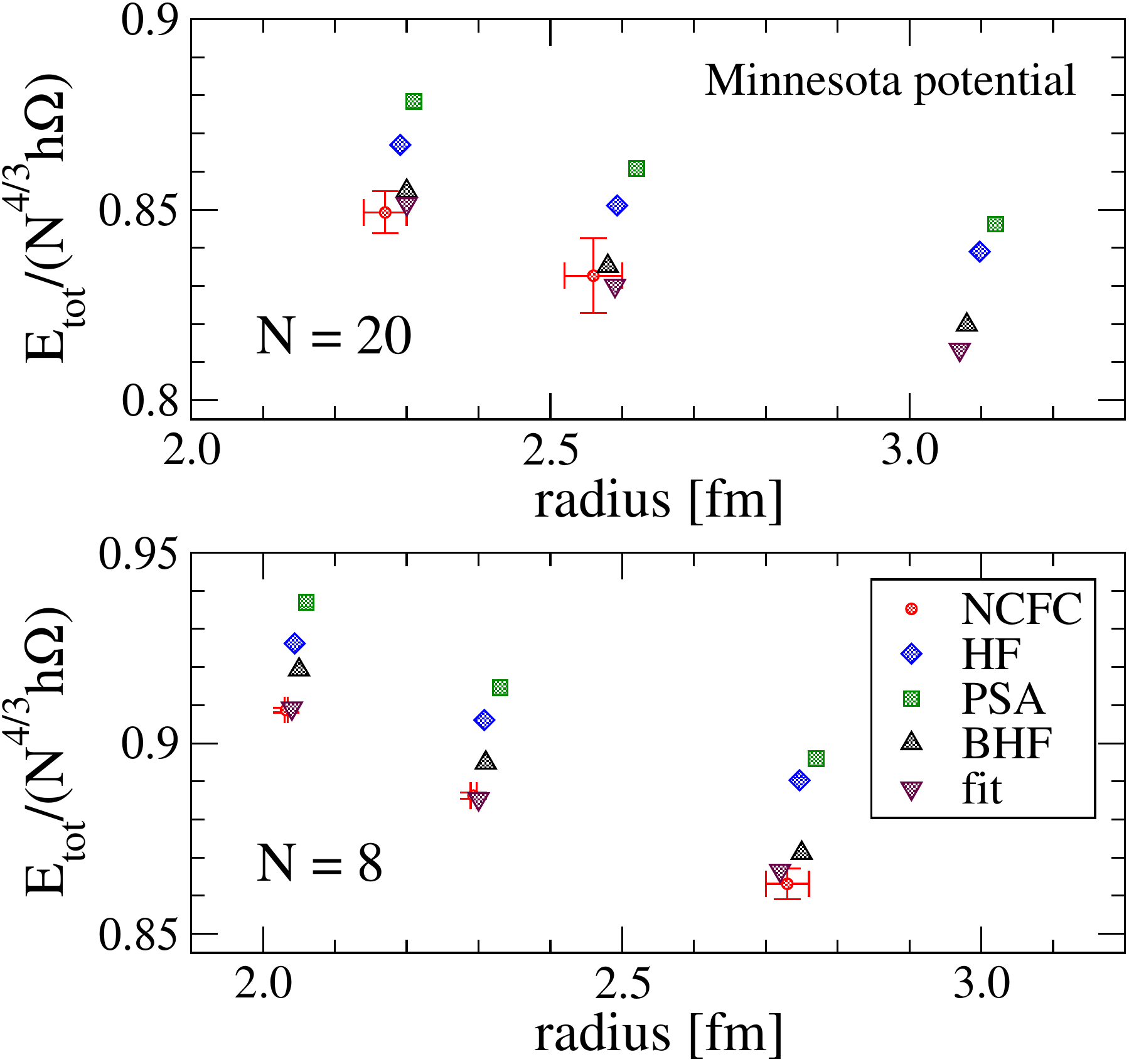}
   \caption{Calculations of the ground-state energies and radii
   of trapped systems of twenty (top) and eight (bottom) neutron drops~\cite{Bogner:2011kp}.
   The different clusters of points are for different harmonic oscillator
   trap frequencies. The NCFC results are exact within the error bars.
   They are compared to various approximate microscopic functions,
   with the BHF and ``fit'' calculations expected to be
    the best approximations.}
   \label{fig:neutron-drops}
\end{center}
\end{figure}

Several DME implementations strategies have been developed, with 
the first tests recently made
against ab-initio calculations using a semi-realistic
interaction (Minnesota) in trapped neutron drops~\cite{Bogner:2011kp}.
Neutron drops are a powerful theoretical laboratory for improving existing
nuclear energy functionals, with particular value in providing 
microscopic input needed for neutron-rich nuclei, where there
are fewer constraints from experiment.
The necessity of an external potential (because the untrapped system
is unbound, with positive pressure) is turned into a virtue by
allowing external control over the environment
(see Fig.~\ref{fig:KS} for a schematic of the trapped neutrons).
Density functional theory, which provides the theoretical underpinning
for the microscopic EDF's, dictates that the same functional applies
for any external potential, which can therefore be varied to
probe and isolate different aspects of the EDF.
Results summarized in Fig.~\ref{fig:neutron-drops} show promising agreement
with the DME functionals and the (essentially exact) ab initio
results using NCFC.
Many more developments in this line will be forthcoming.

\subsection{Summary and survey of open problems}

In these lectures, we've made a whirlwind tool of
atomic nuclei at low resolution.  With the renormalization
group (RG), the strategy has been to lower the resolution 
and track dependence on it.  We've seen how high resolution
leads to coupling of low momenta to high momenta, which hinders solutions
of the many-body problem for low-energy properties.  
With RG evolution, correlations  
in wave functions are reduced dramatically, leading to faster
convergence of many-body methods.  A consequence is that non-local
potentials and many-body operators are induced, so these must
be accommodated. 

Flow equations (SRG) achieve low resolution by \emph{decoupling}.
This can be in the form of band or block diagonalization 
of the Hamiltonian matrix.  The flow equations implement a
series of unitary
transformations, in which observables (measurable quantities)
are not altered but the physics interpretation can change!
In the nuclear case, the usual plan is to evolve until
few-body forces start to grow rapidly, or to use an in-medium
version of the SRG.

With the RG, cutoff dependence becomes a new tool in low-energy
nuclear physics. 
The basic idea is that, in principle, observables should be unchanged
with RG evolution.  In practice, there are approximations in the
RG implementation and in calculating nuclear observables.
These come from  truncation or approximation of ``induced'' many-body    
forces/operators and from many-body approximations.  For nuclei
there can be dramatic changes even with apparently
small changes in the resolution scale.  We can use
these changes as diagnostics of approximations and to estimate
theoretical errors.
Some specific applications of cutoff dependence include:
   \bi
   \I using cutoff dependence at different orders in an EFT expansion,
    which carries over to the corresponding RG-evolved interactions;
   \I using the running of ground-state energies with cutoff in
   few-body systems to estimate errors and identify
   correlations (e.g., the Tjon line);
     
   \I in nuclear matter calculations, validating MBPT convergence
   and setting lower bounds on the errors from 
   uncertainties in many-body interactions;
   
   \I in calculations of finite nuclei,
     diagnosing missing many-body forces;
   
   \I identifying and characterizing scheme-dependent observables,
   such as spectroscopic factors.
  \ei   
The possibilities have really only begun to be explored.

We have seen glimpses of the
many promising applications of RG methods to nuclei.
Configuration interaction and coupled cluster approaches
using softened interactions
converge faster, opening up new possibilities and allowing the
limits of computational feasibility to be extended.
Ground-breaking ab-initio reaction calculations are now possible.
Applications of low-momentum interactions
to microscopic shell model calculations bring new understanding
to phenomenological results, highlighting the role of three-body
forces.  Because many-body perturbation theory (MBPT) is feasible with
the evolved interactions,
the door is opened to constructive nuclear density functional
theory.

There are also many open questions and difficult problems in applying
RG to low-energy nuclear physics.  Here is a subset:
  \bi
    \I Power counting for evolved many-body operators.
    That is, how do we anticipate the size of contributions from
    induced many-body interactions and other operators?
    This is essential if we are to have reliable estimates of
    theoretical errors, because truncations are unavoidable.
    We need both analytic estimates to guide us as
    well as more extensive numerical tests.
    Many of the same issues apply to chiral EFT; can the additional
    information available from SRG flow parameter dependence help with
    analyzing or even constructing EFT's?
    
    \I Only a few possibilities for SRG generators have been
    considered for nuclear systems.
    Can other choices for the SRG $G_s$ operator help to control
    the growth of many-body forces?  Can convergence be improved in the 
    harmonic oscillator basis, which is limited by an infrared
    cutoff as well as an ultraviolet cutoff? 
    Can a generator be found to drive non-local potentials to
    local form, so they can be used with quantum Monte Carlo methods?
    Or can the SRG equations be formulated to directly produce
    a local projection and a perturbative residual interaction?	
	
    \I An apparent close connection between the block-diagonal
    generator SRG and the ``standard'' $\vlowk$ RG has been established
    empirically, but a formal demonstration of the connection and
    its limits has not been made.	
	
    \I What other bases for SRG evolution would be advantageous?
     The need for a momentum-space implementation for evolution
     in the $A=3$ space and beyond is foremost.  Beyond providing
     necessary checks of evolution in the harmonic oscillator
     basis, the evolved interactions in this form could be directly
     applied to test MBPT in infinite matter and to test nuclear
     scaling.  Another possibility is to use hyperspherical coordinates,
     which combine the advantage of a discrete basis with better
     asymptotic behavior
     (and which would be useful for visualization of many-body forces). 
     
    \I There are many open questions and problems involving operators.
    These include formal issues such as the scaling of many-body
    operators and technical issues such as how to handle boosts of
    operators that are not galilean invariant. And there are simply
    many applications that are yet to be made (e.g., electroweak
    processes).
 
    \I The flow to universal form exhibited by two-body interactions
    has been clear from the beginning of RG applications to nuclei,
    but whether this same behavior is expected for many-body interactions
    or for other operators is still open.
	
    \I How can we use more of the power of the RG?

  \ei
There is no shortage of opportunities and challenges!

%%%%%%%%%%%%%%%%%%%%%%%%%%%%%%%%%%%%%%%%%%%%%%%%%%%%%%%%%%%%%%%%%%%%%%%%
%%%%%%%%%%%%%%%%%%%%%%%%%%%%%%%%%%%%%%%%%%%%%%%%%%%%%%%%%%%%%%%%%%%%%%%%



%% References

%\nocite{*}
\bibliographystyle{elsarticle-num-names}
\bibliography{vlowk_refs}






% \end{thebibliography}

\end{document}